\documentclass[A4paper,preprint,12pt,tightenlines,amsmath,amssymb,floatfix,nofootinbib,aps,raggedbottom]{revtex4-2}

\usepackage{graphicx}
\usepackage{bm}
\usepackage{url,multirow,xcolor}
\usepackage[font=footnotesize,margin=24pt,justification=centerlast]{caption}

\begin{document}

\title{ Effects of perturbation for transition operator of double-$\bm{\beta}$ decay on nuclear matrix element, effective axial-vector current coupling, and half-life \vspace{5pt} }

\author{J.\ Terasaki \\ \vspace{0pt}}
\affiliation{ Institute of Experimental and Applied Physics\hbox{,} Czech Technical University in Prague, Husova 240/5, 110\hspace{3pt}00 Prague 1, Czech Republic}

\author{O.\ Civitarese \\ \vspace{0pt}}
\affiliation{ \raisebox{0pt}{Departamento de F\'{i}sica, Universidad Nacional de La Plata, 49 y 115. C.C. 67 (1900),} La Plata, Argentina \\
and \raisebox{0pt}{IFLP-CONICET, diag 115 y 64. La Plata, Provicia de Buenos Aires, Argentina}\vspace{10pt}}


\begin{abstract} 
We calculate the nuclear matrix element (NME), effective axial-vector current coupling $g_A^\mathrm{eff}$, and half-life of the double-$\beta$ ($\beta\beta$) decay using the transition operator perturbed by the nuclear interaction. The correction terms for the NME are obtained by extending the hadron sector to a higher order in terms of the Rayleigh-Schrödinger perturbation theory. The NME calculations are performed for the neutrinoless $\beta\beta$ ($0\nu\beta\beta$) and the two-neutrino $\beta\beta$ ($2\nu\beta\beta$) decays of $^{136}$Xe. The nuclear wave functions are calculated by the quasiparticle random-phase approximation (QRPA) with the Skyrme, the Coulomb, and the contact pairing interactions. Sufficiently large single-particle valence spaces are used. The correction terms for the NME are comparable with the leading term in absolute value, and the sum of the corrections has the opposite sign to that of the leading term. The $g_A^\mathrm{eff}$’s for the $\beta\beta$ NME are calculated by a few methods depending on the truncation of the NME and the half-life referred to. Similarities are found between some of these $g_A^\mathrm{eff}$’s including those of the $0\nu\beta\beta$ NME. This leads to the conclusion that the value of $g_A^\mathrm{eff}$ can indeed be determined by the perturbed transition operator. It is in a comparable range of the $g_A$ for the $2\nu\beta\beta$ NME. The perturbation effect on the $2\nu\beta\beta$ half-life is discussed by comparing the calculated half-lives with the different $g_A$’s and the NME components. 
\end{abstract}

%
\maketitle
\newpage
\section{\label{sec:introduction}Introduction}
More than forty years have passed since the calculation of the nuclear matrix element (NME) of the neutrinoless double-$\beta$ ($0\nu\beta\beta$) decay started. The candidate nuclei of this decay are well established; they are distributed in the range of $A$ = 50$‒$150. The experimental search is vigorously conducted in more than 30 projects, including those at the planning stage \cite{Neu24}. The development of the experiments is extraordinary; the two-neutrino double-$\beta$ ($2\nu\beta\beta$) decay was observed for all the candidate nuclei of the $0\nu\beta\beta$ decay used in the experiments \cite{Bar19}. The $2\nu\beta\beta$ decay is one of the rarest observed decays in physics. 
The $0\nu\beta\beta$ decay is the nuclear phenomenon of the spontaneous change of two neutrons to two protons emitting two electrons \cite{Fur39}. This decay is caused by the weak interaction, but no antineutrino is emitted in contrast to the usual $\beta$ decay. The intermediate state between the first and second emissions of the electrons is a virtual state. This process breaks the lepton number conservation, which is for now preserved. If this conservation is broken, it is certainly an epoch-making discovery. In standard ideas, this decay is possible if and only if the Majorana neutrino exists \cite{Doi85}. This exotic fermion is self-conjugate and can mediate an interaction between nucleons. 

There are two more physical significances of the $0\nu\beta\beta$ decay. There is a theory to explain the matter-antimatter imbalance in the universe using the Majorana and the right-handed neutrinos (the seesaw mechanism) \cite{Fuk86}. Finding the cause of this imbalance is one of the most major problems of particle physics and astrophysics. If the existence of the Majorana neutrino is proven, it would give a great impetus to the studies of the matter-antimatter problem. 

The other significance is related to the neutrino mass. In the early days, the neutrino was thought to be either massless or very lightly massive. It was in the late 20th century that the massiveness of the neutrino was clarified \cite{Fuk98,Ahm02}. However, the neutrino mass scale is not yet determined accurately. There are three mass scale parameters measurable experimentally or observationally, that is, the average of the three eigen masses \cite{Agh20}, the electron-neutrino mass \cite{Ake24}, and the effective neutrino mass (the Majorana mass) \cite{Fur39}. The last one is possible to determine if the half-life of the $0\nu\beta\beta$ decay is measured and the accurate transition matrix element is calculated. The transition matrix element consists of the electron sector and the neutrino-nucleon sector. The former is well established. The latter is the $0\nu\beta\beta$ NME, which is distributed in a range of the minimum-to-maximum ratio of two to three \cite{Ago23}. The reduction of this uncertainty is the crucial task for nuclear theory. The accurate $0\nu\beta\beta$ NME is also crucial for designing the detectors of the next generation. The goal of our research is to solve the uncertainty problem of the $0\nu\beta\beta$ NME. 

A major input to the NME is, needless to say, nuclear wave functions. The lightest candidate nucleus of the $0\nu\beta\beta$ decay for the experiments is $^{48}$Ca; therefore, approximation of the nuclear wave function is essential. So far, the shell model, the quasiparticle random-phase approximation (QRPA), the energy density functional (EDF) approach, and the interacting boson model (IBM) have been used. We investigated the validity of the QRPA for $^{136}$Xe \cite{Ter19} with the Skyrme interaction, the Coulomb, and the contact pairing interactions with the large enough single-particle valence spaces, and it was concluded that this approach is a good approximation for this nucleus. Thus, we concentrate on $^{136}$Xe in this article to narrow the causes of the problems. At the elementary-particle level, the weak interaction has the Gamow-Teller (GT) and the Fermi components, and the strength of the former component is the axial-vector current coupling $g_A$. Usually, the effective $g_A$ ($g_A^\mathrm{eff}$) is used in nuclear physics to reproduce the measured half-life for the decays due to the weak interaction. It turned out in our previous study \cite{Ter19} that $g_A^\mathrm{eff}$ to reproduce the $2\nu\beta\beta$ decay half-life of $^{136}$Xe is 0.42, which is much smaller than the value of $g_A$ for an isolated nucleon which is $g_A$ = 1.267 (the bare value). A tendency is reported that $g_A^\mathrm{eff}$ decreases as mass number $A$ increases \cite{Suh17}. At this stage, we reached the strong necessity to consider the perturbation of the transition operator. That is the effects of the changes of the nucleon states during the decay, due to the nucleon-nucleon (NN) interaction, on the NME. In other words, those are the effects that cannot be incorporated in the perturbation of the initial and final states of the decay. 

The perturbation to the NME of the $\beta$ decay was extensively investigated in 1970s$‒$80s \cite{Ari87,Tow87} mainly for relatively light nuclei. More recently, the chiral effective-field theory ($\chi$EFT) was applied to the $\beta$ decay \cite{Gys19}. Concerning the $\beta\beta$ decay, the approaches with $\chi$EFT are advanced in the modification of the NME. We recognize three branches; one is characterized by the effective NN interaction of the $\chi$EFT with the renormalization of the NME due to the limited wave function space \cite{Cor24}. The second is phenomenological NN interactions with the NME components derived by the $\chi$EFT \cite{Cas24}, and the third one uses both NN interactions and the NME components from $\chi$EFT \cite{Bel21}. The new $\beta\beta$ NME components by the $\chi$EFT were suggested in Ref.\ \cite{Cir19}. 

An important question that can be answered by our study is what the $g_A^\mathrm{eff}$ for the $0\nu\beta\beta$ decay is. A speculation is possible that $g_A^\mathrm{eff}$ for the $0\nu\beta\beta$ decay is quite different from that for the other weak decays because the neutrino momentum of the $0\nu\beta\beta$ decay is quite different from that of the other weak decays; the neutrino of the $0\nu\beta\beta$ decay is a virtual particle. Therefore, the $g_A^\mathrm{eff}$ for the $0\nu\beta\beta$ decay has been unknown at all. This is one of the major causes of the uncertainty problem of the $0\nu\beta\beta$ NME. Once the perturbed NME is obtained, it is possible to obtain $g_A^\mathrm{eff}$ by referring to the perturbed NME simulatively. 

In Sec.\ II, we derive the equations of the NME calculated in this article starting from a review of the usual equation. The general equation of the lowest-order correction to the NME is derived. From this, the equations for the vertex correction (vc) and the two-body current (2bc) correction are derived. Subsequently, the canonical basis is introduced. After this, the analogous equations are derived for the $2\nu\beta\beta$ NME. In Sec.\ III, the results of the calculations are shown, and the significant features are discussed. Section IV is devoted to the summary. 

\section{\label{sec:eqs_bb_nme} DERIVATION OF EQUATIONS OF DOUBLE-$\bm{\beta}$ NUCLEAR MATRIX ELEMENT}
\subsection{\label{subsec:leading}Leading order}

We review the basic formulation of the $\beta\beta$ NME, to consider the extension. Many of the equations are taken from Ref.\ \cite{Doi85}, and our explanation is added. There are variants of equations of the $\beta\beta$ NME, but the basic equation is common except for the approaches of Ref.\ \cite{Ver02} and the $\chi$EFT \cite{Cir18}. The basic common idea is a hybrid model of two approaches:
\begin{itemize}
\item application of the quantum field theory to the leptons,
\item application of the Rayleigh-Schrödinger perturbation theory to nuclei.
\end{itemize}
The validity of the application of the quantum field theory for the weak interaction is proven historically \cite{Com83}. We assume that this application is also valid for the Majorana neutrino. We start our discussion with a general scheme. The initial state of a transition is denoted by $|I\rangle$ with the eigenenergy $E_I$ , and the final state is denoted by $|\widetilde{F}\rangle$ with the eigenenergy $E_{ \widetilde{F} }$. These will be treated as the unperturbed states and energies. Let $W$ be a transition operator and $\widetilde{V}$ be a perturbation interaction. Using the first-order perturbation for the wave functions, we obtain the transition matrix element up to the first order with respect to $\widetilde{V}$
\begin{eqnarray}
   {M}^{(1)} =\langle \widetilde{F}|W | I\rangle+\sum_{\widetilde{B}\neq I} \langle \widetilde{F}|W| \widetilde{B} \rangle  \frac{1}{ E_{ \widetilde{B} }-E_I } \langle \widetilde{B}| (-\widetilde{V}) | I \rangle
    +\sum_{ \widetilde{B}^\prime \neq \widetilde{F} } \langle \widetilde{F}| (-\widetilde{V}) | \widetilde{B}^\prime\rangle \frac{1}{ E_{\widetilde{B}^\prime} -E_{ \widetilde{F} }  } \langle \widetilde{B}^\prime | W | I \rangle. \nonumber\\ 
\label{eq:Mtilde1_general}
\end{eqnarray}
$|\widetilde{B}\rangle$ and $|\widetilde{B}^\prime\rangle$ are intermediate states. We assume $W\neq \widetilde{V}$ because these two operators are conceptually distinguished. The equation for $W=\widetilde{V}$ is discussed below.

Now, we introduce the conditions for the $\beta\beta$ decay.
\begin{itemize}
\item The system consists of a nucleus and leptons, except for the initial state with no lepton. $|\widetilde{B}\rangle$ $(|{\widetilde{B}}^\prime\rangle)$ in Eq. (\ref{eq:Mtilde1_general}) consists of the intermediate nucleus, an electron, and a Majorana neutrino. $|\widetilde{F}\rangle$ consists of the final nucleus and two electrons. The initial, intermediate, and final nuclear states are denoted by $|I\rangle$, $|B\rangle$, $(|B^\prime\rangle)$, and $|F\rangle$, respectively. The corresponding energies are denoted by $E_I$, $E_B$, ($E_{B^\prime}$), and $E_F$.
\item $|I\rangle$ has $Z$ protons and $N$ neutrons [($Z,N$)], and $|F\rangle$ has ($Z+2,N-2)$. We assume that they are the ground states. $|B\rangle$ has ($Z+1,N-1$). $Z$ and $N$ are assumed to be even.
\item We need to set $W=\widetilde{V}=H_W$, where $H_W$ is the weak interaction, to obtain the transition matrix element of the second-order weak process. If this condition is inserted into $M^{(1)}$ of Eq.~(\ref{eq:Mtilde1_general}), a physical double-counting problem occurs; see below. We can use the modified $M^{(1)}$ where the redundant term is removed.
 
\end{itemize}
If $H_W$ is inserted into $W$ and $\widetilde{V}$ in Eq.\ (\ref{eq:Mtilde1_general}), the left $H_W$ of the second term is the transition operator, and the right $H_W$ is the perturbation interaction. In the third term, the left $H_W$ is the perturbation interaction, and the right $H_W$ is the transition operator. Of course, there is no such a distinction in the phenomenon. Keeping both the second and the third terms implies that the same physical process is counted twice (double counting). This results in the redundant factor of two due to the energy conservation of $E_{\widetilde{F}}=E_I$. Thus, one of the two terms is removed. The cause of this problem is the two physically ambivalent conditions. One is the conceptual distinction of $W$ and $\widetilde{V}$ in terms of the transition operator and the interaction, and the other is the identicality of them. 
The $\beta\beta$ decay is distinguished from the sequence of two $\beta$ decays, implying that the intermediate states are virtual states. For the $\beta\beta$ decay, the first term of ${M}^{(1)}$ vanishes;
\begin{eqnarray}
\langle \widetilde{F} | H_W |I\rangle = 0. 
\end{eqnarray}
This procedure results in the modified $M^{(1)}$ as
\begin{eqnarray}
\widetilde{M}^{(1)} = \sum_{\widetilde{B} \neq I} \langle \widetilde{F} | H_W | \widetilde{B}\rangle \frac{1}{ E_I - E_{ \widetilde{B} } }\langle \widetilde{B} | H_W | I \rangle. \label{eq:Mtilde1_bb}
\end{eqnarray}
$H_W$ is the lepton current-hadron current interaction. Its density is given by
\begin{eqnarray}
\mathcal{H}_W=\frac{G}{ \sqrt{2} } j_L^\rho \mathcal{J}_{L\rho}^\dagger + \mathrm{h.c.}
\end{eqnarray}
The coordinate dependence of the density is omitted according to custom, as long as a confusion does not occur. $G$ is the strength of the weak interaction (Fermi constant). 
The left-handed ($V-A$) lepton current density is defined by 
\begin{eqnarray}
j_L^\rho = \overline{e}\gamma^\rho (1-\gamma_5)\nu_e,
\end{eqnarray}
$e$: electron field operator,\\
$\nu_e$: electron-neutrino field operator,\\
$\gamma^\rho$: gamma matrix.
\vspace{10pt}\\
\noindent The hadron current density is originally defined for the quarks, and for nuclear calculations, that is replaced by the nucleon current density 
\begin{eqnarray}
\mathcal{J}_{L\rho}^\dagger = \overline{\Psi}\tau^-(g_V \gamma_\rho - g_A \gamma_\rho \gamma_5 )\Psi,
\end{eqnarray}
$\Psi$: nucleon field operator,\\
$g_V$: vector current coupling constant equal to one,\\
$\tau^-$: operator changing a neutron to a proton keeping other properties of the neutron.
\vspace{10pt}\\
This replacement was made by adding spectator quarks to the lepton-quark interaction analogously to the impulse approximation of the nuclear reaction theory \cite{Doi85}. The coupling $g_A$ is equal to one for the single quark but not so for the nucleon. The basic equations of the operator density are summarized in Appendix A. In this article, the right-handed current is not included. The terms depending on the nucleon recoil (the weak magnetism and the pseudo scalar terms) are neglected. 

\begin{figure}[t]
\includegraphics[width=0.3\columnwidth]{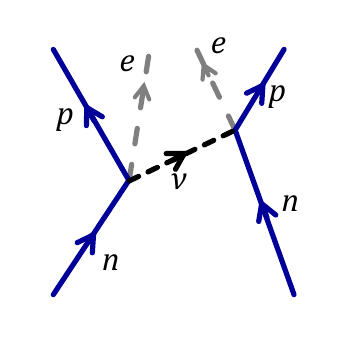}
\vspace{-10pt}
\caption{ \protect \label{fig:diagram_0vld} \baselineskip=13pt 
Diagram of $0\nu\beta\beta$ decay at leading order in terms of vertex. Black short-dashed line expresses the Majorana neutrino. Solid line is used for the nucleon. The two solid lines at the entrance (exit) indicate the neutrons (protons), and these nucleons are included in the initial (final) nucleus. The arrow implies that the nucleon outgoing from (incoming to) any vertex is created (annihilated) at the vertex. Other nucleons in the nucleus are omitted in the diagram, but all nucleons are included in the calculations. Gray long-dashed line stands for the electron.
}
\end{figure}

When the $0\nu\beta\beta$ decay is discussed in this article, the neutrino implies the Majorana neutrino. The leading-order process is illustrated by Fig.\ \ref{fig:diagram_0vld}.  The propagating neutrino is defined for the mass eigenstates $i$ = 1$–$3, and
\begin{eqnarray}
\nu_e= U_{e1}\nu_1 + U_{e2}\nu_2 + |U_{e3}|e^{i\delta_\mathrm{CP}}\nu_3,
\end{eqnarray}
is used for the calculation of the transition matrix element. $U_{ei}$ is the Pontecorvo-Maki-Nakagawa-Sakata (PMNS) matrix element \cite{Bra09}, and $\nu_i$ is the neutrino field operator with the mass eigenstate $i$. $U_{e1}$ and $U_{e2}$ are real, and $\delta_\mathrm{CP}$ denotes the lepton CP violation phase. When the Majorana neutrino is used, the PMNS matrix includes the Majorana phases. Currently, these parameters are unknown. The experimental determination of the other parameters of the PMNS matrix is rapidly in progress \cite{Neu24}. There is a realistic possibility that the PMNS matrix elements have imaginary parts due to the CP violation phases. In this case, the neutrino mass ordering (hierarchy) can be determined. 

After a sequence of mathematical processes, the equation of the decay probability normalized for the half-life $T_{1/2}^{\,0\nu(0)}$  is obtained 
\begin{eqnarray}
\Gamma_{0\nu}^{(0)} = \frac{1}{ T_{1/2}^{\,0\nu(0)} } = G_{0\nu} |M_{0\nu}^{(0)}|^2 \left( \frac{ \langle m_\nu \rangle }{m_e } \right)^2.
\end{eqnarray}
This equation consists of three parts.
\\

\noindent \textit{1. The leading-order $0\nu\beta\beta$ NME}

The product of the two lepton current-hadron current interactions can be rewritten to the product of a scalar derived from the electron part (originally a tensor), the scalar product of the two nucleon currents (originally the two-current tensor), and the neutrino propagator. The NME of the $0\nu\beta\beta$ decay is defined by the nuclear and the neutrino parts, apparently because the neutrino propagator causes the interaction between the two nucleons, i.e., the neutrino potential. The leading-order $0\nu\beta\beta$ NME reads
\begin{eqnarray}
M_{0\nu}^{(0)}&=&\frac{4\pi R}{g_A^2} \int d^3 \bm{x} d^3 \bm{y} \sum_B \int \frac{ d^3 \bm{q} }{ (2\pi)^3 } \frac{1}{|\bm{q}|} \frac{ \mathrm{exp} [ i\bm{q}\cdot (\bm{x}-\bm{y}) ] }{ E_B+|\bm{q}|-\frac{1}{2} (E_I + E_F ) } \langle F| \mathcal{J}^\mu (\bm{x})|B\rangle \nonumber\\
 &&\times\langle B |\mathcal{J}_\mu (\bm{y})| I\rangle,
\label{eq:M00v}
\end{eqnarray}
 
\noindent $R$: nuclear root mean square radius. $R=1.2A^{1/3}$ fm is usually used. \\
$\mathcal{J}_\mu (\bm{x}) \equiv \mathcal{J}_{L\mu}^\dagger (\bm{x})$,\\
$\bm{q}$: momentum of neutrino,\\
$\bm{x}, \bm{y}$: coordinates of vertexes.
\\

Some remarks are noted here. $G_{0\nu}$ is not included in the definition of the NME in the expression of the decay probability. The factor of $4\pi R/g_A^2$  cancels with the counter part in the electron sector (the phase space factor), when the half-life is calculated; see below. The NME is decoupled from the electron sector by approximation. Practically, the $E_B$ in the energy denominator is replaced by an average energy $\overline{E}_C$ (the closure approximation \cite{Hor10,Sim11}). The so called nucleon or vertex form factor \cite{Ver02} is not used in this study because its effect is partially included in the correction terms. 

The principle to determine the sign of the NME is necessary to determine the relative signs between the leading and the correction terms. We fix this sign by the equations of the perturbation formulae, including a negative sign arising from the  lepton sector into the NME. The sign of $M_{0\nu}^{(0)}$ (\ref{eq:M00v}) is consistent with $\widetilde{M}^{(1)}$  (\ref{eq:Mtilde1_bb}).
\\ 

\noindent\textit{2. Phase space factor}

The phase space factor of the $0\nu\beta\beta$ NME is derived from the electron scalar;
\begin{eqnarray}
G_{0\nu} \simeq \frac{ G^4 g_A^4 m_e^2 }{ 2\pi R^2 \mathrm{ln}(2) } \int\frac{ d^3 \bm{p}_1 }{ (2\pi)^3 } \frac{ d^3 \bm{p}_2 }{(2\pi)^3} \delta(\varepsilon_1 + \varepsilon_2 + E_F - E_I ) F_0(Z,\varepsilon_1 ) F_0(Z,\varepsilon_2),
\end{eqnarray}

\noindent\parbox{47pt}{$F_0(Z,\varepsilon_i)$:\vspace{2.3pt}\\}
 \parbox{14cm}{Coulomb correction included in the electron wave function obtained by solving the Dirac equation with the Coulomb field,}\\
$\bm{p}_i$: momentum of electron, $i = 1,2$,\\
$\varepsilon_i = \varepsilon(|\bm{p}_i|)$: electron energy,\\
$m_e$: electron mass.
\\

\noindent $F_0(Z,\varepsilon)$ is given by
\begin{eqnarray}
&&F_0(Z,\varepsilon)=\left[\frac{ \Gamma(3) }{ \Gamma(1)\Gamma(2\gamma_1 + 1) } \right]^2 (2R|\bm{p}|)^{2(\gamma_1 - 1)} |\Gamma(\gamma_1 + iy)|^2 e^{\pi y}, \label{eq:F0Ze}
\\ \nonumber \\
&&\gamma_1 = \sqrt{ 1-(\alpha Z)^2 }, \hspace{12pt} y = \frac{ \alpha Z\varepsilon }{ |\bm{p}| },
\end{eqnarray}

\noindent $\Gamma$: gamma function,\\
$\alpha$: fine-structure constant.
\\

\noindent In some literatures, e.g., Kotila and Iachello \cite{Kot12}, $g_A^4$ is not included in the phase space factor. 
\\

\noindent \textit{3. Effective neutrino mass}

This is defined by 
\begin{eqnarray}
\langle m_\nu \rangle = \left|m_1 U_{e1}^2 + m_2 U_{e2}^2 + m_3 |U_{e3}|^2 e^{2i\delta_\mathrm{CP} } \right|,
\end{eqnarray}

\noindent $m_i:$ eigenmass of neutrino, $i=1$$-$$3$.
\\[-8pt]

\noindent Currently, $\langle m_\nu \rangle$ cannot be obtained from this equation due to the unknown Majorana phases, as mentioned above. 

For convenience, we introduce the compact leading-order NME density not including the neutrino propagator and the associated parts of the equation of $M_{0\nu}^{(0)}$
\begin{eqnarray}
&&\mathcal{M}_{\beta\beta}^{(0)}(\bm{x},\bm{y}) = \sum_{B\neq I} \langle F| \mathcal{J}_\mu (\bm{x}) |B\rangle \frac{1}{ E_{\widetilde{B}} - E_I } \langle B| \mathcal{J}^\mu (\bm{y}))|I\rangle, \label{eq:compact_bb0_density}
\\ \nonumber \\
&&E_{\widetilde{B}} = E_B + |\bm{q}| + \varepsilon.
\end{eqnarray}
Usually, $\varepsilon$ is approximated to be half the energy of the two electrons in the final state, and one uses 
\begin{eqnarray}
E_{\widetilde{B}} - E_I \simeq E_B + |\bm{q}| - \frac{1}{2} (E_I + E_F ). \label{eq:energy_denominator}
\end{eqnarray}
The leading-order $0\nu\beta\beta$ NME can be written 
\begin{eqnarray}
M_{0\nu}^{(0)} = \frac{4\pi R}{g_A^2} \int d^3 \bm{x} d^3 \bm{y} \int \frac{ d^3 \bm{q} }{(2\pi)^3} \frac{1}{|\bm{q}|} \mathrm{exp}[i\bm{q}\cdot (\bm{x} -\bm{y})] \mathcal{M}_{\beta\beta}^{(0)}(\bm{x},\bm{y}). \label{eq:M0v0_w_cmpct_NME_dst}
\end{eqnarray}
Relativistic corrections as the tensor term \cite{Doi85} of the NME are not included in this discussion. The tensor term is around 10 \% of the GT NME \cite{Sim99}. The nucleon wave functions are nonrelativistic ones. The influence of the electron energy on the nuclear sector can be removed, so that $G_{0\nu}$ and $|M_{0\nu}^{(0)}|^2$ are decoupled factors in $\Gamma_{0\nu}^{(0)}$. 

\subsection{\label{subsec:corrections}General equation of lowest-order correction for nuclear matrix element}

We extend the nuclear perturbation of the leading-order NME to obtain the correction terms. Here, the compact NME density (\ref{eq:compact_bb0_density})  is extended. The NME can be obtained analogously to Eq.\  (\ref{eq:M0v0_w_cmpct_NME_dst}). The following procedure is applied:
\begin{itemize}
\item Writing the compact NME density using $W$ and $\widetilde{V}$ with $|I\rangle$ and $|F\rangle$ perturbed in the second order with respect to $\widetilde{V}$. 
\item Applying $\mathcal{J}_\mu (\bm{x})$ for the transition operator $W$. 
\item In each term, one of the two $\widetilde{V}$ is set to $\mathcal{J}^\mu (\bm{y})$ in all possible ways, yielding two terms. 
\item Setting the remaining $\widetilde{V}$ to the residual nuclear interaction $:V:$.
\item Applying $\langle F|\mathcal{J}^\mu (\bm{y})|I\rangle = 0$,  
\item Removing the factorized terms, i.e., those including parts not connected to the neutrino diagrammatically. This is a condition for the corrected transition operator. 
\item Avoiding the double counting of the identical terms due to the identification of the transition operator and one of the perturbation interactions.
\end{itemize}

\noindent This procedure yields the first-order compact NME density for the $\beta\beta$ decay with respect to $:V:$, that is, 
\begin{eqnarray}
&&\mathcal{M}_{\beta\beta}^{(1)}(\bm{x},\bm{y}) = \mathcal{M}_{\beta\beta}^{(JJV)}(\bm{x},\bm{y}) + \mathcal{M}_{\beta\beta}^{(VJJ)}(\bm{x},\bm{y}) + \mathcal{M}_{\beta\beta}^{(JVJ)}(\bm{x},\bm{y}), \label{eq:MbbV_general}
\\ \nonumber \\
&&\mathcal{M}_{\beta\beta}^{(JJV)}(\bm{x},\bm{y}) \equiv -\sum_{C\neq I} \sum_B \langle F|\mathcal{J}_\mu (\bm{x})|B\rangle \frac{1}{ E_B + |\bm{q}| - \frac{1}{2} (E_I + E_F)} \langle B|(-\mathcal{J}^\mu (\bm{y}))|C\rangle \frac{1}{E_C - E_I} \nonumber \\
&&\hspace{80pt}\times\langle C |:(-V):|I\rangle, \label{eq:MbbJJV_general}\\
\nonumber \\
&&\mathcal{M}_{\beta\beta}^{(VJJ)}(\bm{x},\bm{y}) \equiv -\sum_{D\neq F} \sum_B \langle F|:(-V):|D\rangle \frac{1}{E_D - E_F} \langle D|\mathcal{J}_\mu (\bm{x})|B\rangle \frac{1}{ E_B + |\bm{q}| - \frac{1}{2} (E_I + E_F) }  \nonumber \\
&& \hspace{80pt}\times \langle B|(-\mathcal{J}^\mu (\bm{y}))|I\rangle, \label{eq:MbbVJJ_general}\\
\nonumber \\
&&\mathcal{M}_{\beta\beta}^{(JVJ)}(\bm{x},\bm{y}) \equiv -\sum_{BB^\prime} \langle F|\mathcal{J}_\mu (\bm{x})| B^\prime \rangle \frac{1}{E_{B^\prime} + |\bm{q}| -\frac{1}{2} (E_I + E_F) } \langle B^\prime |:(-V):|B\rangle \nonumber \\
&&\hspace{80pt}\times \frac{1}{ E_B + |\bm{q}| - \frac{1}{2} (E_I + E_F) } \langle B|(-\mathcal{J}^\mu (\bm{y}))|I\rangle. \label{eq:MbbJVJ_general}
\end{eqnarray}
$|C\rangle$ and $|D\rangle$ are states of the initial and final nuclei, respectively. $|B^\prime\rangle$ are states with $(Z+1,N-1)$.
The detailed derivation of Eqs.~(\ref{eq:MbbV_general})$-$(\ref{eq:MbbJVJ_general}) is given in Appendix B.

\begin{figure}[t]
\includegraphics[width=1.0\columnwidth]{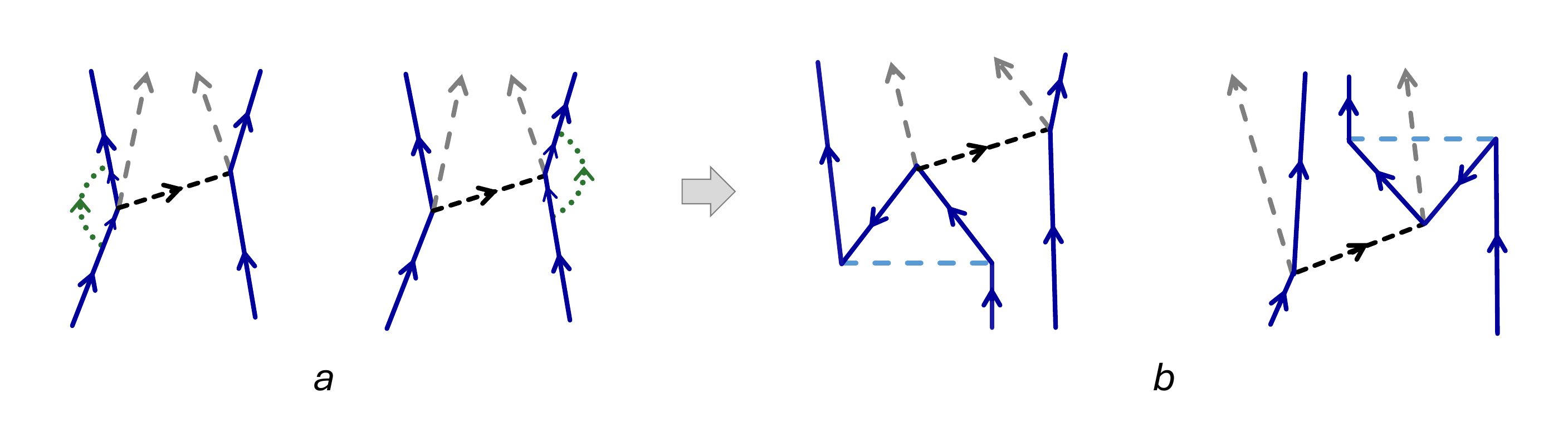}
\vspace{-10pt}
\caption{ \protect \label{fig:diagram_0vvc} \baselineskip=13pt 
Diagrams of $0\nu\beta\beta$ NME with a vertex correction. Diagrams \textit{a}: vertex corrections with a meson exchange (green dots). Diagrams \textit{b}: vertex corrections obtained by replacing the meson of diagrams \textit{a} by an NN potential (light-blue long-dashed line). For the definitions of the other parts, see caption of Fig.\ \ref{fig:diagram_0vld}. 
 }
\end{figure}

\subsection{\label{subsec:vc}Vertex correction}
The vertex corrections calculated in this article are those of the lowest-order with respect to the perturbation interaction. This is illustrated by the diagrams of Fig.\ \ref{fig:diagram_0vvc}. The green dots in the diagrams of Fig.\ \ref{fig:diagram_0vvc}\textit{a} indicate the meson propagation, which causes the perturbation of the transition operator. The diagrams of Fig.\ \ref{fig:diagram_0vvc}\textit{b} are obtained by replacing the mesons in Fig.\ \ref{fig:diagram_0vvc}\textit{a} with the NN interaction. The topologies of the diagrams \textit{a} and \textit{b} are the same one. 

We use the axially symmetric quasiparticle representation for our calculations. The quasiparticle is specified by a label 
\begin{eqnarray}
\alpha \equiv (\tau_z \pi j_z o)_\alpha \equiv (\tau_\alpha^z \pi_\alpha j_\alpha^z o_\alpha),
\end{eqnarray}
$\tau_z$: $z$ component of the isospin, i.e., proton or neutron,\\
$\pi$: parity,\\
$j_z$: $z$ component (along the symmetry axis of the nucleus) of the angular momentum.\\
\parbox{10pt}{$o$:\\\hspace{20pt}\\} \parbox{16cm}{the other label to specify a particle state in the subspace of $(\tau_z \pi j_z )$. The definition is arbitrary. When the quasiparticle basis is used, our choice is the order of the quasiparticle energy.}\\[5pt]

\noindent The indexes $(\tau_z \pi j_z )$ are the good quantum numbers. We use the notation
\begin{eqnarray}
-\alpha \equiv (\tau_z \pi \;\hbox{$-j_z$} o)_\alpha,
\end{eqnarray}
\vspace{-12pt}
\begin{eqnarray}
\tau_{-\alpha}^z = \tau_\alpha^z,\hspace{12pt} \pi_{-\alpha} = \pi_\alpha, \hspace{12pt} j_{-\alpha}^z = -j_\alpha^z, \hspace{12pt} o_{-\alpha} = o_\alpha.
\end{eqnarray}
The bases are defined for the initial and the final states, which are time reversal invariant. For the single particles, we use the canonical basis, that is, the diagonal representation of the one-body density of the nucleus. This basis has the same good quantum numbers as the quasiparticle basis has, and the particle number is also the good quantum number. For the paired system, the order of occupation probability is used for $o$, and for the unpaired system, the order of the Hartree-Fock (HF) energy is used. We also use the notation for the single-particle state
\begin{eqnarray}
i \equiv (\tau_z \pi j_z o)_i \equiv (\tau_i^z \pi_i j_i^z o_i ),
\end{eqnarray}
\vspace{-12pt}
\begin{eqnarray}
-i \equiv (\tau_z \pi \;\hbox{$-j_z$} o)_i,
\end{eqnarray}
and introduce the Hartree-Fock-Bogoliubov (HFB) transformation \cite{Rin80} 
\begin{eqnarray}
\left(
\begin{array}{c}
a_\mu \\ a_{-\mu}^\dagger
\end{array}
\right)
= \sum_j 
\left(
\begin{array}{cc}
U_{j\mu}^\ast & V_{-j\mu}^\ast \\
V_{j-\mu} & U_{-j-\mu} 
\end{array}
\right)
\left(
\begin{array}{c}
c_j \\ 
c_{-j}^\dagger  
\end{array}
\right), 
\end{eqnarray}
\vspace{-12pt}
\begin{eqnarray}
\left(
\begin{array}{c}
c_j \\
c_{-j}^\dagger
\end{array}
\right)
= \sum_\mu 
\left(
\begin{array}{cc}
U_{j\mu} & V_{j-\mu}^\ast \\
V_{-j\mu} & U_{-j-\mu}^\ast 
\end{array}
\right)
\left(
\begin{array}{c}
a_\mu \\
a_{-\mu}^\dagger
\end{array}
\right),
\end{eqnarray}
with\\ 
\vspace{-8pt}\\
${c_i^\dagger,c_i}$: creation and annihilation operators of the single particle $i$, respectively, \\ 
${a_\mu^\dagger,a_\mu}$: creation and annihilation operators of the quasiparticle $\mu$, respectively. \\
\vspace{-8pt}\\
The HFB Hamiltonian is represented as
\begin{eqnarray}
h_\mathrm{HFB} &=& \frac{1}{2} \sum_{ij} (c_i^\dagger \ c_{-i})
\left(
\begin{array}{cc}
h_{ij} & \Delta_{i-j}\\
-\Delta_{-ij}^\ast & -h_{-i-j}^\ast
\end{array}
\right)
\left(
\begin{array}{c}
c_j \\
c_{-j}^\dagger
\end{array}
\right) + \mathrm{const} \nonumber\\
\nonumber\\
&=& \frac{1}{2} \sum_i (c_i^\dagger \ c_{-i} ) \sum_{\mu\nu} 
\left(
\begin{array}{cc}
U_{i\nu} & V_{i-\nu}^\ast \\
V_{-i\nu} & U_{-i-\nu}^\ast 
\end{array}
\right)
\left(
\begin{array}{cc}
\varepsilon_\mu & 0 \\
0 & -\varepsilon_{-\mu}
\end{array}
\right)
\delta_{\mu\nu} \sum_j 
\left(
\begin{array}{cc}
U_{j\mu}^\ast & V_{-j\mu}^\ast \\
V_{j-\mu}&U_{-j-\mu}
\end{array}
\right)
\left(
\begin{array}{c}
c_j \\
c_{-j}^\dagger
\end{array}
\right) \nonumber\\
&&+ \mathrm{const}.
\end{eqnarray}
The matrixes of $h$ and $\Delta$ are the self-consistent one-body and pair fields, respectively. The residual interaction in the nuclear Hamiltonian is expressed as

\begin{eqnarray}
:V: &=& \frac{1}{2} \sum_{ijkl} V_{ij,kl}: c_i^\dagger c_j^\dagger c_l c_k: = V_{2\mathrm{c}2\mathrm{a}} + V_{3\mathrm{c}1\mathrm{a}} + V_{1\mathrm{c}3\mathrm{a}} + V_{4\mathrm{c}} + V_{4\mathrm{a}},
\end{eqnarray}
\vspace{-10pt}
\begin{eqnarray}
V_{2\mathrm{c}2\mathrm{a}} &=& \frac{1}{2} \sum_{ijkl} V_{ij,kl}  \sum_{\mu\nu\kappa\lambda} \bigl\{ U_{i\kappa}^\ast U_{j\mu}^\ast U_{l\lambda} U_{k\nu} a_\kappa^\dagger a_\mu^\dagger a_\lambda a_\nu + U_{i\kappa}^\ast  V_{j-\mu} U_{l\lambda} V_{k-\nu}^\ast a_\kappa^\dagger a_{-\nu}^\dagger a_{-\mu} a_\lambda \nonumber\\
&&-U_{i\kappa}^\ast V_{j-\mu} V_{l-\lambda}^\ast U_{k\nu} a_\kappa^\dagger a_{-\lambda}^\dagger a_{-\mu} a_\nu - V_{i-\kappa} U_{j\mu}^\ast U_{l\lambda} V_{k-\nu}^\ast a_\mu^\dagger a_{-\nu}^\dagger a_{-\kappa} a_\lambda \nonumber \\
&&+V_{i-\kappa} U_{j\mu}^\ast V_{l-\lambda}^\ast U_{k\nu} a_\mu^\dagger a_{-\lambda}^\dagger a_{-\kappa} a_\nu + V_{i-\kappa} V_{j-\mu} V_{l-\lambda}^\ast  V_{k-\nu}^\ast a_{-\lambda}^\dagger a_{-\nu}^\dagger a_{-\kappa} a_{-\mu}\bigr\},
\end{eqnarray}
\vspace{-10pt}
\begin{eqnarray}
V_{3\mathrm{c}1\mathrm{a}} &=& \frac{1}{2} \sum_{ijkl} V_{ij,kl}  \sum_{\mu\nu\kappa\lambda} \bigl\{ -U_{i\kappa}^\ast U_{j\mu}^\ast U_{l\lambda} V_{k-\nu}^\ast a_\kappa^\dagger a_\mu^\dagger a_{-\nu}^\dagger a_\lambda + U_{i\kappa}^\ast U_{j\mu}^\ast  V_{l-\lambda}^\ast U_{k\nu} a_\kappa^\dagger a_\mu^\dagger a_{-\lambda}^\dagger a_\nu \nonumber \\
&&+U_{i\kappa}^\ast V_{j-\mu} V_{l-\lambda}^\ast V_{k-\nu}^\ast  a_\kappa^\dagger a_{-\lambda}^\dagger a_{-\nu}^\dagger a_{-\mu} - V_{i-\kappa} U_{j\mu}^\ast V_{l-\lambda}^\ast V_{k-\nu}^\ast  a_\mu^\dagger a_{-\lambda}^\dagger a_{-\nu}^\dagger a_{-\kappa} \bigr\},
\end{eqnarray}
\vspace{-10pt}
\begin{eqnarray}
V_{1\mathrm{c}3\mathrm{a}} &=& \frac{1}{2} \sum_{ijkl} V_{ij,kl}  \sum_{\mu\nu\kappa\lambda} \bigl\{ U_{i\kappa}^\ast V_{j-\mu}  U_{l\lambda} U_{k\nu} a_\kappa^\dagger a_{-\mu} a_\lambda a_\nu - V_{i-\kappa} U_{j\mu}^\ast U_{l\lambda} U_{k\nu} a_\mu^\dagger a_{-\kappa} a_\lambda a_\nu \nonumber\\
&&-V_{i-\kappa} V_{j-\mu} U_{l\lambda} V_{k-\nu}^\ast a_{-\nu}^\dagger a_{-\kappa} a_{-\mu} a_\lambda + V_{i-\kappa} V_{j-\mu}  V_{l-\lambda}^\ast U_{k\nu} a_{-\lambda}^\dagger a_{-\kappa} a_{-\mu} a_\nu \bigr\},
\end{eqnarray}
\vspace{-10pt}
\begin{eqnarray}
V_{4\mathrm{c}} &=& \frac{1}{2} \sum_{ijkl} V_{ij,kl} \sum_{\mu\nu\kappa\lambda} U_{i\kappa}^\ast U_{j\mu}^\ast V_{l-\lambda}^\ast V_{k-\nu}^\ast a_\kappa^\dagger a_\mu^\dagger a_{-\lambda}^\dagger a_{-\nu}^\dagger,\label{eq:V4c}\\
\nonumber\\
V_{4\mathrm{a}} &=& \frac{1}{2} \sum_{ijkl} V_{ij,kl} \sum_{\mu\nu\kappa\lambda} V_{i-\kappa} V_{j-\mu} U_{l\lambda}  U_{k\nu} a_{-\kappa} a_{-\mu} a_\lambda a_\nu.
\end{eqnarray}
The normal ordering is defined with the quasiparticle basis. $V_{ij,kl} \equiv \langle ij|V|kl\rangle$ is the matrix element of the two-body interaction. There are 16 types of terms in $:V:$, and they are classified in terms of the creation and annihilation operators. In this first article of our study of the vertex correction, we restrict the correction terms to the simplest ones, in which $|I\rangle$ and $|F\rangle$ are the HFB ground states, except for the overlap calculation of two QRPA intermediate states. Many of the vertex correction terms arising from the ground-state correlations have the energy denominators with subtractions of the single-particle or quasiparticle energies. It is another major subject to investigate whether reliable calculation of these terms is possible. In this study, we calculate the vertex correction by $V_{4\mathrm{c}}$, which does not have the possibility of the divergence. We speculate that those terms neglected in this study have only small contributions because the QRPA is a good approximation for $^{136}$Xe. 

First, we consider $\mathcal{M}_{\beta\beta}^{(JJV)}(\bm{x},\bm{y})$. An approximation for the feasibility of the calculation and a constraint to choose the vertex correction are introduced. $|C\rangle$ is also assumed to be an HFB state
\begin{eqnarray}
|C\rangle = a_\kappa^\dagger a_\mu^\dagger a_{-\lambda}^\dagger a_{-\nu}^\dagger |I\rangle.
\end{eqnarray}
Here, the quasiparticle suffixes refer to those used for the summation in the equation of $V_{4\mathrm{c}}$ (\ref{eq:V4c}). The part including $V_{4\mathrm{c}}$ in the equation of $\mathcal{M}_{\beta\beta}^{(JJV)}(\bm{x},\bm{y})$ (\ref{eq:MbbJJV_general}) can be written
\begin{eqnarray}
\lefteqn{ |C\rangle \frac{1}{E_C - E_I} \langle C|:(-V_{4\mathrm{c}}):|I\rangle }\nonumber \\
 &&= \sum_{ijkl} \sum_{ \substack{\mu\nu\kappa\lambda \\ \mathrm{all\ different}} } a_\kappa^\dagger a_\mu^\dagger a_{-\lambda}^\dagger  a_{-\nu}^\dagger |I\rangle \frac{1}{E_C - E_I} (-) \frac{1}{2} V_{ij,kl} U_{i\kappa}^\ast U_{j\mu}^\ast V_{l-\lambda}^\ast V_{k-\nu}^\ast.
\end{eqnarray}
Suppose that $|C\rangle$ is represented by the quasiparticle configuration mixings. The only component contributing to $\langle C|\!:\!(-V_{4\mathrm{c}} )\!:\!|I\rangle$ is four-quasiparticle states. That approximation for $|C\rangle$ is a simplification based on this property. 

We use the limited components of $\mathcal{J}^\mu (\bm{y})$ in Eq.\ (\ref{eq:MbbJJV_general}) as
\begin{eqnarray}
\mathcal{J}^\mu (\bm{y}) \rightarrow \sum_{j^\prime l^\prime} \langle j^\prime |\mathcal{J}^\mu (\bm{y})| l^\prime \rangle  \left(V_{j^\prime -\lambda} U_{l^\prime \mu} -V_{j^\prime \mu} U_{l^\prime -\lambda} \right) a_{-\lambda} a_\mu, \label{eq:J-constraint_1}
\end{eqnarray}
where $\lambda$ and $\mu$ refer to those used for the summation in Eq.\ (\ref{eq:V4c}). The diagrams of Fig.\ \ref{fig:diagram_0vvc}\textit{b} indicate this constraint. In the actual calculation, the expression
\begin{eqnarray}
V = \frac{1}{4} \sum_{ijkl} \langle ij|V|kl\rangle_\mathrm{as}  c_i^\dagger c_j^\dagger c_l c_k, \nonumber
\end{eqnarray}
\vspace{-12pt}
\begin{eqnarray}
\langle ij|V|kl\rangle _\mathrm{as} \equiv \langle ij|V|kl\rangle - \langle ij|V|lk\rangle \equiv V_{ij,kl} - V_{ij,lk}, \label{eq:V_as}
\end{eqnarray}
is applied. It is seen from the indexes in Eqs.\ (\ref{eq:V4c}) and (\ref{eq:J-constraint_1}) that the two-particle combinations of $(jl)$ and $(ik)$ in Eq.\ (\ref{eq:V_as}) have good quantum numbers originating from $\mathcal{J}^\mu(\bm{y})$. The first term of $\langle ij|V|kl\rangle$ in Eq.\ (\ref{eq:V_as}) is the direct term, and $-\langle ij|V|lk\rangle$ is the exchange term. The diagrams of Fig.\ \ref{fig:diagram_0vvc}\textit{b} use the exchange matrix elements of the perturbation interaction. The direct matrix elements are not used in the calculations of this article because we use the Skyrme interaction, which does not have the charge exchange component. 

Another constraint to $\mathcal{J}^\mu (\bm{y})$ is also possible; 
\begin{eqnarray}
\mathcal{J}^\mu (\bm{y}) \rightarrow \sum_{j^\prime l^\prime} \langle j^\prime |\mathcal{J}^\mu (\bm{y})| l^\prime \rangle  \left(V_{j^\prime - \nu} U_{l^\prime \kappa} - V_{j^\prime \kappa} U_{l^\prime -\nu} \right) a_{-\nu} a_\kappa.
\end{eqnarray}
The vertex correction obtained from this is identical to that with the first constraint, causing a factor of two. Using the second constraint is equivalent to using the first one with $\langle ji|V|lk\rangle_\mathrm{as}$; see Eq.\ (\ref{eq:MbbJJV_constraint}) below. Therefore, this seeming redundancy is compensated by the coefficient of 1/2 included in the expression of $V$. Thus, the factor of two is included. 

Hereafter, we distinguish the protons and neutrons by the convention of particle suffixes as \\
\indent $a,b,c,\cdots$: proton,\\
\indent $i,j,k,\cdots$: neutron,\\
\indent $\alpha,\beta,\gamma,\cdots$: proton quasiparticle,\\
\indent $\mu,\nu,\kappa,\cdots$: neutron quasiparticle,\\
\vspace{-8pt}\\
to clarify the charge changes in the equation of the NME, otherwise mentioned. It follows that
\begin{eqnarray}
\lefteqn{ \mathcal{M}_{\beta\beta}^{(JJV)}(\bm{x},\bm{y}) }\nonumber \\
&\simeq& -\sum_B \sum_{ai} \langle a|\mathcal{J}(\bm{x})|i\rangle \sum_{\alpha\mu} V_{a-\alpha} U_{i\mu} \langle F|a_{-\alpha} a_\mu |B\rangle \frac{1}{\overline{E}_B + |\bm{q}| - \frac{1}{2} (E_I + E_F ) } \sum_{bj} \langle b|\mathcal{J}(\bm{y})|j\rangle \nonumber \\
 &&\times \sum_{ \substack{\beta\gamma\nu\kappa\\ \mathrm{all\ different}} } \sum_{cdkl} \frac{1}{2} \langle dk|V|lc\rangle_\mathrm{as} \bigl( V_{b-\beta} U_{j\nu} U_{d\gamma}^\ast U_{k\nu}^\ast V_{c-\beta}^\ast V_{l-\kappa}^\ast \langle B|a_\gamma^\dagger a_{-\kappa}^\dagger |I\rangle \nonumber \\
 &&- V_{b\beta} U_{j-\nu} U_{k\kappa}^\ast U_{d\beta}^\ast V_{l-\nu}^\ast V_{c-\gamma}^\ast \langle B|a_\kappa^\dagger a_{-\gamma}^\dagger |I\rangle \bigr) \frac{1}{\varepsilon_\gamma + \varepsilon_\nu + \varepsilon_\beta + \varepsilon_\kappa}. \label{eq:MbbJJV_constraint}
\end{eqnarray}
Here, the time reversal invariance of the quasiparticle basis is used. When the greek indexes are used for the quasiparticles, we use an abbreviation
\begin{eqnarray}
\mathcal{J}(\bm{x})\mathcal{J}(\bm{y}) \equiv \mathcal{J}_\mu(\bm{x}) \mathcal{J}^\mu(\bm{y}).
\end{eqnarray}
Concerning the intermediate states, the replacement is made as 
\begin{eqnarray}
\sum_B |B\rangle\langle B| \rightarrow \sum_{B_F B_I} |B_F \rangle\langle B_F| B_I\rangle\langle B_I |,
\end{eqnarray}
in accordance with the usual QRPA approach, e.g.\ \cite{Sim11}. $|B_I\rangle$ and $|B_F\rangle$ are the QRPA states based on the initial and final states, respectively. For the QRPA overlap $\langle B_F| B_I\rangle$, see Ref.\ \cite{Ter13}. An average energy $\overline{E}_B$ is used for the intermediate-state energy in the energy denominator. This is the usual closure approximation for the $0\nu\beta\beta$ NME. 

The QRPA states are created as 
\begin{eqnarray}
|B_I \rangle = O_{BI}^\dagger |I_\mathrm{QRPA} \rangle,
\end{eqnarray}
\vspace{-16pt}
\begin{eqnarray}
O_{BI}^\dagger = \sum_{\mu\nu} \bigr( X_{\alpha\mu}^{BI} a_\alpha^{I\dagger} a_\mu^{I\dagger} - Y_{-\alpha-\mu}^{BI} a_{-\mu}^I a_{-\alpha}^I \bigr),
\end{eqnarray}
\vspace{-16pt}
\begin{eqnarray}
O_{BI} |I_\mathrm{QRPA} \rangle = 0.
\end{eqnarray}
This method to obtain the nuclear states is the proton-neutron QRPA \cite{Rin80}, which is called the QRPA in this article. The forward QRPA amplitude $X_{\alpha\mu}^{BI}$ and the backward QRPA amplitude $Y_{-\alpha -\mu}^{BI}$ are obtained by solving the QRPA equation. $|I_\mathrm{QRPA} \rangle$ is the QRPA initial ground state and is used for $|I\rangle$ in the leading term and the QRPA overlap calculation. The suffix $I$ on the operators and the QRPA amplitudes indicates that they are associated with the initial state. We use this convention whenever necessary. If those are associated with the final state, the suffix $F$ is used. 

The charge selection rule can be seen for the parts of the equation of $\mathcal{M}_{\beta\beta}^{(JJV)}(\bm{x},\bm{y})$ (\ref{eq:MbbJJV_constraint}); e.g., only the proton-neutron interactions are used. Similarly, the selection rule can be seen with respect to $\pi$ and $j_z$. In the exchange matrix element $\langle dk|V|cl\rangle$ in Eq.\ (\ref{eq:MbbJJV_constraint}), $d$ and $l$ are coupled to $\pi_d \pi_l = \pi_B$ and $j_d^z - j_l^z = K_B$ for the term including $\langle B|a_\gamma^\dagger a_{-\kappa}^\dagger |I\rangle$, where $\pi_B$ and $K_B$ are the parity and the $K$ quantum number of $|B\rangle$, because of $\pi_\gamma \pi_\kappa = \pi_B$ and $j_\gamma^z - j_\kappa^z = K_B$. In accordance with this coupling, $(k\;\hbox{$-c$})$ has $\pi_B$ and $-K_B$. From $\langle F|a_{-\alpha} a_\mu |B\rangle$, $(-\alpha \mu)$ has $\pi_B$ and $K_B$, thus, $\pi_a \pi_i = \pi_B$ and  $j_a^z - j_i^z = -K_B$.

The vertex correction to the $0\nu\beta\beta$ NME with $\mathcal{M}_{\beta\beta}^{(JJV)}(\bm{x},\bm{y})$ reads
\begin{eqnarray}
M_{0\nu}^{(JJV)} &=& \frac{4\pi R}{g_A^2} \int d^3\bm{x} d^3\bm{y} \sum_{B_I B_F} \int \frac{d^3 \bm{q}}{(2\pi)^3} \frac{1}{|\bm{q}|}
\frac{\mathrm{exp}[i\bm{q}\cdot(\bm{x} -\bm{y})]}{\overline{E}_B + |\bm{q}| - \frac{1}{2}(E_I + E_F) } \sum_{bj} \psi_b^{I\dagger}(\bm{x})J(\bm{x}) \psi_j^I(\bm{x}) \nonumber\\
&& \times \sum_{ai} \psi_a^{F\dagger}(\bm{y})J(\bm{y}) \psi_i^F(\bm{y}) \sum_{ \substack{\beta\gamma\nu\kappa\\ \mathrm{all\  different}} } \sum_{cdkl} \bigl( V_{b-\beta}^I U_{j\nu}^I U_{d\gamma}^{I\ast} U_{k\nu}^{I\ast} V_{c-\beta}^{I\ast} V_{l-\kappa}^{I\ast} \langle B_I |a_\gamma^{I\dagger} a_{-\kappa}^{I\dagger} |I\rangle \nonumber\\
&& - V_{b\beta}^I U_{j-\nu}^I U_{k\kappa}^{I\ast} U_{d\beta}^{I\ast} V_{l-\nu}^{I\ast} V_{c-\gamma}^{I\ast} \langle B_I |a_\kappa^{I\dagger} a_{-\gamma}^{I\dagger} |I\rangle \bigr) \frac{1}{2} \langle dk|V|cl\rangle_I  \frac{1}{ \varepsilon_\kappa^I + \varepsilon_\beta^I + \varepsilon _\nu^I + \varepsilon_\gamma^I } \nonumber\\
&& \times\sum_{\alpha\mu} V_{a-\alpha}^F U_{i\mu}^F \langle F|a_{-\alpha}^F a_\mu^F | B_F \rangle \langle B_F|B_I \rangle,
\label{eq:M0vJJV} 
\end{eqnarray}
where $J(\bm{x})\equiv J_\mu(\bm{x})$ is the four-vector nucleon current operator in the nonrelativistic approximation, and one can use
\begin{eqnarray}
\lefteqn{ \psi_b^{I\dagger}(\bm{x}) J_\mu(\bm{x}) \psi_j^I(\bm{x}) \psi_a^{F\dagger}(\bm{y}) J^\mu(\bm{y})\psi_i^F(\bm{y}) } \nonumber\\
&=& g_V^2 \bigl\{\psi_b^{I\dagger}(\bm{x}) \tau^- \psi_j^I(\bm{x})\bigr\}\bigl\{\psi_a^{F\dagger}(\bm{y}) \tau^- \psi_i^F(\bm{y})\bigr\} - g_A^2 \sum_{k=1-3} \bigl\{\psi_b^{I\dagger}(\bm{x}) \sigma_k \tau^-  \psi_j^I(\bm{x})\bigr\}\nonumber\\
&&\times\bigl\{\psi_a^{F\dagger}(\bm{y}) \sigma_k \tau^- \psi_i^F(\bm{y})\bigr\},
\end{eqnarray}
$\sigma_k$: spin Pauli matrix, $k$ is 1$‒$3,\\
$\psi_n^I(\bm{x})$, $\psi_n^F(\bm{y})$: single-particle wave functions. \\
\vspace{-8pt}\\

By the analogous derivation, the vertex correction for the $0\nu\beta\beta$ NME from $\mathcal{M}_{\beta\beta}^{(VJJ)}(\bm{x},\bm{y})$ is obtained
\begin{eqnarray}
M_{0\nu}^{(VJJ)} &=& -\frac{4\pi R}{g_A^2} \int d^3\bm{x} d^3\bm{y} \sum_{B_I B_F} \int\frac{d^3 \bm{q}}{(2\pi)^3} \frac{1}{|\bm{q}|} \frac{\mathrm{exp}[i\bm{q}\cdot(\bm{x} - \bm{y})]}{ \overline{E}_B + |\bm{q}| - \frac{1}{2} (E_I + E_F) } \sum_{bj}  \psi_b^{F\dagger}(\bm{x})J(\bm{x})\psi_j^F(\bm{x}) \nonumber\\
&& \times\sum_{ai}  \psi_a^{I\dagger}(\bm{y})J(\bm{y})\psi_i^I(\bm{y})\sum_{ \substack{\beta\gamma\nu\kappa\\ \mathrm{all\ different}} } \sum_{cdlk} \bigl( V_{c-\beta}^F V_{l-\kappa}^F U_{d\gamma}^F U_{k\nu}^F U_{b-\beta}^{F\ast} V_{j\nu}^{F\ast} \langle F|a_{-\kappa}^F a_\gamma^F | B_F \rangle \nonumber\\
&& - V_{l-\nu}^F V_{c-\gamma}^F U_{k\kappa}^F U_{d\beta}^F U_{b\beta}^{F\ast} V_{j-\nu}^{F\ast} \langle F|a_{-\gamma}^F a_\kappa^F | B_F \rangle \bigr) \frac{1}{ \varepsilon_\kappa^F + \varepsilon_\beta^F + \varepsilon_\gamma^F + \varepsilon_\nu^F } \frac{1}{2} \langle lc|V|kd\rangle_F \nonumber\\
&& \times\sum_{\alpha\mu} U_{a\alpha}^{I\ast} V_{i-\mu}^{I\ast} \langle B_F |B_I\rangle\langle B_I |a_\alpha^{I\dagger} a_{-\mu}^{I\dagger} |I\rangle. \label{eq:M0vVJJ}
\end{eqnarray}
$\mathcal{M}_{\beta\beta}^{(JVJ)}(\bm{x},\bm{y})$ (\ref{eq:MbbJVJ_general}) is finite under the HFB approximation for the nuclear states. 

\subsection{\label{subsec:canonical_basis}Canonical basis}
We use the canonical-quasiparticle basis \cite{Rin80} defined by 
\begin{eqnarray}
\left(
\begin{array}{c}
b_i^\dagger \\
b_{-i} 
\end{array}
\right) = 
\left(
\begin{array}{cc}
u_i & -\mathrm{sgn}(j_i^z )v_i \\
\mathrm{sgn}(j_i^z )v_i & u_i 
\end{array}
\right)
\left(
\begin{array}{c}
c_i^\dagger \\
c_{-i}
\end{array}
\right), 
\end{eqnarray}
for the QRPA calculations. This equation is applied for both the proton and the neutron. Two sets of operators $\{a_\mu^\dagger\}$ and $\{b_i^\dagger\}$ are connected by a unitary transformation. Here, $v_i$ is the square root of the eigen value of diagonalization of the one-body density of the system, and 
\begin{eqnarray}
u_i = \sqrt{ 1 - v_i^2 }.
\end{eqnarray}
The $\mathrm{sgn}(j_i^z)$ is due to our phase convention associated with the time reversal. This basis is defined for both the initial and final nuclei separately. The reason for using this basis is that it is efficient for the calculation of the interaction matrix elements of the QRPA Hamiltonian \cite{Ter05}.

The transformation is necessary from the canonical-quasiparticle basis to the quasiparticle basis to calculate $M_{0\nu}^{(VJJ)}$ and $M_{0\nu}^{(JJV)}$. It turned out that this transformation between the two-body bases needed a very large dimension to maintain the unitarity; the sufficient dimension is estimated to be much larger than 300$\,$000. We cannot perform the matrix element calculations of the interaction and the two-body $\beta\beta$ transition operator with this dimension because the necessary resource is too large. All those matrix elements are finite. We decided to use the canonical-quasiparticle basis to calculate the vertex corrections approximately instead of the quasiparticle basis. The quasiparticle energies in the energy denominator are replaced by the diagonal elements of the HFB Hamiltonian matrix with the canonical-quasiparticle basis. 

The operators used for the NME are originally defined with the single-particle (canonical) basis, and one can express these operators with the canonical-quasiparticle basis. The equations of the NME with this basis can be obtained from Eqs.\ (\ref{eq:M0vJJV}) and (\ref{eq:M0vVJJ}) by a conventional procedure as 
\begin{eqnarray}
V_{a-\alpha}^F = -V_{-a\alpha}^{F\ast} \rightarrow \mathrm{sgn}(j_a^z) v_a^F, \hspace{20pt} U_{i\beta}^F \rightarrow u_i^F, \nonumber
\end{eqnarray}
\vspace{-20pt}
\begin{eqnarray}
a_{-\alpha}^F \rightarrow b_{-a}^F,\hspace{20pt} a_\beta^F \rightarrow b_i^F.
\end{eqnarray}
Here, the arrow indicates the replacement. The summations with respect to $\alpha$ and $\beta$ are removed from the previous equation. In addition, we use the approximation
\begin{eqnarray}
\varepsilon_\kappa^I \simeq \widetilde{\varepsilon}_{kk}^{\,I},
\end{eqnarray}
as mentioned above. Here, $\widetilde{\varepsilon}_{kk}^{\,I}$ is the diagonal element of the HFB Hamiltonian matrix in the canonical-quasiparticle basis. The index $\kappa$ is replaced by $k$ if the term includes $U_{k\kappa}^I$ or $V_{k-\kappa}^I$; see Eq.\ (\ref{eq:M0vJJV}). The other indexes are treated in the same manner.

We define the two-body matrix of the neutrino potential
\begin{eqnarray}
W_{-ai,-bj}^{\mathrm{GT},FI} &\equiv& \int d^3\bm{x} d^3\bm{y} \int \frac{d^3\bm{q}}{(2\pi)^3} \frac{1}{|\bm{q}|} \frac{ \mathrm{exp}[i\bm{q}\cdot(\bm{x} - \bm{y})] }{ \overline{E}_B + |\bm{q}| - \frac{1}{2} (E_I + E_F ) } \sum_{k=1-3} \psi_{-a}^{F\dagger}(\bm{x}) \sigma_k \tau^- \psi_i^F(\bm{x}) \nonumber\\
&& \times\psi_{-b}^{I\dagger} (\bm{y}) \sigma_k \tau^- \psi_j^I(\bm{y}),
\end{eqnarray}
\vspace{-18pt}
\begin{eqnarray}
W_{-ai,-bj}^{\mathrm{F},FI} &\equiv& \int d^3\bm{x} d^3\bm{y} \int \frac{d^3\bm{q}}{(2\pi)^3} \frac{1}{|\bm{q}|} \frac{ \mathrm{exp}[i\bm{q}\cdot(\bm{x} - \bm{y})] }{ \overline{E}_B + |\bm{q}| - \frac{1}{2} (E_I + E_F ) } \psi_{-a}^{F\dagger}(\bm{x}) \tau^- \psi_i^F(\bm{x}) \nonumber\\
&& \times\psi_{-b}^{I\dagger} (\bm{y}) \tau^- \psi_j^I(\bm{y}).
\end{eqnarray}
The term with the operator of $\sigma_k \tau^-$ is the GT component, and the one with only $\tau^-$ is the Fermi component. Below, all the NME components are classified into these two components. The single-particle wave functions are now those of the canonical basis. 

The final forms of the equations for the calculations read 
\begin{eqnarray}
M_{0\nu}^{(0)} = M_{0\nu}^{\mathrm{GT}(0)} - \frac{g_V^2}{g_A^2} M_{0\nu}^{\mathrm{F}(0)}, \label{eq:M0v0}
\end{eqnarray}
\vspace{-12pt}
\begin{eqnarray}
M_{0\nu}^{\mathrm{GT}(0)} &=& 4\pi R \sum_{B_I B_F} \sum_{abij}  W_{-ai,b-j}^{ \mathrm{GT},FI } \bigl\{\mathrm{sgn}(j_a^z ) v_a u_i X_{ai}^{BF} + u_a \mathrm{sgn}(j_i^z) v_i Y_{-a-i}^{BF} \bigr\}\langle B_F|B_I \rangle \nonumber\\
&& \times\bigl\{ \mathrm{sgn}(j_b^z) v_b u_j Y_{-b-j}^{BI} + u_b \mathrm{sgn}(j_j^z) v_j X_{bj}^{BI} \bigr\},
\end{eqnarray}
\vspace{-12pt}
\begin{eqnarray}
M_{0\nu}^{\mathrm{F}(0)} &=& 4\pi R \sum_{B_I B_F} \sum_{abij}  W_{-ai,b-j}^{ \mathrm{F},FI } \bigl\{\mathrm{sgn}(j_a^z ) v_a u_i X_{ai}^{BF} + u_a \mathrm{sgn}(j_i^z) v_i Y_{-a-i}^{BF} \bigr\}\langle B_F|B_I \rangle \nonumber\\
&& \times\bigl\{ \mathrm{sgn}(j_b^z) v_b u_j Y_{-b-j}^{BI} + u_b \mathrm{sgn}(j_j^z) v_j X_{bj}^{BI} \bigr\},
\end{eqnarray}
\vspace{-12pt}
\begin{eqnarray}
M_{0\nu}^{(JJV)} = M_{0\nu}^{\mathrm{GT}(JJV)} - \frac{g_V^2}{g_A^2} M_{0\nu}^{\mathrm{F}(JJV)} ,
\end{eqnarray}
\vspace{-12pt}
\begin{eqnarray}
M_{0\nu}^{\mathrm{GT}(JJV)} &\simeq& -4\pi R \sum_{B_I B_F}  \sum_{\substack{ bcjk\\ \mathrm{all\ different} } } \sum_{ai} \mathrm{sgn}(j_b^z) v_b^I u_j^I (W^{\mathrm{GT},FI})_{-bj,-ai}^T\mathrm{sgn}(j_a^z) v_a^F u_i^F X_{ai}^{BF} \langle B_F|B_I
\rangle \nonumber\\
&& \times X_{ck}^{BI} \Bigl(\overline{\mathcal{V}}_{ck,jb}^{(4a1)I} +\overline{\mathcal{V}}_{ck,jb}^{(4a2)I} \Bigr), \label{eq:M0vGTJJV}
\end{eqnarray}
\vspace{-12pt}
\begin{eqnarray}
M_{0\nu}^{\mathrm{F}(JJV)} &\simeq& -4\pi R \sum_{B_I B_F}  \sum_{\substack{ bcjk\\ \mathrm{all\ different} } } \sum_{ai} \mathrm{sgn}(j_b^z) v_b^I u_j^I (W^{\mathrm{F},FI})_{-bj,-ai}^T\mathrm{sgn}(j_a^z) v_a^F u_i^F X_{ai}^{BF} \langle B_F|B_I
\rangle \nonumber\\
&& \times X_{ck}^{BI} \Bigl(\overline{\mathcal{V}}_{ck,jb}^{(4a1)I} +\overline{\mathcal{V}}_{ck,jb}^{(4a2)I} \Bigr), \label{eq:M0vFJJV}
\end{eqnarray}
\vspace{-12pt}
\begin{eqnarray}
\overline{\mathcal{V}}_{ck,jb}^{(4a1)I} = \frac{1}{2} u_c^I \mathrm{sgn}(j_k^z) v_k^I \langle cj|V|-b-k\rangle_I u_j^I \mathrm{sgn}(j_b^z) v_b^I \frac{1}{ \widetilde{\varepsilon}_{cc}^I + \widetilde{\varepsilon}_{jj}^I + \widetilde{\varepsilon}_{bb}^I +\widetilde{\varepsilon}_{kk}^I }, \label{eq:V4a1I}
\end{eqnarray}
\vspace{-12pt}
\begin{eqnarray}
\overline{\mathcal{V}}_{ck,jb}^{(4a2)I} = \frac{1}{2} v_c^I \mathrm{sgn}(j_k^z) u_k^I \langle cj|V|-b-k\rangle_I \mathrm{sgn}(j_b^z) u_b^I v_j^I \frac{1}{ \widetilde{\varepsilon}_{cc}^I + \widetilde{\varepsilon}_{jj}^I + \widetilde{\varepsilon}_{bb}^I + \widetilde{\varepsilon}_{kk}^I }, \label{eq:V4a2I}
\end{eqnarray}
\vspace{-12pt}
\begin{eqnarray}
(W^{\mathrm{GT},FI})_{-bj,-ai}^T = W_{-ai,-bj}^{\mathrm{GT},FI}, \hspace{20pt}
(W^{\mathrm{F},FI} )_{-bj,-ai}^T = W_{-ai,-bj}^{\mathrm{F},FI},
\end{eqnarray}
\vspace{-12pt}
\begin{eqnarray}
M_{0\nu}^{(VJJ)} = M_{0\nu}^{\mathrm{GT}(VJJ)} - \frac{g_V^2}{g_A^2} M_{0\nu}^{\mathrm{F}(VJJ)},
\end{eqnarray}
\vspace{-12pt}
\begin{eqnarray}
M_{0\nu}^{\mathrm{GT}(VJJ)} &\simeq& -4\pi R \sum_{B_I B_F}  \sum_{ \substack{ bcjk \\ \mathrm{all\ different} } } \Bigl( \overline{\mathcal{V}}_{jb,ck}^{(4b1)F} + \overline{\mathcal{V}}_{jb,ck}^{(4b2)F} \Bigr) X_{ck}^{BF} \langle B_F|B_I \rangle \sum_{ai} u_a^I \mathrm{sgn}(j_i^z) v_i^I X_{ai}^{BI} \nonumber\\
&& \times(W^{\mathrm{GT},FI} )_{a-i,b-j}^T u_b^F \mathrm{sgn}(j_j^z) v_j^F,\label{eq:M0vGTVJJ}
\end{eqnarray}
\vspace{-12pt}
\begin{eqnarray}
M_{0\nu}^{\mathrm{F}(VJJ)} &\simeq& -4\pi R \sum_{B_I B_F}  \sum_{ \substack{ bcjk \\ \mathrm{all\ different} } } \Bigl( \overline{\mathcal{V}}_{jb,ck}^{(4b1)F} + \overline{\mathcal{V}}_{jb,ck}^{(4b2)F} \Bigr) X_{ck}^{BF} \langle B_F|B_I \rangle \sum_{ai} u_a^I \mathrm{sgn}(j_i^z) v_i^I X_{ai}^{BI} \nonumber\\
&& \times(W^{\mathrm{F},FI} )_{a-i,b-j}^T u_b^F \mathrm{sgn}(j_j^z) v_j^F,\label{eq:M0vFVJJ}
\end{eqnarray}
\vspace{-12pt}
\begin{eqnarray}
\overline{\mathcal{V}}_{jb,ck}^{(4b1)F} = \frac{1}{2} v_j^F \mathrm{sgn}(j_b^z ) u_b^F \langle cj|V|-b-k\rangle_F\,  v_c^F \mathrm{sgn}(j_k^z) u_k^F  \frac{1}{ \widetilde{\varepsilon}_{cc}^F + \widetilde{\varepsilon}_{jj}^F + \widetilde{\varepsilon}_{bb}^F + \widetilde{\varepsilon}_{kk}^F }, \label{eq:V4b1F}
\end{eqnarray}
\vspace{-12pt}
\begin{eqnarray}
\overline{\mathcal{V}}_{jb,ck}^{(4b2)F} = \frac{1}{2} \mathrm{sgn}(j_b^z ) v_b^F u_j^F  \langle cj|V|-b-k\rangle _F  \,\mathrm{sgn}(j_k^z ) v_k^F u_c^F  \frac{1}{ \widetilde{\varepsilon}_{cc}^F + \widetilde{\varepsilon}_{jj}^F + \widetilde{\varepsilon}_{bb}^F + \widetilde{\varepsilon}_{kk}^F }. \label{eq:V4b2F}
\end{eqnarray}
The vertex corrections are obtained by the calculation of the trace of the products of matrixes. We refer to the sum of $M_{0\nu}^{(JJV)}$ and $M_{0\nu}^{(VJJ)}$ as the vertex correction;
\begin{eqnarray}
M_{0\nu}^{(\mathrm{vc})} =  M_{0\nu}^{(VJJ)} + M_{0\nu}^{(JJV)} = M_{0\nu}^{\mathrm{GT(vc)}} - \frac{g_V^2}{g_A^2} M_{0\nu}^{\mathrm{F(vc)}} , \label{eq:M0vvc}
\end{eqnarray}
\vspace{-12pt}
\begin{eqnarray}
M_{0\nu}^{\mathrm{GT(vc)}} = M_{0\nu}^{\mathrm{GT}(VJJ)} + M_{0\nu}^{\mathrm{GT}(JJV)}, \hspace{20pt} M_{0\nu}^{\mathrm{F(vc)}} = M_{0\nu}^{\mathrm{F}(VJJ)} + M_{0\nu}^{\mathrm{F}(JJV)} .
\end{eqnarray}
The origin of $M_{0\nu}^{(\mathrm{vc})}$ is Eqs.~(\ref{eq:MbbJJV_general}) and (\ref{eq:MbbVJJ_general}).

\subsection{ \label{subsec:2bc} Two-body current correction }
\begin{figure}[t]
\includegraphics[width=0.3\columnwidth]{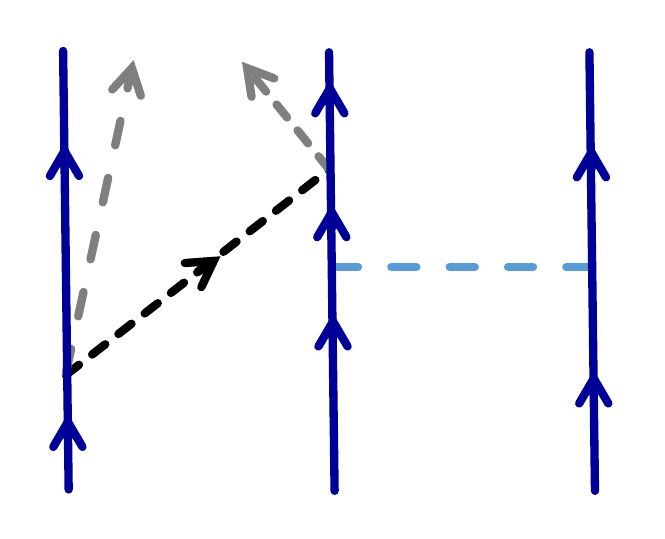}
\vspace{-10pt}
\caption{ \protect \label{fig:diagram_0v2bc} \baselineskip=13pt
Diagram of two-body current correction. For the definitions of the parts, see captions of Figs.\ \ref{fig:diagram_0vld} and \ref{fig:diagram_0vvc}. }
\end{figure}

This correction is defined diagrammatically by the diagram of Fig.\ \ref{fig:diagram_0v2bc}, and the equation is obtained by choosing the components corresponding to the diagram from $\mathcal{M}_{\beta\beta}^{(JVJ)}(\bm{x},\bm{y})$ (\ref{eq:MbbJVJ_general}). The central force has a contribution to our 2bc term. The result of the derivation is 
\begin{eqnarray}
M_{0\nu}^{\mathrm{(2bc)}} = M_{0\nu}^{\mathrm{GT(2bc)}} - \frac{ g_V^2}{g_A^2} M_{0\nu}^{\mathrm{F(2bc)}}, \label{eq:M0v2bc}
\end{eqnarray}
\vspace{-12pt}
\begin{eqnarray}
M_{0\nu}^{\mathrm{GT(2bc)}} &=& 8\pi R \sum_m \sum_{ai} \delta_{im} \sum_{B_F^\prime} \mathrm{sgn}(j_{-a}^z) v_a^F u_i^F X_{ai}^{B^\prime F} \sum_{bj} X_{bj}^{B^\prime F\ast} \delta_{mj} \sum_{ck} V_{c-j,-bk} \,\mathrm{sgn}(j_k^z)  \mathrm{sgn}(j_j^z) \nonumber\\
&& \times ( u_j^F v_b^F v_c^F u_k^F + u_b^F v_j^F v_k^F u_c^F ) \sum_{B_F} X_{ck}^{BF} \sum_{B_I} \langle B_F|B_I \rangle \sum_{dl} X_{dl}^{BI\ast} u_d^I \,\mathrm{sgn}(j_l^z) v_l^I \nonumber\\ 
&& \times W_{-ai,d-l}^{\mathrm{GT},0\nu 2b}, \label{eq:M0vGT2bc}
\end{eqnarray}
\vspace{-12pt}
\begin{eqnarray}
M_{0\nu}^{\mathrm{F(2bc)}} &=& 8\pi R \sum_m \sum_{ai} \delta_{im} \sum_{B_F^\prime} \mathrm{sgn}(j_{-a}^z) v_a^F u_i^F X_{ai}^{B^\prime F} \sum_{bj} X_{bj}^{B^\prime F\ast} \delta_{mj} \sum_{ck} V_{c-j,-bk} \,\mathrm{sgn}(j_k^z)  \mathrm{sgn}(j_j^z) \nonumber\\
&& \times ( u_j^F v_b^F v_c^F u_k^F + u_b^F v_j^F v_k^F u_c^F ) \sum_{B_F} X_{ck}^{BF} \sum_{B_I} \langle B_F|B_I \rangle \sum_{dl} X_{dl}^{BI\ast} u_d^I \,\mathrm{sgn}(j_l^z) v_l^I \nonumber\\ 
&& \times W_{-ai,d-l}^{\mathrm{F},0\nu 2b}, \label{eq:M0vF2bc}
\end{eqnarray}
\vspace{-12pt}
\begin{eqnarray}
W_{-ai,d-l}^{\mathrm{GT},0\nu 2b} &\equiv& \int d^3\bm{x} d^3\bm{y} \int \frac{d^3 \bm{q}}{(2\pi)^3} \frac{1}{|\bm{q}|} \frac{ \mathrm{exp}[i\bm{q}\cdot (\bm{x} - \bm{y})]}{ \bigl\{\overline{E}_B + |\bm{q}| - \frac{1}{2}(E_I + E_F)\bigr\}^2 } \sum_{k=1-3}\psi_{-a}^{F\dagger}(\bm{x}) \sigma_k \tau^- \psi_i^F (\bm{x}) \nonumber\\
&& \times\psi_d^{I\dagger}(\bm{y}) \sigma_k \tau^- \psi_{-l}^I (\bm{y}), \label{eq:WGT0v2b}
\end{eqnarray}
\vspace{-12pt}
\begin{eqnarray}
W_{-ai,d-l}^{\mathrm{F},0\nu 2b} &\equiv& \int d^3\bm{x} d^3\bm{y} \int \frac{d^3 \bm{q}}{(2\pi)^3} \frac{1}{|\bm{q}|} \frac{ \mathrm{exp}[i\bm{q}\cdot (\bm{x} - \bm{y})]}{ \bigl\{\overline{E}_B + |\bm{q}| - \frac{1}{2}(E_I + E_F)\bigr\}^2 } \psi_{-a}^{F\dagger}(\bm{x}) \tau^- \psi_i^F (\bm{x}) \nonumber\\
&& \times\psi_d^{I\dagger}(\bm{y}) \tau^- \psi_{-l}^I (\bm{y}). \label{eq:WF0v2b}
\end{eqnarray}
Two sets of the intermediate states of $B$ and $B^\prime$ are used, and both are created by the QRPA. $B$ is represented by $B_I$ based on the initial state and $B_F$ based on the final state. $B^\prime \equiv B_F^\prime$ is obtained using the final state. Because of this structure, the approximation of the quasiparticle energy in the vc term is not necessary here. Only the exchange matrix elements of the proton-neutron interaction are used for $V_{c-j,-bk}$ because the origin of this matrix element is $\langle B_F^\prime|\!:\!V\!:\!|B_F \rangle$; see Eq.\ (\ref{eq:MbbJVJ_general}). The pairing interaction is not used for the 2bc term because the transition operator of the $\beta\beta$ decay is of the particle-hole type. 

The calculation of the 2bc correction includes the term 
\begin{eqnarray}
\langle B_F^\prime |:V:|B_F \rangle \sim \langle F|(a_\mu^F a_\alpha^F) a_\gamma^{F\dagger} a_\kappa^{F\dagger} a_\delta^F a_\lambda^F (a_\beta^{F\dagger} a_\nu^{F\dagger})|F\rangle.
\end{eqnarray}
The nonvanishing terms can be classified into two groups. One group is the terms satisfying 
\begin{eqnarray}
a_\delta^F a_\lambda^F (a_\beta^{F\dagger} a_\nu^{F\dagger})|F\rangle = \pm|F\rangle,
\end{eqnarray}
\vspace{-12pt}
\begin{eqnarray}
\langle F|(a_\mu^F a_\alpha^F) a_\gamma^{F\dagger} a_\kappa^{F\dagger} = (\pm \hbox{ or } \mp)\langle F|.
\end{eqnarray}
The other group is the terms not satisfying this condition. The terms of this group have one more condition for the indexes to have nonvanishing terms compared to the first group. One more condition implies many fewer terms. Thus, we ignore the terms of this group. 

For computation of $M_{0\nu}^{(2bc)}$, the trace is calculated of the product of the matrixes depending on the canonical quasiparticle state $m$; see Eqs.\ (\ref{eq:M0vGT2bc}) and (\ref{eq:M0vF2bc}). This state $m$ connects the left edge of the NN interaction and the upper vertex of the weak interaction in Fig.\ \ref{fig:diagram_0v2bc}. The 2bc term is obtained by the summation of the traces with respect to $m$. 

Finally, we have the $0\nu\beta\beta$ NME by the perturbed transition operator
\begin{eqnarray}
M_{0\nu} = M_{0\nu}^{(0)} + M_{0\nu}^\mathrm{(vc)} + M_{0\nu}^{(2bc)},
\end{eqnarray}
which is used for the decay probability normalized for the half-life $T_{1/2}^{0\nu}$ 
\begin{eqnarray}
\Gamma_{0\nu} = \frac{1}{T_{1/2}^{\,0\nu}} = G_{0\nu} |M_{0\nu}|^2 \left( \frac{\langle m_\nu \rangle}{m_e} \right)^2.
\end{eqnarray}
For convenience, we define
\begin{eqnarray}
M_{0\nu}^\mathrm{GT} = M_{0\nu}^\mathrm{GT(0)} + M_{0\nu}^\mathrm{GT(vc)} + M_{0\nu}^\mathrm{GT(2bc)} ,
\end{eqnarray}
\vspace{-12pt}
\begin{eqnarray}
M_{0\nu}^F = M_{0\nu}^\mathrm{F(0)} + M_{0\nu}^\mathrm{F(vc)} + M_{0\nu}^\mathrm{F(2bc)} .
\end{eqnarray}
One can also write
\begin{eqnarray}
M_{0\nu} = M_{0\nu}^\mathrm{GT} - \frac{g_V^2}{g_A^2} M_{0\nu}^\mathrm{F}.
\end{eqnarray}

\subsection{\label{subsec:2vbb_NME} Two-neutrino double-$\bm{\beta}$ nuclear matrix element }
The NME of the $2\nu\beta\beta$ decay is decoupled from the lepton sector by approximation. Technically, this NME can be obtained by replacing the neutrino potential in the $0\nu\beta\beta$ NME with the reciprocal of the energy denominator for the $2\nu\beta\beta$ NME and adjusting the constant factor at the beginning of the equation. The neutrino potential is defined by
\begin{eqnarray}
\widetilde{h}_+ (r) = \int \frac{d^3\bm{q}}{(2\pi)^3} \frac{1}{|\bm{q}|} \frac{ \mathrm{exp}[i\bm{q}\cdot(\bm{x} - \bm{y})] }{ \overline{E}_B + |\bm{q}| - \frac{1}{2}(E_I + E_F) }, \label{eq:neutrino_potential}
\end{eqnarray}
\vspace{-12pt}
\begin{eqnarray}
r = |\bm{r}|, \hspace{20pt} \bm{r} = \bm{x} - \bm{y},
\end{eqnarray}
which is used for $M_{0\nu}^{(0)}$, $M_{0\nu}^{(VJJ)}$, and $M_{0\nu}^{(JJV)}$. That for $M_{0\nu}^\mathrm{(2bc)}$ can be found in Eqs.\ (\ref{eq:WGT0v2b}) and (\ref{eq:WF0v2b}). The energy denominator of the $2\nu\beta\beta$ NME is intermediate-state dependent and independent of the lepton energies approximately. The leading term of the $2\nu\beta\beta$ NME is given by 
\begin{eqnarray}
M_{2\nu}^{(0)} = M_{2\nu}^\mathrm{GT(0)} - \frac{g_V^2}{g_A^2} M_{2\nu}^\mathrm{F(0)}, \label{eq:M2v0}
\end{eqnarray}
\vspace{-12pt}
\begin{eqnarray}
M_{2\nu}^\mathrm{GT(0)} &=& -m_e \sum_{B_I B_F} \frac{1}{ E_B - \frac{1}{2}(E_I + E_F) } \sum_{ai}\bigl\{u_a^F \mathrm{sgn}(j_i^z) v_i^F Y_{-a-i}^{BF} + \mathrm{sgn}(j_a^z) v_a^F u_i^F X_{ai}^{BF} \bigr\} (\tau^- \bm{\sigma})_{-ai} \nonumber\\
&&  \cdot \sum_{bj} (\tau^- \bm{\sigma})_{b-j} \bigl\{ X_{bj}^{BI} u_b^I \mathrm{sgn}(j_j^z) v_j^I + Y_{-b-j}^{BI} \,\mathrm{sgn}(j_b^z) v_b^I u_j^I \bigr\}\langle B_F|B_I \rangle,
\end{eqnarray}
\vspace{-12pt}
\begin{eqnarray}
M_{2\nu}^\mathrm{F(0)} &=& -m_e \sum_{B_I B_F} \frac{1}{ E_B - \frac{1}{2}(E_I + E_F) } \sum_{ai}\bigl\{u_a^F \mathrm{sgn}(j_i^z) v_i^F Y_{-a-i}^{BF} + \mathrm{sgn}(j_a^z) v_a^F u_i^F X_{ai}^{BF} \bigr\} (\tau^-)_{-ai} \nonumber\\
&&  \times \sum_{bj} (\tau^-)_{b-j} \bigl\{ X_{bj}^{BI} u_b^I \mathrm{sgn}(j_j^z) v_j^I + Y_{-b-j}^{BI} \,\mathrm{sgn}(j_b^z) v_b^I u_j^I \bigr\}\langle B_F|B_I \rangle,
\end{eqnarray}
\vspace{-12pt}
\begin{eqnarray}
(\tau^- \bm{\sigma})_{-ai} = \int d^3 \bm{x} \psi_{-a}^\dagger (\bm{x}) \tau^- \bm{\sigma} \psi_i (\bm{x}).
\end{eqnarray}
The angular momentum and the parity ($J^\pi$) of the intermediate states are limited to $1^+$ for the GT component and $0^+$ for the Fermi component. $E_B$ in the denominator is either $E_{BI}$ or $E_{BF}$. The QRPA is a good approximation for $^{136}$Xe, so that the difference in the two results is negligible. 

\begin{figure}[t]
\includegraphics[width=0.3\columnwidth]{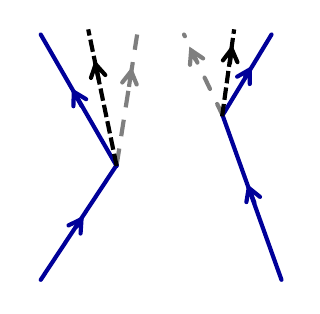}
\vspace{-10pt}
\caption{ \protect \label{fig:diagram_2vld} \baselineskip=13pt 
Diagram of leading term of $2\nu\beta\beta$ NME. Black short-dashed line shows the Dirac antineutrino. The other parts are the same as those of Fig.\ \ref{fig:diagram_0vld}.  }
\end{figure}

\begin{figure}[t]
\includegraphics[width=1.0\columnwidth]{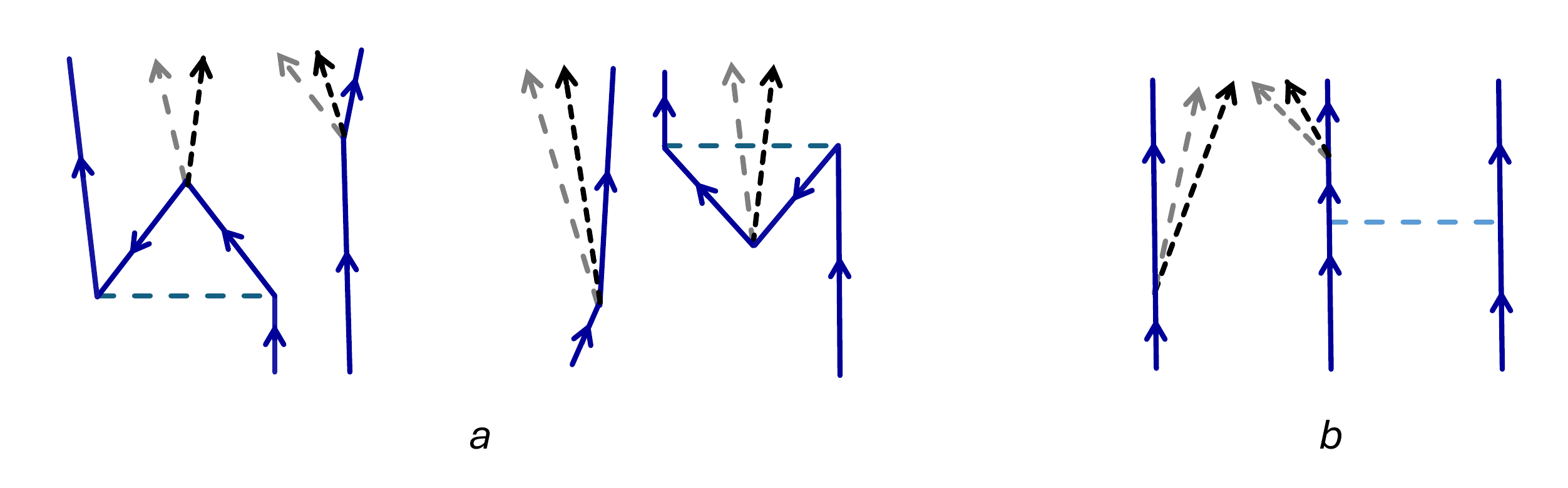}
\vspace{-10pt}
\caption{ \protect \label{fig:diagram_2vcr} \baselineskip=13pt 
Diagrams of correction terms for $2\nu\beta\beta$ NME. Diagram $a$: vertex corrections. Diagram $b$: two-body current term. For the definitions of the parts, see captions of Figs.\ \ref{fig:diagram_0vvc} and \ref{fig:diagram_2vld}. }
\end{figure}

The equations of the vertex correction for the $2\nu\beta\beta$ NME are as follows:
\begin{eqnarray}
M_{2\nu}^{(JJV)} =M_{2\nu}^{\mathrm{GT}(JJV)} - \frac{g_V^2}{g_A^2} M_{2\nu}^{\mathrm{F}(JJV)},
\end{eqnarray}
\vspace{-12pt}
\begin{eqnarray}
M_{2\nu}^{\mathrm{GT}(JJV)} &\simeq& -m_e \sum_{B_I B_F} \sum_{ \substack{bcjk \\ \mathrm{all\  different} } }  \sum_{ai} \mathrm{sgn}(j_b^z) v_b^I u_j^I (W_{2\nu}^{\mathrm{GT},FI})_{-bj,-ai}^T \,\mathrm{sgn}(j_a^z) v_a^F u_i^F X_{ai}^{BF} \nonumber\\
&& \times\langle B_F|B_I \rangle X_{ck}^{BI} \Bigl( \overline{\mathcal{V}}_{ck,jb}^{(4a1)I} + \overline{\mathcal{V}}_{ck,jb}^{(4a2)I} \Bigr),
\end{eqnarray}
\vspace{-12pt}
\begin{eqnarray}
M_{2\nu}^{\mathrm{F}(JJV)} &\simeq& -m_e \sum_{B_I B_F} \sum_{ \substack{bcjk \\ \mathrm{all\  different} } }  \sum_{ai} \mathrm{sgn}(j_b^z) v_b^I u_j^I (W_{2\nu}^{\mathrm{F},FI})_{-bj,-ai}^T \,\mathrm{sgn}(j_a^z) v_a^F u_i^F X_{ai}^{BF} \nonumber\\
&& \times\langle B_F|B_I \rangle X_{ck}^{BI} \Bigl( \overline{\mathcal{V}}_{ck,jb}^{(4a1)I} + \overline{\mathcal{V}}_{ck,jb}^{(4a2)I} \Bigr),
\end{eqnarray}
\vspace{-12pt}
\begin{eqnarray}
 (W_{2\nu}^{\mathrm{GT},FI})_{-bj,-ai}^T &=& (W_{2\nu}^{\mathrm{GT},FI})_{-ai,-bj} \nonumber\\
&=& \frac{1}{ E_B - \frac{1}{2} (E_I + E_F) } \int d^3\bm{x} d^3\bm{y} \sum_{k=1-3} \psi_{-a}^{F\dagger} (\bm{x})\sigma_k \tau^- \psi_i^F(\bm{x}) \nonumber\\
&& \times\psi_{-b}^{I\dagger}(\bm{y}) \sigma_k \tau^- \psi_j^I (\bm{y}),
\end{eqnarray}
\vspace{-12pt}
\begin{eqnarray}
 (W_{2\nu}^{\mathrm{F},FI})_{-bj,-ai}^T &=& (W_{2\nu}^{\mathrm{F},FI})_{-ai,-bj} \nonumber\\
&=& \frac{1}{ E_B - \frac{1}{2} (E_I + E_F) } \int d^3\bm{x} d^3\bm{y} \psi_{-a}^{F\dagger} (\bm{x}) \tau^- \psi_i^F(\bm{x}) \psi_{-b}^{I\dagger}(\bm{y}) \tau^- \psi_j^I (\bm{y}),
\end{eqnarray}
($E_B$ is either $E_{BI}$ or $E_{BF}$),
\begin{eqnarray}
M_{2\nu}^{(VJJ)} = M_{2\nu}^{\mathrm{GT}(VJJ)} - \frac{g_V^2}{g_A^2} M_{2\nu}^{\mathrm{F}(VJJ)} ,
\end{eqnarray}
\vspace{-12pt}
\begin{eqnarray}
M_{2\nu}^{\mathrm{GT}(VJJ)} &\simeq& -m_e \sum_{B_I B_F} \sum_{ \substack{ bcjk \\ \mathrm{all\  different} } } \Bigl( \overline{\mathcal{V}}_{jb,ck}^{(4b1)F} + \overline{\mathcal{V}}_{jb,ck}^{(4b2)F} \Bigr) X_{ck}^{BF} \langle B_F|B_I \rangle \sum_{ai} u_a^I \mathrm{sgn}(j_i^z) v_i^I X_{ai}^{BI} \nonumber\\
&& \times(W_{2\nu}^{\mathrm{GT},FI})_{a-i,b-j}^T u_b^F \mathrm{sgn}(j_j^z) v_j^F,
\end{eqnarray}
\vspace{-12pt}
\begin{eqnarray}
M_{2\nu}^{\mathrm{F}(VJJ)} &\simeq& -m_e \sum_{B_I B_F} \sum_{ \substack{ bcjk \\ \mathrm{all\  different} } } \Bigl( \overline{\mathcal{V}}_{jb,ck}^{(4b1)F} + \overline{\mathcal{V}}_{jb,ck}^{(4b2)F} \Bigr) X_{ck}^{BF} \langle B_F|B_I \rangle \sum_{ai} u_a^I \mathrm{sgn}(j_i^z) v_i^I X_{ai}^{BI} \nonumber\\
&& \times(W_{2\nu}^{\mathrm{F},FI})_{a-i,b-j}^T u_b^F \mathrm{sgn}(j_j^z) v_j^F.
\end{eqnarray}
$\overline{\mathcal{V}}_{ck,jb}^{(4a1)I}$, $\overline{\mathcal{V}}_{ck,jb}^{(4a2)I}$, $\overline{\mathcal{V}}_{jb,ck}^{(4b1)F}$, and $\overline{\mathcal{V}}_{jb,ck}^{(4b2)F}$ used for the $0\nu\beta\beta$ NME can be also used here; see Eqs.\ (\ref{eq:V4a1I}), (\ref{eq:V4a2I}), (\ref{eq:V4b1F}), and (\ref{eq:V4b2F}). As the $0\nu\beta\beta$ NME, we define
\begin{eqnarray}
M_{2\nu}^\mathrm{(vc)} =  M_{2\nu}^{(VJJ)} + M_{2\nu}^{(JJV)} = M_{2\nu}^\mathrm{GT(vc)} - \frac{g_V^2}{g_A^2} M_{2\nu}^\mathrm{F(vc)}, \label{eq:M2vvc}
\end{eqnarray}
\vspace{-12pt}
\begin{eqnarray}
M_{2\nu}^\mathrm{GT(vc)} = M_{2\nu}^{\mathrm{GT}(VJJ)} + M_{2\nu}^{\mathrm{GT}(JJV)}, \hspace{20pt} M_{2\nu}^\mathrm{F(vc)} = M_{2\nu}^{\mathrm{F}(VJJ)} + M_{2\nu}^{\mathrm{F}(JJV)}.
\end{eqnarray}
For the two-body current term, we have
\begin{eqnarray}
M_{2\nu}^\mathrm{(2bc)} = M_{2\nu}^\mathrm{GT(2bc)} - \frac{g_V^2}{g_A^2} M_{2\nu}^\mathrm{F(2bc)} , \label{eq:M2v2bc}
\end{eqnarray}
\vspace{-12pt}
\begin{eqnarray}
M_{2\nu}^\mathrm{GT(2bc)} &=& 2m_e \sum_m \sum_{ai} \delta_{im} \sum_{B_F^\prime} \mathrm{sgn}(j_{-a}^z) v_a^F u_i^F X_{ai}^{B^\prime F} \sum_{bj} X_{bj}^{B^\prime F\ast} \delta_{mj} \sum_{ck} V_{c-j,-bk} \,\mathrm{sgn}(j_k^z) \mathrm{sgn}(j_j^z) \nonumber \\
&& \times( u_j^F v_b^F v_c^F u_k^F + u_b^F v_j^F v_k^F u_c^F ) \sum_{B_F} X_{ck}^{BF} \sum_{B_I} \langle B_F|B_I \rangle \sum_{dl} X_{dl}^{BI\ast} u_d^I \,\mathrm{sgn}(j_l^z) v_l^I W_{-ai,d-l}^{\mathrm{GT},2\nu 2b}, \nonumber\\ \label{M2vGT2bc}
\end{eqnarray}
\vspace{-12pt}
\begin{eqnarray}
M_{2\nu}^\mathrm{F(2bc)} &=& 2m_e \sum_m \sum_{ai} \delta_{im} \sum_{B_F^\prime} \mathrm{sgn}(j_{-a}^z) v_a^F u_i^F X_{ai}^{B^\prime F} \sum_{bj} X_{bj}^{B^\prime F\ast} \delta_{mj} \sum_{ck} V_{c-j,-bk} \,\mathrm{sgn}(j_k^z) \mathrm{sgn}(j_j^z) \nonumber \\
&& \times( u_j^F v_b^F v_c^F u_k^F + u_b^F v_j^F v_k^F u_c^F ) \sum_{B_F} X_{ck}^{BF} \sum_{B_I} \langle B_F|B_I \rangle \sum_{dl} X_{dl}^{BI\ast} u_d^I \,\mathrm{sgn}(j_l^z) v_l^I W_{-ai,d-l}^{\mathrm{F},2\nu 2b}, \nonumber\\ \label{M2vGT2bc}
\end{eqnarray}
\vspace{-12pt}
\begin{eqnarray}
W_{-ai,d-l}^{\mathrm{GT},2\nu 2b} \equiv \int d^3\bm{x} d^3\bm{y} \sum_{k=1-3} \frac{ \psi_{-a}^{F\dagger}(\bm{x}) \sigma_k \tau^- \psi_i^F(\bm{x}) \psi_d^{I\dagger}(\bm{y}) \sigma_k \tau^- \psi_{-l}^I(\bm{y}) }{ \bigl\{ E_{BF} - \frac{1}{2} (E_I + E_F) \bigr\} \bigl\{ E_{BI} - \frac{1}{2}(E_I + E_F) \bigr\} },
\end{eqnarray}
\vspace{-12pt}
\begin{eqnarray}
W_{-ai,d-l}^{\mathrm{F},2\nu 2b} \equiv \int d^3\bm{x} d^3\bm{y} \frac{ \psi_{-a}^{F\dagger}(\bm{x}) \tau^- \psi_i^F(\bm{x}) \psi_d^{I\dagger}(\bm{y}) \tau^- \psi_{-l}^I(\bm{y}) }{ \bigl\{ E_{BF} - \frac{1}{2} (E_I + E_F) \bigr\} \bigl\{ E_{BI} - \frac{1}{2}(E_I + E_F) \bigr\} }.
\end{eqnarray}
$M_{2\nu}^{(0)}$ is illustrated by the diagram of Fig.\ \ref{fig:diagram_2vld}, and the correction terms are illustrated in Fig.\ \ref{fig:diagram_2vcr}; the two diagrams of Fig.\ 5$a$ express $M_{2\nu}^{(JJV)}$ (left) and $M_{2\nu}^{(VJJ)}$ (right), and diagram of Fig.\ 5$b$ corresponds to $M_{2\nu}^\mathrm{(2bc)}$. The vertical interval between the two weak-interaction vertexes implies a virtual nuclear state.

The phase space factor for the $2\nu\beta\beta$ decay \cite{Doi85} is derived by the phase space integral of the squared transition matrix element of the lepton currents; 
\begin{eqnarray}
G_{2\nu} &\simeq& \frac{\pi}{2} \frac{(Gg_A )^4}{m_e^2 \,\mathrm{ln}(2)} \int \frac{d^3 \bm{p}_1}{(2\pi)^3} \frac{d^3 \bm{p}_2}{(2\pi)^3} \frac{d^3 \bm{q}_1}{(2\pi)^3} \frac{d^3 \bm{q}_2}{(2\pi)^3} F_0(Z,\varepsilon_1 )F_0(Z,\varepsilon_2 )\langle \mu_B \rangle^2 \left( \frac{\langle K_B \rangle + \langle L_B \rangle}{2} \right)^2 \nonumber\\
&& \times\delta( \varepsilon_1 + \varepsilon_2 + \omega_1 + \omega_2 + E_F - E_I ),
\end{eqnarray}
\vspace{-12pt}
\begin{eqnarray}
\varepsilon_k = \sqrt{ p_k^2 + m_e^2 }, \hspace{20pt} \omega_k = q_k, \hspace{20pt}(k=1,2),
\end{eqnarray}
\vspace{-12pt}
\begin{eqnarray}
\langle K_B \rangle = \left[ \langle \mu_B \rangle + \frac{ \varepsilon_1 + \omega_1 -\varepsilon_2 - \omega_2 }{2m_e} \right]^{-1} + \left[ \langle \mu_B \rangle - \frac{ \varepsilon_1 + \omega_1 - \varepsilon_2 - \omega_2 }{2m_e} \right]^{-1},
\end{eqnarray}
\vspace{-12pt}
\begin{eqnarray}
\langle L_B \rangle = \langle K_B \rangle ( \omega_1 \leftrightarrow \omega_2 ).
\end{eqnarray}
Here, $\langle \mu_B \rangle$ is the average value of
\begin{eqnarray}
\mu_B = \frac{1}{m_e} \left( E_B - \frac{E_I+E_F}{2} \right),
\end{eqnarray}
with respect to the intermediate state $B$. Two emitted electrons have the energies $\varepsilon_k$ and the momenta $\bm{p}_k$ $(k=1,2)$, and two emitted Dirac antineutrinos have the energies $\omega_k$ and the momenta $\bm{q}_k$. For $F_0(Z,\varepsilon)$, see Eq.\ (\ref{eq:F0Ze}). The factor 
\begin{eqnarray}
\langle \mu_B \rangle^2 \left( \frac{\langle K_B \rangle + \langle L_B \rangle}{2} \right)^2,
\end{eqnarray}
is the approximate correction bringing the $\langle \mu_B\rangle$ dependence to $G_{2\nu}$; in fact, this dependence is small \cite{Doi85}. Originally, the transition matrix element is not factorized to the lepton and the hadron sectors. Keeping an influence of the coupling may be the motivation for this factor. 

Finally, the half-life $T_{1/2}^{2\nu}$ can be determined by
\begin{eqnarray}
\frac{1}{ T_{1/2}^{2\nu} } = G_{2\nu} |M_{2\nu}|^2, 
\end{eqnarray}
\vspace{-12pt}
\begin{eqnarray}
M_{2\nu} = M_{2\nu}^{(0)} + M_{2\nu}^{(\mathrm{vc})} +M_{2\nu}^{(\mathrm{2bc})}. 
\end{eqnarray}

\subsection{\label{subsec:other_diagrams}Other diagrams}
\begin{figure}[t]
\includegraphics[width=0.9\columnwidth]{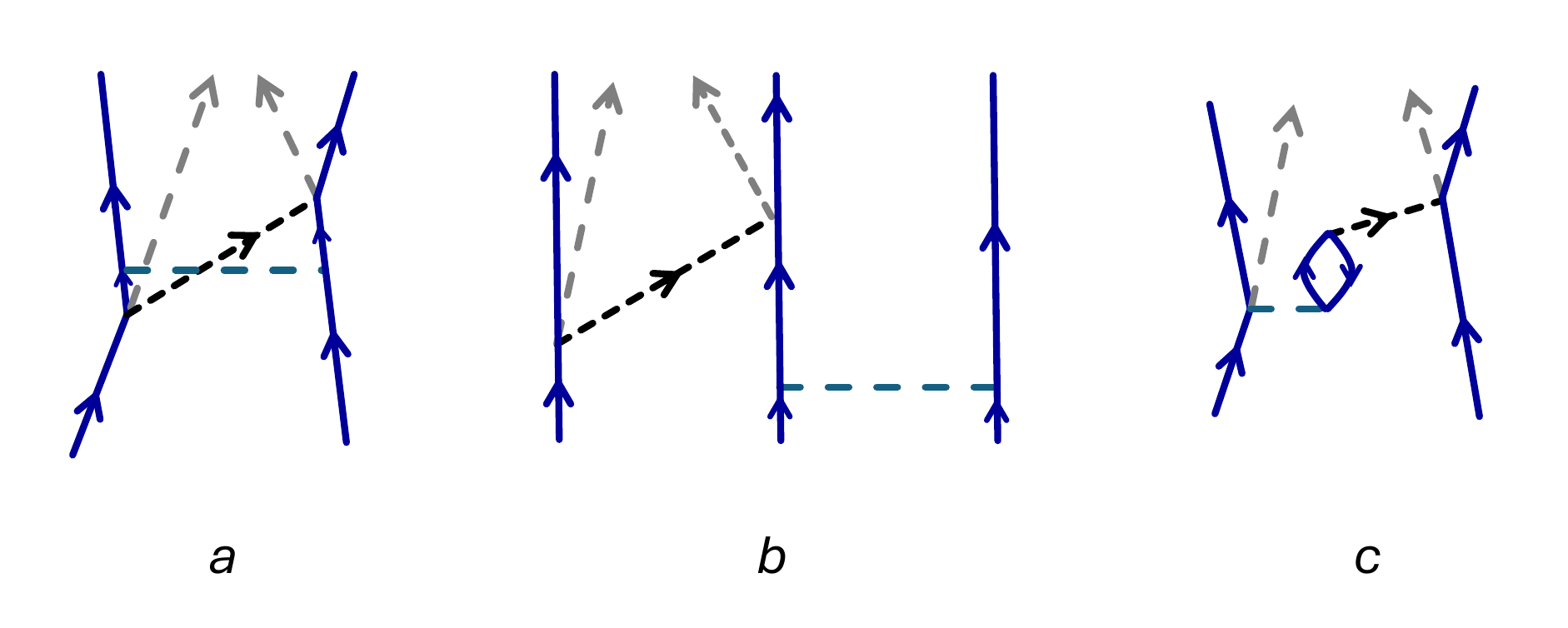}
\vspace{-10pt}
\caption{ \protect \label{fig:diagram_others} \baselineskip=13pt 
Reference diagrams for discussion. For the parts, see captions of Figs.\ \ref{fig:diagram_0vld} and \ref{fig:diagram_0vvc}.  }
\end{figure}

We study the lowest-order correction to the transition operator with respect to the vertex. Still, a few diagrams of this order are not calculated in this study, and the reasons are worthy of discussing. Figure \ref{fig:diagram_others}$a$ shows the diagram with the NN interaction activated during the Majorana neutrino propagation; the two interactions act on the same pair of nucleons. The neutrino speed is very close to the light speed. Suppose that the neutrino reaches from one nucleon to the other instantly. This implies that the NN interaction and the neutrino exchange interaction occur simultaneously between the two nucleons. This possibility may be negligible. If the same assumption is applied for the diagram of Fig.\ \ref{fig:diagram_0v2bc}, a three-body interaction is constructed. This is possible; thus, the 2bc term is realistic. 

The nucleon basis used in the calculation is determined by solving the HFB equation. The HFB ground state has the lowest energy, so the effects of the interaction are maximally included in the basis. The nucleons in the initial and final states are not necessary to modify. Thus, the diagram of Fig.\ \ref{fig:diagram_others}$b$ is not calculated. The self-energy term of the nucleon is not calculated due to the same reason. 

The diagram of Fig.\ \ref{fig:diagram_others}$c$ is not included in our calculation because the interaction that we use does not have a charge exchange component, as mentioned above. The perturbation interaction at the current order is identical to that used for the proton-neutron QRPA. 

The diagrams including the charged-pion exchange between a nucleon and a lepton \cite{Ver02} are not included in our current calculation because we consider only the perturbation due to the NN interaction in this article; the pion propagator of these diagrams cannot be replaced by the NN interaction. 

We summarize the applicability of the exclusions of those diagrams. The exclusion of the diagram 6a is general. The diagram 6b is excluded, as long as the quasiparticle or single-particle basis is obtained consistently with the interaction. The diagram 6c does not occur if the perturbation interaction does not have a charge exchange component.

\section{ \label{sec:result_discussion} RESULT OF NUMERICAL CALCULATION AND DISCUSSION }
The calculations were performed for $^{136}$Xe $\rightarrow$ $^{136}$Ba using the Skyrme interaction (EDF) for the perturbation interaction and the QRPA Hamiltonian. The contact isoscalar and isovector pairing interactions \cite{Ter20} were also used for the QRPA Hamiltonian. We used two parameter sets of the Skyrme interaction SkM$^\ast$ \cite{Bar82} and SGII \cite{Gia81}. The first step of the calculation is to solve the HFB equation. The equation is formulated with a coordinate mesh within a cylindrical box \cite{Ter03,Bla05,Obe07}. The solutions have the axial symmetry, i.e., the M scheme is used. For the QRPA states used for the intermediate states, max($K$) = 8. The QRPA solutions are obtained in the $K\pi$ spaces \cite{Ter10}. In the calculations using the Skyrme interaction SkM$^\ast$, the dimension of the two-canonical-quasiparticle basis for the QRPA Hamiltonian is 35$\,$000 for $K = 0$ and 15$\,$000 for $K = 8$. The dimensions of the single-particle basis for (proton, neutron) is (4$\,$500, 3$\,$900) including both $j_z>0$ and $j_z<0$. The dimension of the two-body $\beta\beta$ transition operator matrix is 16$\,$000 for $K = 0$ and 6$\,$400 for $K = 8$. These dimensions were determined in the leading-order $0\nu\beta\beta$ NME calculations to obtain the convergence of the result. The size parameters of the calculation with SGII are similar to those of the SkM$^\ast$ calculation. 

For convenience in discussion, we call the sum of the vc and the 2bc correction terms the summed correction term, and the sum of the leading and the summed correction terms is called the perturbed term. First, we shall discuss the result with SkM$^\ast$. The leading and the two correction terms for the $0\nu\beta\beta$ and $2\nu\beta\beta$ NME are shown in Table \ref{tab:NME0v2v}. The absolute values of the GT correction terms are of the same order as those of the leading terms. For the Fermi component, the absolute values of the vc and 2bc correction terms are larger than those of the leading terms. The correction terms are as important as the leading term. The sign of the summed correction term is opposite to the corresponding leading term except for the Fermi component of the $2\nu\beta\beta$ NME. A simple rule can be conjectured. The NN interaction is attractive on average; therefore, the summed correction terms have the opposite sign to that of the leading term except for the Fermi component of the $2\nu\beta\beta$ NME. 

\begin{table}
\caption{\label{tab:NME0v2v} The calculated GT and Fermi components of the $0\nu\beta\beta$ and the $2\nu\beta\beta$ NMEs with SkM$^\ast$. The items of leading, vc, and 2bc refer to 
Eqs.\ (\ref{eq:M0v0}), (\ref{eq:M0vvc}), and (\ref{eq:M0v2bc}) of the $0\nu\beta\beta$ NME, respectively, and Eqs.\ (\ref{eq:M2v0}), (\ref{eq:M2vvc}), and (\ref{eq:M2v2bc}) of the $2\nu\beta\beta$ NME, respectively. The sum is the perturbed NME. The sign is chosen so that the GT leading terms are positive.  }
\begin{ruledtabular}
\begin{tabular}{ccccc}
\multicolumn{5}{c}{SkM$^\ast$} \\
\hline\\[-10pt]
 & \multicolumn{2}{c}{$0\nu\beta\beta$ NME} & \multicolumn{2}{c}{$2\nu\beta\beta$ NME} \\
\cline{2-3} \cline{4-5} \\[-11pt]
 & GT & Fermi & GT & Fermi \\
\hline \\[-10pt]
Leading & $\;\;\:$3.095 &      $-$0.467 & $\;\;\:$0.102 &      $-$0.002 \\
vc      & $\;\;\:$1.332 &      $-$0.984 & $\;\;\:$0.055 &      $-$0.033 \\
2bc     &      $-$2.731 & $\;\;\:$1.758 &      $-$0.192 & $\;\;\:$0.030 \\
Sum     & $\;\;\:$1.696 & $\;\;\:$0.307 &      $-$0.035 &      $-$0.005 \\
\end{tabular}
\end{ruledtabular}
\end{table}
%

The $2\nu\beta\beta$ Fermi components of the vc and the 2bc terms cancel by 90 \%. In fact, we chose the term of the 2bc so that the cancellation occurs. The Fermi component of the $2\nu\beta\beta$ NME should be small under the approximate conservation of the isospin. The symmetry-conserving truncation is appropriate in the expansion-truncation approximation. In practice, this condition is not satisfied automatically in arbitrary truncations. If other terms are introduced, the counter terms should also be included. We calculate the corresponding terms of the $0\nu\beta\beta$ NME for consistency. 

\begin{figure}[t]
\includegraphics[width=0.333\columnwidth]{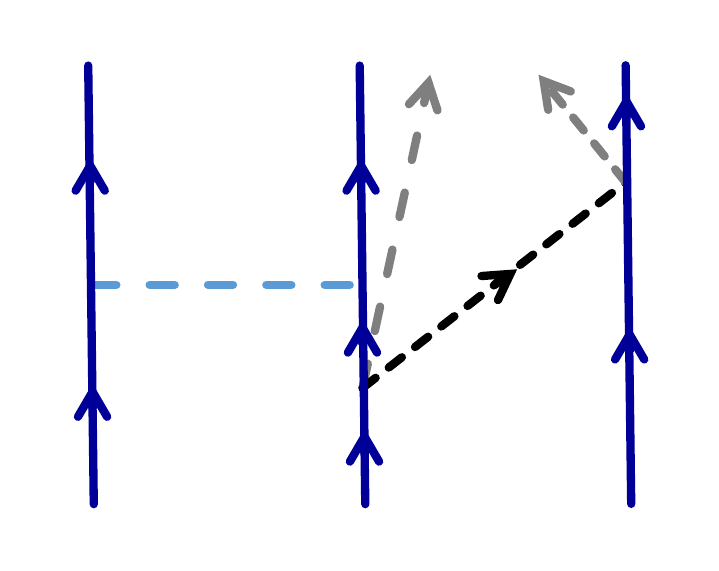}
\vspace{-10pt}
\caption{ \protect \label{fig:diagram_0v2bc_notused} \baselineskip=13pt 
Another possible diagram. For the definitions of the parts, see captions of Figs.\ \ref{fig:diagram_0vld} and \ref{fig:diagram_0vvc}. }
\end{figure}

Another diagram of the 2bc correction is possible; see Fig.\ \ref{fig:diagram_0v2bc_notused}. The Fermi component of $2\nu\beta\beta$ NME with SkM$^\ast$ is $‒$0.0115, and the GT component is 0.0708. We could not find the counter term to that Fermi component in our current calculation, so this term is not included in our result due to the symmetry-conserving truncation. 

If the usual presentation method with the bare $g_A$ of 1.267 is used, the leading and the perturbed $0\nu\beta\beta$ NME are $M_{0\nu}^{(0)}$ = 3.384, and $M_{0\nu}$ = 1.505, respectively. Thus, the NME is reduced by a half due to the correction terms. The corresponding $0\nu\beta\beta$ half-life ($\langle m_\nu \rangle$ = 1 meV assumed) with $M_{0\nu}^{(0)}$ is $6.01\times 10^{29}$ y, and that with the perturbed $M_{0\nu}$ is $3.04\times 10^{30}$ y; the bare $g_A$ is used for both the half-lives. The decrease in the Fermi component of the $0\nu\beta\beta$ NME is so large that its sign is inverted; 
\begin{eqnarray}
M_{0\nu}^\mathrm{F(0)} = ‒0.467, \hspace{20pt} M_{0\nu}^F = 0.307.
\end{eqnarray}

The effective coupling $g_A^\mathrm{eff}$ is used in many calculations involving spin transition operators. The most common definition of $g_A^\mathrm{eff}$ is to reproduce the experimental half-life, e.g., \cite{Bro85}. The NME is calculated by the integrals by the coordinates and momentum; see, e.g., Eq.\ (\ref{eq:M00v}). Therefore, it is possible to define a momentum dependent $g_A^\mathrm{eff}(|\bm{q}|)$ associated with the NME density in the $\bm{q}$ space. Likewise, it is also possible to define the coordinate dependent $g_A^\mathrm{eff}(|\bm{r}|)$. In this article, we consider the constant $g_A^\mathrm{eff}$ associated with the NME. Since we have the GT and Fermi components of the leading and the perturbed terms, it is possible to calculate $g_A^\mathrm{eff}$ by different methods. One of them is to refer to the half-life calculated from the perturbed NME with the bare $g_A$ to determine the $g_A^\mathrm{eff}$ for the leading term of the $0\nu\beta\beta$ NME $g_{A,0\nu}^\mathrm{eff}(\mathrm{ld;pert})$ and that of the $2\nu\beta\beta$ NME $g_{A,2\nu}^\mathrm{eff}(\mathrm{ld;pert})$. This method is equivalent to using 
\begin{eqnarray}
\bigl\{ g_{A,0\nu}^\mathrm{eff}(\mathrm{ld;pert}) \bigr\}^2 M_{0\nu}^\mathrm{GT(0)} - g_V^2 M_{0\nu}^\mathrm{F(0)} = g_A^2 M_{0\nu}^\mathrm{GT} - g_V^2 M_{0\nu}^\mathrm{F}, \label{eq:eq_to_determine_gAeff0v}
\end{eqnarray}
\vspace{-12pt}
\begin{eqnarray}
\bigl\{ g_{A,2\nu}^\mathrm{eff}(\mathrm{ld;pert}) \bigr\}^2 M_{2\nu}^\mathrm{GT(0)} - g_V^2 M_{2\nu}^\mathrm{F(0)} = g_A^2 M_{2\nu}^\mathrm{GT} - g_V^2 M_{2\nu}^\mathrm{F}. \label{eq:eq_to_determine_gAeff2v}
\end{eqnarray}
When the different $g_A^\mathrm{eff}$ are discussed, the half-life is better than the NME for comparison because of their $g_A^\mathrm{eff}$ dependence. The right-hand sides of Eqs.\ (\ref{eq:eq_to_determine_gAeff0v}) and (\ref{eq:eq_to_determine_gAeff2v}) are the simulation for the exact value. We use $g_V = 1$ as usual. For the $2\nu\beta\beta$ NME, it is also possible to refer to the experimental half-life to determine $g_A^\mathrm{eff}$. This method can be applied for both the leading and the perturbed terms. The $g_A^\mathrm{eff}$ obtained by these methods are summarized in Table \ref{tab:gAeff0v2v}. 

\begin{table}
\caption{\label{tab:gAeff0v2v} Effective axial-vector current coupling $g_A^\mathrm{eff}$ obtained by different methods. These methods are defined by the second through the fourth columns. The second column shows the considered decay. The third column specifies the GT and Fermi components used with the $g_A^\mathrm{eff}$, and those are either the leading (ld) or the perturbed (pert) term, indicated by the first argument of $g_A^\mathrm{eff}$ in the first column. The reference to reproduce is either the half-life with the perturbed NME and $g_A$ = 1.267 (pert) [see Eqs.\ (\ref{eq:eq_to_determine_gAeff0v}) and (\ref{eq:eq_to_determine_gAeff2v})] or the experimental half-life (phen), indicated by the second argument of $g_A^\mathrm{eff}$ in the first column. The experimental half-life for the $2\nu\beta\beta$ decay of $^{136}$Xe is $2.18\times 10^{21}$ y \cite{Bar19}. For the equations of the NME, see caption of Table \ref{tab:NME0v2v}. }
\begin{ruledtabular}
\begin{tabular}{ccccc}
\multicolumn{5}{c}{SkM$^\ast$} \\
\hline\\[-10pt]
 & & Used GT and Fermi & & \\
 Specified $g_A^\mathrm{eff}$ & Decay & components & Reference to reproduce  & Value \\
\hline \\[-10pt]
\multirow{2}{*}{$g_{A,0\nu}^\mathrm{eff}$(ld; pert)}   & 
\multirow{2}{*}{$0\nu\beta\beta$}                      & 
\multirow{2}{*}{Leading}                               &
Half-life with perturbed  &  
\multirow{2}{*}{0.796}\\[-3pt]
                                      &                  &           & NME and bare $g_A$ &  \\[3pt]
\multirow{2}{*}{$g_{A,2\nu}^\mathrm{eff}$(ld; pert)}   &
\multirow{2}{*}{$2\nu\beta\beta$}                      &
\multirow{2}{*}{Leading}                               &
 Half-life with perturbed &  
\multirow{2}{*}{0.696}\\[-3pt]
                                      &                  &           & NME and bare $g_A$     & \\[3pt]
$g_{A,2\nu}^\mathrm{eff}$(ld; phen)   & $2\nu\beta\beta$ & Leading   & Experimental half-life & 0.422 \\
$g_{A,2\nu}^\mathrm{eff}$(pert; phen) & $2\nu\beta\beta$ & Perturbed & Experimental half-life & 0.806 \\
\end{tabular}
\end{ruledtabular}
\end{table}
%

The simulated $g_{A,0\nu}^\mathrm{eff}\mathrm{(ld; pert)}$ for the $0\nu\beta\beta$ decay and $g_{A,2\nu}^\mathrm{eff}\mathrm{(ld; pert)}$ for the $2\nu\beta\beta$ decay are different only by 14 \%. This indicates the possibility that $g_A^\mathrm{eff}$ for the $2\nu\beta\beta$ decay can be used approximately for the $0\nu\beta\beta$ decay. There is a potential speculation that the $g_A^\mathrm{eff}$ for the $0\nu\beta\beta$ and the $2\nu\beta\beta$ decays are quite different because the relevant neutrino momentum is different. This speculation does not apply to the SkM$^\ast$ calculation. This is an important finding of this study. 

For the half-life calculation, we always use the values of the phase space factor of Ref.\ \cite{Kot12}. The $2\nu\beta\beta$ half-life obtained from the perturbed terms and the bare $g_A$ is $2.62\times 10^{20}$ y, and the experimental value is $2.18\times 10^{21}$ y \cite{Bar19}. Still, there is a difference of an order of magnitude. The half-life calculated with the leading term and the bare $g_A$ is $3.26\times 10^{19}$ y. Thus, the half-life of the perturbed NME is much closer to the experimental value. The more the perturbation effect, the longer the half-life. If the experimental half-life is reproduced with the bare $g_A$ of 1.267, strongly perturbed NME components would be necessary. When our perturbed NME is used, the $g_{A,2\nu}^\mathrm{eff}$ to reproduce the experimental half-life is $g_{A,2\nu}^\mathrm{eff}\mathrm{(pert; phen)}$ = 0.806. That for the leading NME is $g_{A,2\nu}^\mathrm{eff}\mathrm{(ld; phen)}$ = 0.422. The smaller $g_A^\mathrm{eff}$, the larger perturbation effect the $g_A^\mathrm{eff}$ has. The less perturbed NME components are sufficient for the smaller $g_A^\mathrm{eff}$. 

It is shown by Table \ref{tab:gAeff0v2v} that $g_{A,2\nu}^\mathrm{eff}\mathrm{(ld; pert)}$ (0.696) and $g_{A,2\nu}^\mathrm{eff}\mathrm{(pert; phen)}$ (0.806) are close. This indicates that the half-life obtained from the combination of $M_{2\nu}^\mathrm{GT}$, $M_{2\nu}^\mathrm{F}$, and $g_{A,2\nu}^\mathrm{eff}\mathrm{(ld; pert)}$ is close to the experimental value. Also the value for the effective $g_A$ for the $0\nu\beta\beta$ NME (0.796) is quite close to those values. Originally, $g_{A,2\nu}^\mathrm{eff}\mathrm{(ld; pert)}$ is intended to be used with $M_{2\nu}^\mathrm{GT(0)}$  and $M_{2\nu}^\mathrm{F(0)}$. Nevertheless, we attempted to use $M_{2\nu}^\mathrm{GT}$ and $M_{2\nu}^\mathrm{F}$, and the half-life by this trial combination turned out to be $4.77\times 10^{21}$ y, which is larger than the experimental value only by a factor of 2.2. Based on our status that the discrepancy by a factor of ten is not unusual for the half-life, it is possible to state that the approximate prediction for the $2\nu\beta\beta$ half-life is possible by this trial combination method. The half-lives discussed here are summarized in Table \ref{tab:halflife}. 

\begin{table}
\caption{\label{tab:halflife} Experimental and calculated half-lives of the $2\nu\beta\beta$ decay. The used components of the NME are shown in Table \ref{tab:NME0v2v}. For the definitions of the  $g_A^\mathrm{eff}$, see caption of Table \ref{tab:gAeff0v2v}. }
\begin{ruledtabular}
\begin{tabular}{ccc}
 & & Half-life ($10^{21}$ y) \\
\hline
\multicolumn{2}{c}{Experimental} & 2.18 \\
\hline
\multicolumn{3}{c}{SkM$^\ast$} \\
\hline
Used GT and Fermi & & \\[-4pt]
components & Used $g_A$ & \\
\hline\\[-10pt]
Leading   & 1.267  &  0.03 \\
Perturbed & 1.267  &  0.26 \\
Leading   & $g_{A,2\nu}^\mathrm{eff}\mathrm{(ld; pert)}$  &  0.34 \\
Perturbed & $g_{A,2\nu}^\mathrm{eff}\mathrm{(ld; pert)}$  &  4.77 \\
\end{tabular}
\end{ruledtabular}
\end{table}
%
\begin{table}
\caption{\label{tab:NME0v2v_SGII} Calculated GT and the Fermi components of the $0\nu\beta\beta$ and the $2\nu\beta\beta$ NME. The Skyrme parameter set SGII is used. For the used equations of the NME, see caption of Table \ref{tab:NME0v2v}.  }
\begin{ruledtabular}
\begin{tabular}{ccccc}
\multicolumn{5}{c}{SGII} \\
\hline\\[-10pt]
 & \multicolumn{2}{c}{$0\nu\beta\beta$ NME} & \multicolumn{2}{c}{$2\nu\beta\beta$ NME} \\
\cline{2-3} \cline{4-5} \\[-11pt]
 & GT & Fermi & GT & Fermi \\
\hline \\[-10pt]
Leading & $\;\;\:$2.947 &      $-$0.460 & $\;\;\:$0.052 &      $-$0.001 \\
vc      & $\;\;\:$1.456 &      $-$0.714 & $\;\;\:$0.037 &      $-$0.022 \\
2bc     &      $-$3.497 & $\;\;\:$1.568 &      $-$0.067 & $\;\;\:$0.021 \\
Sum     & $\;\;\:$0.906 & $\;\;\:$0.394 & $\;\;\:$0.022 &      $-$0.002 \\
\end{tabular}
\end{ruledtabular}
\end{table}
%
\begin{table}[h]
\caption{\label{tab:gAeff0v2v_SGII} The same as Table \ref{tab:gAeff0v2v} but for the Skyrme interaction SGII. See caption of Table \ref{tab:gAeff0v2v}. }
\begin{ruledtabular}
\begin{tabular}{ccccc}
\multicolumn{5}{c}{SGII} \\
\hline\\[-10pt]
 & & Used GT and Fermi & & \\
 Specified $g_A^\mathrm{eff}$ & Decay & components & Reference to reproduce  & Value \\
\hline \\[-10pt]
\multirow{2}{*}{$g_{A,0\nu}^\mathrm{eff}$(ld; pert)}   & 
\multirow{2}{*}{$0\nu\beta\beta$}                      & 
\multirow{2}{*}{Leading}                               &
Half-life with perturbed  &  
\multirow{2}{*}{0.454}\\[-3pt]
                                      &                  &           & NME and bare $g_A$ &  \\[3pt]
\multirow{2}{*}{$g_{A,2\nu}^\mathrm{eff}$(ld; pert)}   &
\multirow{2}{*}{$2\nu\beta\beta$}                      &
\multirow{2}{*}{Leading}                               &
 Half-life with perturbed &  
\multirow{2}{*}{0.847}\\[-3pt]
                                      &                  &           & NME and bare $g_A$     & \\[3pt]
$g_{A,2\nu}^\mathrm{eff}$(ld; phen)   & $2\nu\beta\beta$ & Leading   & Experimental half-life & 0.563 \\
$g_{A,2\nu}^\mathrm{eff}$(pert; phen) & $2\nu\beta\beta$ & Perturbed & Experimental half-life & 0.833 \\
\end{tabular}
\end{ruledtabular}
\end{table}
%
\begin{table}[h]
\caption{\label{tab:halflife_SGII} The same as Table \ref{tab:halflife} but for the Skyrme interaction SGII. See also caption of Table \ref{tab:NME0v2v}. }
\begin{ruledtabular}
\begin{tabular}{ccc}
 & & Half-life ($10^{21}$ y) \\
\hline
\multicolumn{2}{c}{Experimental} & 2.18 \\
\hline
\multicolumn{3}{c}{SGII} \\
\hline
Used GT and Fermi & & \\[-4pt]
components & Used $g_A$ & \\
\hline\\[-10pt]
Leading   & 1.267  & 0.09 \\
Perturbed & 1.267  & 0.47 \\
Leading   & $g_{A,2\nu}^\mathrm{eff}\mathrm{(ld; pert)}$  & 0.47 \\
Perturbed & $g_{A,2\nu}^\mathrm{eff}\mathrm{(ld; pert)}$  & 2.05 \\
\end{tabular}
\end{ruledtabular}
\end{table}
%

One may ask which $g_{A,2\nu}^\mathrm{eff}$ can be compared with the bare value. It is possible to obtain the same half-life by changing the NME components and $g_{A,2\nu}^\mathrm{eff}$ in a coordinated manner. Thus, it is necessary to unify the NME components for the comparison. Let us define the alternative effective $g_{A,2\nu}$ to reproduce the theoretical half-life of $4.77\times 10^{21}$ y with the true GT NME, which is defined by the bare $g_A$ and the experimental half-life. We denote this effective $g_{A,2\nu}$ as $g_{A,2\nu}^\mathrm{alt}$. The ratio of $g_{A,2\nu}^\mathrm{alt}$ to the bare $g_A$ can be obtained 
\begin{eqnarray}
\frac{ (2.18\times 10^{21})^{1/4} }{ (4.77\times 10^{21})^{1/4} } = 0.82.
\end{eqnarray}
Thus, $g_{A,2\nu}^\mathrm{alt}$ is equal to 1.04; this is the value to compare with 1.267. The shorter the calculated half-life, the larger $g_{A,2\nu}^\mathrm{alt}$. This implies that $g_{A,2\nu}^\mathrm{alt}$ is larger for calculations with less the perturbation effect of the transition operator. This is opposite to the usual $g_{A,2\nu}^\mathrm{eff}$ reproducing the experimental half-life. Thus, $g_{A,2\nu}^\mathrm{alt}$ cannot be compared to $g_{A,2\nu}^\mathrm{eff}$.

Next, we discuss the SGII calculation briefly in comparison with the SkM$^\ast$ calculation. The calculated GT and the Fermi components of the $0\nu\beta\beta$ and the $2\nu\beta\beta$ NME are shown in Table \ref{tab:NME0v2v_SGII}. The different $g_A^\mathrm{eff}$ and the half-lives are summarized in Tables \ref{tab:gAeff0v2v_SGII} and \ref{tab:halflife_SGII}, respectively. The ratio of $g_{A,0\nu}^\mathrm{eff}\mathrm{(ld; pert)}$ and $g_{A,2\nu}^\mathrm{eff}\mathrm{(ld; pert)}$ is 0.454/0.847 = 0.536 (see Table \ref{tab:gAeff0v2v_SGII}), thus, the $g_A^\mathrm{eff}$ for the $0\nu\beta\beta$ and the $2\nu\beta\beta$ decays are not close to each other in the SGII calculation in contrast to the SkM$^\ast$ calculation. $M_{0\nu}$ with $g_A$ = 1.267 is found to be 0.662, while the corresponding $M_{0\nu}$ of the SkM$^\ast$ calculation is 1.505. This indicates that the perturbation effect of SGII is stronger than that of SkM$^\ast$. The experimental half-life of the $2\nu\beta\beta$ decay is reproduced well by $M_{0\nu}$ with $g_{A,2\nu}^\mathrm{eff}\mathrm{(ld; pert)}$; see Table \ref{tab:halflife_SGII}. This calculation does not use phenomenology. 
The reason for this success is seen in Table \ref{tab:gAeff0v2v_SGII}; $g_{A,2\nu}^\mathrm{eff}\mathrm{(ld; pert)}$ is very close to $g_{A,2\nu}^\mathrm{eff}\mathrm{(pert; phen)}$. The argument ld of $g_{A,2\nu}^\mathrm{eff}\mathrm{(ld; pert)}$ physically implies the large perturbation effect, and the argument phen of $g_{A,2\nu}^\mathrm{eff}\mathrm{(pert; phen)}$ also implies the large perturbation effect. Thus, these two $g_A^\mathrm{eff}$ have close values; this tendency is clear in both the SkM$^\ast$ and the SGII calculations. 

Two interactions show rather different perturbed NME (the sum in Tables \ref{tab:NME0v2v} and \ref{tab:NME0v2v_SGII}). The question is which interaction is better. We speculate that the SkM$^\ast$ calculation is physically better than the SGII calculation because we have examined the performance of the SkM$^\ast$ by the theoretical consistency checks and the comparisons with the experimental data \cite{Ter20}. The investigated items are as follows:

\begin{enumerate}
\item The self-check of the dual intermediate states ($B_F$ and $B_I$) using the $2\nu\beta\beta$ decay; the two sets of intermediate-state energies obtained by the QRPA based on the initial state and those based on the final state give nearly the same $2\nu\beta\beta$ NME. 
\item The convergence of the NME with respect to the dimension of the valence single-particle space. 
\item The comparison of the two spectra of the intermediate nucleus $^{136}$Cs obtained by the QRPA based on the initial and the final states and comparison of them with the experimental data. 
\item The GT sum rule. 
\item The comparison of the GT$^-$ (neutron $\rightarrow$ proton) strength with the data of the charge-change reaction.
\item The comparison of the $\beta$ decay spectrum and the GT transition strengths of $^{138}$Xe with the experimental data. 
\item Test using a new quantity expressing a higher-order effect in the $2\nu\beta\beta$ NME; \cite{Sim18} \cite{Gan19}.
\end{enumerate}

The QRPA with SkM$^\ast$ turns out to be excellent for $^{136}$Xe. The physical difference between SkM$^\ast$ and SGII is most evidently shown by the binding energy. The binding energy per nucleon of $^{136}$Xe is 8.396 MeV (experimental data \cite{nndc25}), 8.415 MeV (SkM$^\ast$), and 8.603 MeV (SGII). The calculated values are those of the HFB approximation for the ground state. The deviation of the SkM$^\ast$ value from the experimental data is one order of magnitude smaller than that of the SGII value. The attractive force of the SGII is stronger than that of the SkM$^\ast$. This indicates that the perturbation effect of SGII is stronger. Generally speaking, SkM$^\ast$ is better than SGII in spite of the success of SGII in the half-life of the $2\nu\beta\beta$ decay.

\begin{figure}[t]
\includegraphics[width=0.45\columnwidth]{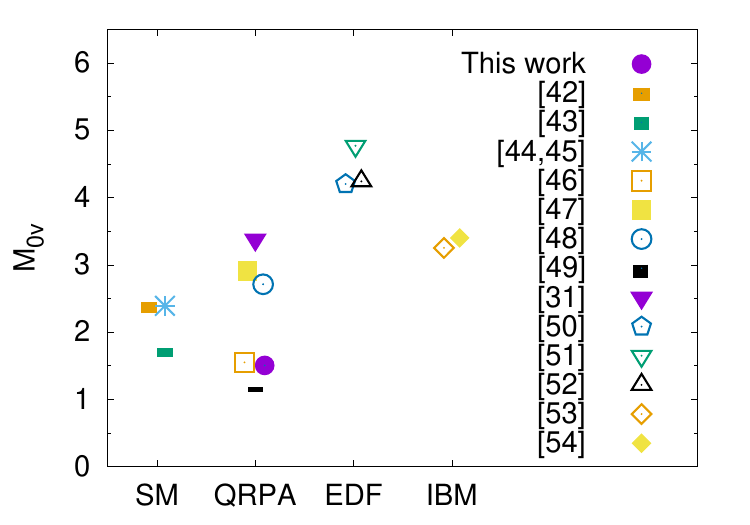}
\vspace{-10pt}
\caption{ \protect \label{fig:NME0v_groups} \baselineskip=13pt 
Calculated $0\nu\beta\beta$ NME by us and other groups. The bare $g_A$ is used. SM denotes the shell model. The reference numbers are assigned to the symbols. Our perturbed $M_{0\nu}$ is denoted as this work, and our leading $M_{0\nu}^{(0)}$ is that of Ref.\ \cite{Ter20}. The SM calculations are those of Refs.\ \cite{Men18,Hor16,Cor20b,Cor22}, the QRPA calculations are those of Refs.\ \cite{Ter20,Mus13,Hyv15,Sim18b,Fan18} and this work, the EDF calculations are those of Refs.\ \cite{Rod10,Vaq13,Son17}, and the IBM calculations are given by Refs.\ \cite{Bar15,Dep20}. The symbols of \cite{Men18}, \cite{Hor16}, and \cite{Fan18} show the vertical ranges; otherwise, the single values are shown. The EDF approach is methodologically characterized by the generator coordinate method \cite{Rin80}. In the $\beta\beta$ decay studies, this method is called EDF by custom. Also refer to the talk of I. Shimizu in Conf.\ \cite{Neu24}. }
\end{figure}

Figure \ref{fig:NME0v_groups} depicts the $0\nu\beta\beta$ NME calculated by us and other groups with the bare $g_A$. Our $M_{0\nu}^{(0)}$ is at the top of the QRPA values, and our perturbed $M_{0\nu}$ is near the bottom. Thus, the variation due to the perturbation of the transition operator is significant in the distribution of all calculations. The QRPA approach has the largest number of calculations and the largest variation. The method dependence of the $0\nu\beta\beta$ NME can be recognized. The EDF approach has the largest NME, next is the IBM, and the lowest group is the shell model. The QRPA distribution is similar to the combined distribution of the IBM and the shell model. This picture is not changed by our perturbation calculation. 

\subsection{ \label{subsec:sensitivity} Strong sensitivity of $\bm{\beta\beta}$ nuclear matrix element to perturbation }
In this and the next sections, we attempt qualitative discussions on the significant consequences brought by the perturbed transition operators. Tables I and IV indicate that the effect of the perturbation is significant for the NME. $^{136}$Xe has the magic neutron number of 82, and $Z$ = 54 is much closer to the magic number of 50 than the midshell. Thus, the QRPA is a good approximation for this nucleus, so that the effect of the residual interaction is weak. The question is why the perturbation effect is so large for the $\beta\beta$ NME of this nucleus. A possible reason is that the GT$^+$ transition from $^{136}$Ba to $^{136}$Cs is very small so that the ratio of the variation of the transition strength to the unperturbed strength is large. 

\begin{figure}[t]
\includegraphics[width=1.0\columnwidth]{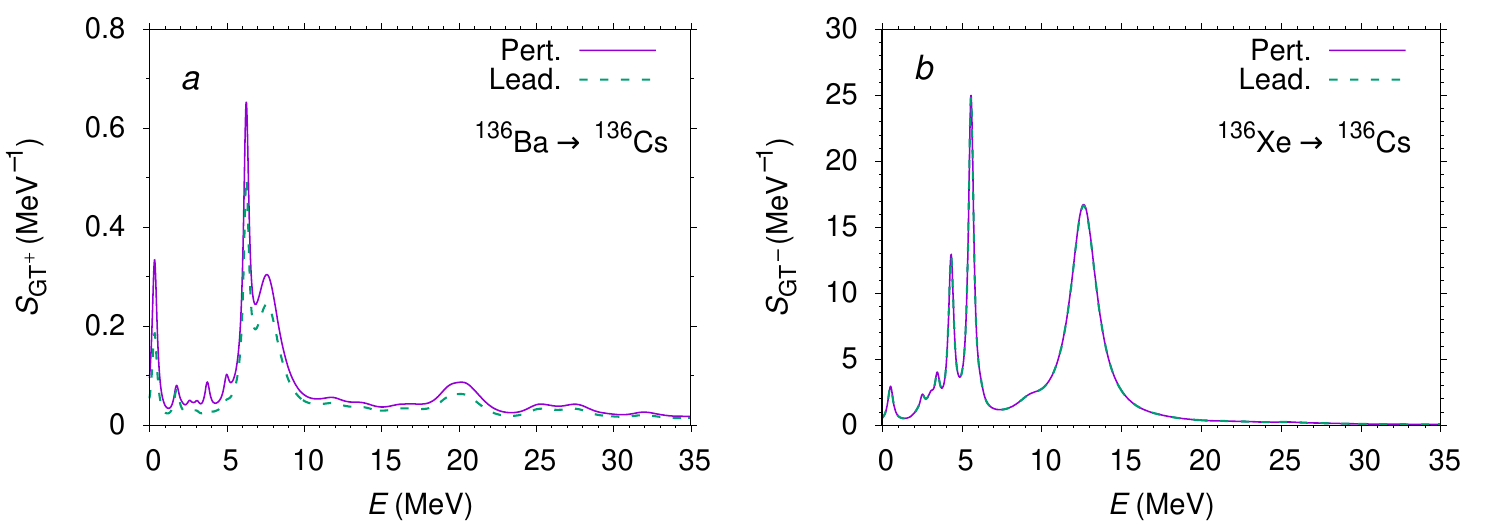}
\vspace{-10pt}
\caption{ \protect \label{fig:strfn_gt} \baselineskip=13pt 
Panel $a$: perturbed (Pert.) and leading (Lead.) GT$^+$  strength functions for $^{136}$Ba $\rightarrow$ $^{136}$Cs as functions of excitation energy of $^{136}$Cs. Panel $b$: GT$^-$ strength function for $^{136}$Xe $\rightarrow$ $^{136}$Cs; two lines are almost identical. The perturbation is caused only by the vertex correction. }
\end{figure}

For confirming this scenario, it is necessary to calculate the perturbed single-charge change transition strength. However, the method to calculate the 2bc correction for this transition is not yet clear. If the virtual Majorana neutrino in the diagram of Fig.\ \ref{fig:diagram_0v2bc} is replaced by the real Dirac antineutrino emitted from the middle nucleon and going out of the nucleus, that simpler diagram seems to be a perturbation for the initial state. In this section, we refer to the perturbation by the vc for discussion. Figure \ref{fig:strfn_gt} shows that, indeed, the variation of the GT$^+$ strength (panel $a$) relative to the leading one is larger than that of the GT$^-$ strength (panel $b$); the variation in the latter is negligible. The strength of the GT$^+$ transition is two orders of magnitude smaller than that of the GT$^-$ transition because these nuclei are neutron rich. The $\beta\beta$ NME and the single-charge change strength have a schematic relation 
\begin{eqnarray}
\beta\beta\ \mathrm{NME} \sim \sqrt{ \mathrm{GT}^+\ \mathrm{strength}\times \mathrm{GT}^-\ \mathrm{strength} }.
\end{eqnarray}
The ratio of the perturbed $\beta\beta$ NME to the leading NME is strongly affected by the ratio of the perturbed GT$^+$ strength to the leading one. Thus, the transition between the final and the intermediate states is the main cause of the strong sensitivity of the $\beta\beta$ NME to the perturbation. 

\subsection{ \label{subsec:similarity} Similarity between $\bm{g_A^\mathrm{eff}}$’s of neutrinoless and two-neutrino $\bm{\beta\beta}$ nuclear matrix elements}
In the SkM$^\ast$ calculation, the two $g_A^\mathrm{eff}$’s of the $0\nu\beta\beta$ and $2\nu\beta\beta$ NME are close to each other. This is remarkable because this implies that the neutrino potential does not significantly affect $g_A^\mathrm{eff}$. One can rewrite Eq.\ (\ref{eq:eq_to_determine_gAeff0v}) to
\begin{eqnarray}
\bigl\{ g_{A,0\nu}^\mathrm{eff}\mathrm{(ld;pert)}\bigr\}^2 = g_A^2 \Biggl( 1 + \frac{ M_{0\nu}^\mathrm{GT(vc)} }{ M_{0\nu}^\mathrm{GT(0)} } + \frac{ M_{0\nu}^\mathrm{GT(2bc)} }{M_{0\nu}^\mathrm{GT(0)} } \Biggr) - g_V^2 \Biggl( \frac{ M_{0\nu}^\mathrm{F(vc)} }{ M_{0\nu}^\mathrm{GT(0)} } + \frac{ M_{0\nu}^\mathrm{F(2bc)} }{ M_{0\nu}^\mathrm{GT(0)} } \Biggr).
\end{eqnarray}
Let us investigate $M_{0\nu}^\mathrm{GT(vc)}/M_{0\nu}^\mathrm{GT(0)}$ as a sample. The vc term  consists of the $VJJ$ and the $JJV$ terms, and the former is dominant as shown by Table \ref{tab:JJV_VJJ}. With the $VJJ$ term, we have 
\begin{eqnarray}
\frac{ M_{0\nu}^\mathrm{GT(vc)} }{ M_{0\nu}^\mathrm{GT(0)} } \simeq \frac{\mathcal{A}}{\mathcal{B}}, \label{eq:ratio_M0vGTvc_M0vGT0}
\end{eqnarray}
\vspace{-12pt}
\begin{eqnarray}
\mathcal{A} &=& -\sum \Bigl( \overline{\mathcal{V}}_{jb,ck}^{(4b1)F} + \overline{\mathcal{V}}_{jb,ck}^{(4b2)F} \Bigr) X_{ck}^{BF} \langle B_F|B_I \rangle u_a^I \mathrm{sgn}(j_i^z) v_i^I X_{ai}^{BI} (W^{\mathrm{GT},FI})_{a-i,b-j}^T \nonumber\\
&& \times u_b^F \mathrm{sgn}(j_j^z) v_j^F,
\end{eqnarray}
\vspace{-12pt}
\begin{eqnarray}
\mathcal{B} = \sum_{B_I B_F} \sum_{abij} W_{-ai,b-j}^{\mathrm{GT},FI} \mathrm{sgn}(j_a^z) v_a u_i X_{ai}^{BF} \langle B_F|B_I \rangle u_b \mathrm{sgn}(j_j^z) v_j X_{bj}^{BI},
\end{eqnarray}
where the summation in $\mathcal{A}$ is taken with respect to the intermediate states $\{B_F, B_I\}$, and the single particle states $\{a,b,c,i,j,k\}$. These particle states are all different states. For convenience in discussion, we define a notation
\begin{eqnarray}
\mathcal{G}_{b-j,a-i}(\bm{R},\bm{r}) = \sum_{k=1-3} \psi_b^{F\dagger}(\bm{x}) \sigma_k \tau^- \psi_{-j}^F (\bm{x}) \psi_a^{I\dagger} (\bm{y}) \sigma_k \tau^- \psi_{-i}^I (\bm{y}),
\end{eqnarray}
\vspace{-12pt}
\begin{eqnarray}
\bm{R} = (\bm{x} + \bm{y})/2.
\end{eqnarray}
The matrix element of the the neutrino potential can be written
\begin{eqnarray}
W_{b-j,a-i}^{\mathrm{GT},FI} = \int dr r^2 d\Omega \,d^3\!\bm{R} \,\widetilde{h}_+ (r) \mathcal{G}_{b-j,a-i} (\bm{R},\bm{r}),
\end{eqnarray}
\vspace{-12pt}
\begin{eqnarray}
d\Omega = d\theta d\varphi \,\mathrm{sin}\theta.
\end{eqnarray}
The relative coordinate of two nucleons is denoted by $\bm{r} = (r,\theta,\varphi)$. For $\widetilde{h}_+ (r)$, see Eq.\ (\ref{eq:neutrino_potential}). 

\begin{table}
{}\hspace{-25pt}\parbox{0.6\textwidth}{
\caption{\label{tab:JJV_VJJ} Vertex correction terms $M_{0\nu}^{\mathrm{GT}(JJV)}$, $M_{0\nu}^{\mathrm{GT}(VJJ)}$ (second column), $M_{0\nu}^{\mathrm{F}(JJV)}$, and $M_{0\nu}^{\mathrm{F}(VJJ)}$  (third column) obtained by the SkM$^\ast$ calculation. See Eqs.\ (\ref{eq:M0vGTJJV}), (\ref{eq:M0vFJJV}), (\ref{eq:M0vGTVJJ}), and (\ref{eq:M0vFVJJ}).  }
}
\begin{minipage}{0.5\textwidth}
\begin{ruledtabular}
\begin{tabular}{ccc}
Term & GT & Fermi \\
\hline \\[-10pt]
$JJV$ & 0.094 & $-$0.049 \\
$VJJ$ & 1.238 & $-$0.934
\end{tabular}
\end{ruledtabular}
\end{minipage}
\end{table}
%

\begin{figure}[t]
\includegraphics[width=0.6\columnwidth]{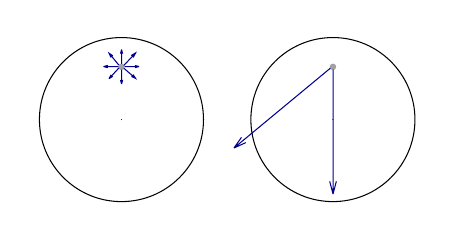}
\vspace{-10pt}
\caption{ \protect \label{fig:angles} \baselineskip=13pt 
Illustration of angles of relative coordinate of two nucleons with fixed radial variable. The small filled circle illustrates a nucleon, and the endpoints of the arrows indicate other nucleons if they are in the nucleus (large circle). The left is a model for $r$ = 1 fm, and the right is a model for twice the nuclear root-mean-square radius. 
}
\end{figure}

We introduce a help important for our discussion from Ref.\ \cite{Sim18}. The authors calculated the $\beta\beta$ NME density $\mathcal{M}_{0\nu}^\mathrm{GT}(r)$ and $\mathcal{M}_{2\nu}^I(r)$, ($I$ = GT, Fermi), defined by 
\begin{eqnarray}
M_{0\nu}^\mathrm{GT} = \int dr \mathcal{M}_{0\nu}^\mathrm{GT}(r),\hspace{20pt}   M_{2\nu}^I = \int dr \mathcal{M}_{2\nu}^I (r).
\end{eqnarray}
Their NME is the leading term in our terminology. The three $\beta\beta$ NME densities vanish at $r$ = 0, increase up to a high peak at $r$ = $r_c$ = 1 fm, and decrease with the full width at half maximum of $\Delta r_c$ = 1.5 fm. After this oscillation, the densities rapidly attenuate with oscillations and vanish. Why are the parameters $r_c$ and $\Delta r_c$ independent of the kind of the NME term? This highly dynamics independent property indicates a geometrical reason. The NME density is obtained by the angular integral with respect to the relative coordinate. The $\beta\beta$ decays occur in the surface region. Thus, it is possible to consider the geometry of the two nucleons illustrated by Fig.\ \ref{fig:angles}. If $r$ = 1 fm, another nucleon can be found at any angles. If the $r$ is much larger, the angles to find another nucleon are limited. $W_{b-j,a-i}^{\mathrm{GT},FI}$ can be approximated as 
\begin{eqnarray}
W_{b-j,a-i}^{\mathrm{GT},FI} \simeq \Delta r_c r_c^2 \,\widetilde{h}_+ (r_c) \int d\Omega \,d^3\!\bm{R} \,\mathcal{G}_{b-j,a-i}(\bm{R},r_c,\theta,\varphi),
\end{eqnarray}
and, because of the geometrical nature, it is easy to speculate that $r_c$ and $\Delta r_c$ are common also for the NME term including the perturbation interaction. Therefore, $M_{0\nu}^\mathrm{GT(vc)}$ and $M_{0\nu}^\mathrm{GT(0)}$ in Eq.\ (\ref{eq:ratio_M0vGTvc_M0vGT0}) may have similar $\Delta r_c r_c^2 \,\widetilde{h}_+(r_c)$, which cancel. In this way, the neutrino potential can be approximately removed from $g_A^\mathrm{eff}$. Therefore, it is not surprising that $g_A^\mathrm{eff}$’s of the $0\nu\beta\beta$ and $2\nu\beta\beta$ NMEs are similar. After $\widetilde{h}_+(r)$ is removed, there are still two differences between the $g_A^\mathrm{eff}$’s of the two NMEs. One is the number of the intermediate states, and the other is the energy denominator. The numerical calculation is necessary to find the similarity of the two $g_A^\mathrm{eff}$. 

\section{ \label{sec:summary} SUMMARY }
We extended the formulation of the $0\nu\beta\beta$ and $2\nu\beta\beta$ NMEs to include the perturbation of the transition operator. This is the extension of the leading term broadly used by using the Rayleigh-Schr\"{o}dinger perturbation theory. Our previous studies strongly indicate this necessity. The extension is analogous for the two NMEs, and the phase space factors are not affected. Our calculation is based on the canonical-quasiparticle basis, and the nuclear wave functions are obtained by the QRPA. The basis is self-consistently determined for the Hamiltonian, thus, we do not perturb the initial and final states. The appropriate correction terms of the NME were chosen from the general equation by referring to the diagrams; the vc and the 2bc terms linear with respect to the residual NN interaction were used. 

Concerning the study of the effective transition operator for the $\beta\beta$ decay, the $\chi$EFT \cite{Cir18} and the shell model \cite{Cor24,Men11} are advanced. Here, we state the differences of our approach from those ones. First, the matrix elements of our neutrino potentials do not have a divergence in contrast to that of the $\chi$EFT. Thus, we do not have a cutoff parameter or a counter term. Second, concerning the two-body current, we do not use a three-body interaction because the weak interaction Lagrangian that we use does not have a coupling of the weak and the strong interactions. An effective three-body interaction of the coupling of these two interactions is possible in higher-order diagrams not considered in this study (extensions of the diagrams of Fig.\  \ref{fig:diagram_0vvc}$b$). Third, the offshell effects of the very high-energy, e.g., GeV, region are not considered because the phenomenological nuclear interactions include those effects implicitly, if any. Fourth, our valence single-particle space is always large enough; there is no excluded space. Thus, our perturbed transition operator does not have a component to take into account the excluded single-particle space. Finally, we have fewer terms of the NME than the $\chi$EFT approach has. There are two reasons; one is that our phenomenological interactions do not have the charge-exchange components. The other reason is that our nucleon states are determined by solving the HFB equation self-consistently. Thus, we focus on the perturbation of the transition operator.

The decay instance of $^{136}$Xe $\rightarrow$ $^{136}$Ba was used for our application because the QRPA is a good approximation for these nuclei. It turned out that the correction terms are comparable with the leading term, and the NME was reduced significantly. The symmetry conserving truncation was achieved in the Fermi component of the $2\nu\beta\beta$ NME. Two parameter sets of the Skyrme interaction were used in our calculation. SkM$^\ast$ is highly reliable in particular in terms of the binding energy, which is important because the correction terms include the interaction explicitly. We also calculated $g_A^\mathrm{eff}$ for the leading term by referring to the perturbed NME with the bare $g_A$. The most important finding of our study is that these $g_A^\mathrm{eff}$ for the $0\nu\beta\beta$ and $2\nu\beta\beta$ NMEs are similar. They are of the order of unity (approximately 0.8; see Table \ref{tab:gAeff0v2v}). This is remarkable because $g_A^\mathrm{eff}$ for the $0\nu\beta\beta$ NME was unknown at all, and this lack was one of the major causes of the uncertainty problem of the $0\nu\beta\beta$ NME. The possibility appears that $g_A^\mathrm{eff}$ for the $2\nu\beta\beta$ NME to reproduce the experimental half-life can be approximately used for the $0\nu\beta\beta$ NME. Last but not least, we proposed $g_{A,2\nu}^\mathrm{alt}$ to compare with the bare $g_A$. This $g_{A,2\nu}^\mathrm{alt}$ is close to 1.0 when the calculated half-life is twice the experimental one.

We discussed analytically the strong sensitivity of the $\beta\beta$ NME to the perturbation and the similarity of the $g_A^\mathrm{eff}$’s of the two NMEs. The sensitivity originates from the quite small GT$^+$ transition strength between the intermediate and the final states. Thus, if this strength is measured, it would be useful for obtaining the correct NME. We showed that the neutrino potential can be removed approximately from the analytical equation of $g_A^\mathrm{eff}$. Thus, the similarity between the two $g_A^\mathrm{eff}$’s is not surprising. 

Our perturbed $0\nu\beta\beta$ NME with the bare $g_A$ is still in the distribution range of the results of other groups. The $2\nu\beta\beta$ half-life obtained from our perturbed NME with the bare $g_A$ is much closer to the experimental value than that from the leading NME, but our perturbation is not yet sufficient. 
The validity of the perturbative calculation is the essential problem. The large effect of the 2bc was reported by Refs.~\cite{Men11, Gys19} before our calculation, and no one yet calculates the higher-order terms for the $\beta\beta$ decay. The question is how one can handle the effects non-perturbatively. The approximation of $g_{A,0\nu}^\mathrm{eff}$ by the $g_{A,2\nu}^\mathrm{eff}$ that reproduces the experimental $2\nu\beta\beta$ half-life is a possible solution because the experimental data are non-perturbative. 

\begin{acknowledgments}  
This study was supported by the Czech Science Foundation (GA\v{C}R), project No. 24-10180S. The computation for this study was performed by Karolina (OPEN-30-65), IT4Innovations supported by the Ministry of Education, Youth and Sports of the Czech Republic through the e-INFRA CZ (ID:90254); the computers of MetaCentrum provided by the e-INFRA CZ project (ID:90254), supported by the Ministry of Education, Youth and Sports of the Czech Republic; Yukawa-21 at Yukawa Institute for Theoretical Physics, Kyoto University. 
\end{acknowledgments}

\hspace{0pt}\\
\hspace{0pt}\\
\noindent \textbf{Appendix A. Operator density}
\hspace{0pt}\\

For an operator $P(\hat{\bm{x}})$ with $\hat{\bm{x}}$ being the coordinate operator, the operator density can be written
\begin{eqnarray}
\mathcal{P}(\bm{x}) = P(\hat{\bm{x}})\delta(\hat{\bm{x}} - \bm{x}),
\end{eqnarray}
where $\bm{x}$ is the coordinate variable. Therefore, one has
\begin{eqnarray}
\langle i|\mathcal{P}(\bm{x})|j\rangle = \psi_i^\dagger(\bm{x}) P(\bm{x}) \psi_j(\bm{x}),
\end{eqnarray}
where $i$ and $j$ denote nucleon states, and $\psi_i(\bm{x})$ and $\psi_j(\bm{x})$ are the corresponding wave functions, respectively. 

The expression for an $N$-body system is 
\begin{eqnarray}
\mathcal{P}(\bm{x}) = \sum_{m=1-N} P(\hat{\bm{x}}_m) \delta(\hat{\bm{x}}_m - \bm{x}),
\end{eqnarray}
and $\mathcal{P}(\bm{x})$ in the second-quantization is 
\begin{eqnarray}
\sum_{ij} \langle i|\mathcal{P}(\bm{x})|j\rangle c_i^\dagger c_j = \sum_{ij} \psi_i^\dagger(\bm{x}) P(\bm{x}) \psi_j(\bm{x}) c_i^\dagger c_j,
\end{eqnarray}
with the creation (annihilation) operator $c_i^\dagger$  ($c_j$).

\hspace{0pt}\\
\hspace{0pt}\\
\noindent \textbf{Appendix B. First-order correction of nuclear matrix element}
\hspace{0pt}\\

 In this appendix, we derive Eqs.~(\ref{eq:MbbV_general})$-$(\ref{eq:MbbJVJ_general}) using the Rayleigh-Schr\"{o}dinger perturbation theory. The notations of Sec.~\ref{sec:eqs_bb_nme} are used, unless otherwise mentioned. The first-order component of the perturbed initial wave function is given by
\begin{eqnarray}
|I^{(1)} \rangle =\sum_{\mathcal{B}\neq I} \frac{1}{E_\mathcal{B} -E_I}|\mathcal{B}\rangle\langle \mathcal{B}|(-\mathcal{V})|I\rangle.
\end{eqnarray}
$\mathcal{V}$ is a perturbation interaction unspecified to derive the general equation. This interaction is determined to obtain the $0\nu\beta\beta$ NME below in a manner similar to the discussion for the lowest-order NME. Due to this, the nucleus in the state $|\mathcal{B}\rangle$ is unspecified, and there is a possibility of including the leptons, depending on $\mathcal{V}$. The second-order component without the normalization factor is derived
\begin{eqnarray}
|I^{(2)}\rangle &=& \sum_{\mathcal{B},\mathcal{C}\neq I} |\mathcal{C}\rangle  \frac{1}{E_\mathcal{C} -E_I} \langle \mathcal{C}|(-\mathcal{V})|\mathcal{B}\rangle \frac{1}{E_\mathcal{B} -E_I} \langle \mathcal{B}|(-\mathcal{V})|I\rangle \nonumber \\
&& +\langle I|\mathcal{V}|I\rangle \sum_{\mathcal{B}\neq I} |\mathcal{B}\rangle \frac{1}{(E_\mathcal{B} -E_I)^2} \langle \mathcal{B}|(-\mathcal{V})|I\rangle \nonumber \\
&& -\frac{1}{2} \sum_{\mathcal{B}\neq I} \langle I|(-\mathcal{V})|\mathcal{B}\rangle \frac{1}{(E_\mathcal{B} -E_I)^2} \langle \mathcal{B}|(-\mathcal{V})|I\rangle |I\rangle.
\label{eq:comp2_perturbed_wf}
\end{eqnarray}
The perturbation characteristic in the second order is caused by the residual interaction defined by the HFB calculation, and our quasiparticle basis is obtained self-consistently with the Hamiltonian. Thus, the correction of our basis is not included to avoid the double counting of the physical effects, implying that the factorized terms are not included in our calculation. This applies to the second term of Eq.~(\ref{eq:comp2_perturbed_wf}), and the normalization factor is not included due to this reason. This is also for considering only the perturbation of the transition operators. The perturbed components of the final state are introduced analogously and denoted by $|\widetilde{F}^{(1)}\rangle$ and $|\widetilde{F}^{(2)}\rangle$. 

The next step is to take the second-order terms with respect to $\mathcal{V}$ from the transition matrix element 
\begin{eqnarray}
\large\left(\langle \widetilde{F}| +\langle \widetilde{F}^{(1)}| +\langle \widetilde{F}^{(2)}|\large\right) H_W \large\left(|I\rangle +|I^{(1)} \rangle +|I^{(2)}\rangle \large\right).
\end{eqnarray}
This equation includes $\langle \widetilde{F}|H_W |I\rangle$, which vanishes. Excluding the factorized terms, one obtains
\begin{eqnarray}
M_\mathrm{step2}^{(2)} &=& \sum_{\mathcal{C}\neq I} \langle \widetilde{F}|H_W |\mathcal{C}\rangle \frac{1}{E_\mathcal{C} -E_I} \sum_{\mathcal{B}\neq I} \langle \mathcal{C}|(-\mathcal{V})|\mathcal{B}\rangle \frac{1}{E_\mathcal{B} -E_I} \langle \mathcal{B}|(-\mathcal{V})|I\rangle \nonumber \\
&& +\sum_{\mathcal{C}\neq \widetilde{F}} \frac{1}{E_\mathcal{C} -E_{\widetilde{F}}} \langle \widetilde{F}|(-\mathcal{V})|\mathcal{C}\rangle \sum_{\mathcal{B}\neq I} \frac{1}{E_\mathcal{B} -E_I} \langle \mathcal{C}|H_W |\mathcal{B}\rangle\langle \mathcal{B}|(-\mathcal{V})|I\rangle \nonumber \\
&& +\sum_{\mathcal{C}\neq \widetilde{F}} \frac{1}{E_\mathcal{C} -E_{\widetilde{F}}} \langle \widetilde{F}|(-\mathcal{V})|\mathcal{C}\rangle \sum_{\mathcal{B}\neq \widetilde{F}} \frac{1}{E_\mathcal{B} -E_{\widetilde{F}}} \langle \mathcal{C}|(-\mathcal{V})|\mathcal{B}\rangle\langle \mathcal{B}|H_W |I\rangle.
\end{eqnarray}

The next step is to derive the $\beta\beta$ transition matrix element from $M_\mathrm{step2}^{(2)}$. We set $\mathcal{V}$ to $:V:+H_W$ and take the terms with one $:V:$ and two $H_W$. This is equivalent to the procedure noted in Sec.~\ref{subsec:corrections}. Now, each intermediate state can be specified. This step needs the auxiliary process to avoid the physically double counting, as discussed for the lowest-order NME; $H_W$ for the transition and that for the perturbation are not distinguished physically. From Eq.~(\ref{eq:energy_denominator}) and $E_{\widetilde{F}} = E_I$,  one can use
\begin{eqnarray} 
E_{\widetilde{B}} -E_{\widetilde{F}} = E_{\widetilde{B}} -E_I = E_B+|\bm{q}|-\frac{1}{2} (E_I +E_F ).
\end{eqnarray}
Here, we use the notation of Sec.~\ref{sec:eqs_bb_nme}; $\widetilde{B}$ is the state including the intermediate nucleus, a neutrino (momentum $\bm{q}$), and an electron. $B$ denotes the nuclear state. The result of this step is 
\begin{eqnarray}
\lefteqn{ M_\mathrm{step3}^{(2)} } \nonumber \\
&=& \sum_{B} \langle \widetilde{F} |H_W |\widetilde{B}\rangle \frac{1}{E_B+|\bm{q}|-\frac{1}{2}(E_I +E_F)} \sum_{C\neq I} \langle \widetilde{B}|(-H_W)|C\rangle \frac{1}{E_C-E_I} \langle C|(-:V:)|I\rangle \nonumber \\
&& +\sum_{D\neq F} \frac{1}{E_D-E_F} \langle \widetilde{F}|(-:V:)| \widetilde{D}\rangle \sum_B \frac{1}{E_B+|\bm{q}|-\frac{1}{2}(E_I+E_F)} \langle \widetilde{D}|H_W| \widetilde{B}\rangle \langle \widetilde{B}|(-H_W)|I\rangle \nonumber \\
&& +\sum_{B^\prime} \frac{1}{E_{B^\prime}+|\bm{q}|-\frac{1}{2}(E_I+E_F)} \langle \widetilde{F}|(-H_W)| \widetilde{B}^\prime\rangle \sum_B \frac{1}{E_B+|\bm{q}|-\frac{1}{2}(E_I+E_F)} \langle \widetilde{B}^\prime|(-:V:)|\widetilde{B}\rangle\nonumber \\
&& \times\langle \widetilde{B}|H_W|I\rangle.
\end{eqnarray}
$C$ and $D$ denote the states of the initial and final nuclei, respectively, and $\widetilde{D}$ denotes the state of $D$ with the two electrons.

The final step is adjustments. We remove the lepton transition matrix element, implicitly included, from $M_\mathrm{step3}^{(2)}$, replace the operators with the operator density, and adjust the overall sign in accordance with the leading-order NME. This step leads to the compact NME density of Eqs.~(\ref{eq:MbbV_general})$-$(\ref{eq:MbbJVJ_general}).


\begin{thebibliography}{55}%
\makeatletter
\providecommand \@ifxundefined [1]{%
 \@ifx{#1\undefined}
}%
\providecommand \@ifnum [1]{%
 \ifnum #1\expandafter \@firstoftwo
 \else \expandafter \@secondoftwo
 \fi
}%
\providecommand \@ifx [1]{%
 \ifx #1\expandafter \@firstoftwo
 \else \expandafter \@secondoftwo
 \fi
}%
\providecommand \natexlab [1]{#1}%
\providecommand \enquote  [1]{``#1''}%
\providecommand \bibnamefont  [1]{#1}%
\providecommand \bibfnamefont [1]{#1}%
\providecommand \citenamefont [1]{#1}%
\providecommand \href@noop [0]{\@secondoftwo}%
\providecommand \href [0]{\begingroup \@sanitize@url \@href}%
\providecommand \@href[1]{\@@startlink{#1}\@@href}%
\providecommand \@@href[1]{\endgroup#1\@@endlink}%
\providecommand \@sanitize@url [0]{\catcode `\\12\catcode `\$12\catcode
  `\&12\catcode `\#12\catcode `\^12\catcode `\_12\catcode `\%12\relax}%
\providecommand \@@startlink[1]{}%
\providecommand \@@endlink[0]{}%
\providecommand \url  [0]{\begingroup\@sanitize@url \@url }%
\providecommand \@url [1]{\endgroup\@href {#1}{\urlprefix }}%
\providecommand \urlprefix  [0]{URL }%
\providecommand \Eprint [0]{\href }%
\providecommand \doibase [0]{https://doi.org/}%
\providecommand \selectlanguage [0]{\@gobble}%
\providecommand \bibinfo  [0]{\@secondoftwo}%
\providecommand \bibfield  [0]{\@secondoftwo}%
\providecommand \translation [1]{[#1]}%
\providecommand \BibitemOpen [0]{}%
\providecommand \bibitemStop [0]{}%
\providecommand \bibitemNoStop [0]{.\EOS\space}%
\providecommand \EOS [0]{\spacefactor3000\relax}%
\providecommand \BibitemShut  [1]{\csname bibitem#1\endcsname}%
\let\auto@bib@innerbib\@empty
\bibitem [{Neu()}]{Neu24}%
  \BibitemOpen
  \href@noop {} {}\bibinfo {howpublished} {XXXI International Conference on
  Neutrino Physics and Astrophysics, Milano (Italy), June 16$-$22, 2024
  \url{(https://neutrino2024.org)}}\BibitemShut {NoStop}%
\bibitem [{\citenamefont {Barabash}(2019)}]{Bar19}%
  \BibitemOpen
  \bibfield  {author} {\bibinfo {author} {\bibfnamefont {A.~S.}\ \bibnamefont
  {Barabash}},\ }in\ \href@noop {} {\emph {\bibinfo {booktitle} {Workshop on
  Calculation of Double-beta-decay Matrix Elements (MEDEX'19)}}},\ \bibinfo
  {editor} {edited by\ \bibinfo {editor} {\bibfnamefont {O.}~\bibnamefont
  {Civitarese}}, \bibinfo {editor} {\bibfnamefont {I.}~\bibnamefont {Stekl}},\
  and\ \bibinfo {editor} {\bibfnamefont {J.}~\bibnamefont {Suhonen}}}\
  (\bibinfo  {publisher} {AIP Publishing},\ \bibinfo {address} {Melville},\
  \bibinfo {year} {2019})\ p.\ \bibinfo {pages}
  {020002{\textendash}1}\BibitemShut {NoStop}%
\bibitem [{\citenamefont {Furry}(1939)}]{Fur39}%
  \BibitemOpen
  \bibfield  {author} {\bibinfo {author} {\bibfnamefont {W.~H.}\ \bibnamefont
  {Furry}},\ }\href@noop {} {\bibfield  {journal} {\bibinfo  {journal} {Phys.
  Rev.}\ }\textbf {\bibinfo {volume} {56}},\ \bibinfo {pages} {1184} (\bibinfo
  {year} {1939})}\BibitemShut {NoStop}%
\bibitem [{\citenamefont {Doi}\ \emph {et~al.}(1985)\citenamefont {Doi},
  \citenamefont {Kotani},\ and\ \citenamefont {Takasugi}}]{Doi85}%
  \BibitemOpen
  \bibfield  {author} {\bibinfo {author} {\bibfnamefont {M.}~\bibnamefont
  {Doi}}, \bibinfo {author} {\bibfnamefont {T.}~\bibnamefont {Kotani}},\ and\
  \bibinfo {author} {\bibfnamefont {E.}~\bibnamefont {Takasugi}},\ }\href@noop
  {} {\bibfield  {journal} {\bibinfo  {journal} {Prog. Theor. Phys. Suppl.}\
  }\textbf {\bibinfo {volume} {83}},\ \bibinfo {pages} {1} (\bibinfo {year}
  {1985})}\BibitemShut {NoStop}%
\bibitem [{\citenamefont {Fukugita}\ and\ \citenamefont
  {Yanagida}(1986)}]{Fuk86}%
  \BibitemOpen
  \bibfield  {author} {\bibinfo {author} {\bibfnamefont {M.}~\bibnamefont
  {Fukugita}}\ and\ \bibinfo {author} {\bibfnamefont {T.}~\bibnamefont
  {Yanagida}},\ }\href@noop {} {\bibfield  {journal} {\bibinfo  {journal}
  {Phys. Lett. B}\ }\textbf {\bibinfo {volume} {174}},\ \bibinfo {pages} {45}
  (\bibinfo {year} {1986})}\BibitemShut {NoStop}%
\bibitem [{\citenamefont {Fukuda}\ \emph {et~al.}(1998)\citenamefont {Fukuda},
  \citenamefont {Hayakawa}, \citenamefont {Ichihara}, \citenamefont {Inoue},
  \citenamefont {Ishihara}, \citenamefont {Ishino}, \citenamefont {Itow},
  \citenamefont {Kajita}, \citenamefont {Kameda}, \citenamefont {Kasuga} \emph
  {et~al.}}]{Fuk98}%
  \BibitemOpen
  \bibfield  {author} {\bibinfo {author} {\bibfnamefont {Y.}~\bibnamefont
  {Fukuda}}, \bibinfo {author} {\bibfnamefont {T.}~\bibnamefont {Hayakawa}},
  \bibinfo {author} {\bibfnamefont {E.}~\bibnamefont {Ichihara}}, \bibinfo
  {author} {\bibfnamefont {K.}~\bibnamefont {Inoue}}, \bibinfo {author}
  {\bibfnamefont {K.}~\bibnamefont {Ishihara}}, \bibinfo {author}
  {\bibfnamefont {H.}~\bibnamefont {Ishino}}, \bibinfo {author} {\bibfnamefont
  {Y.}~\bibnamefont {Itow}}, \bibinfo {author} {\bibfnamefont {T.}~\bibnamefont
  {Kajita}}, \bibinfo {author} {\bibfnamefont {J.}~\bibnamefont {Kameda}},
  \bibinfo {author} {\bibfnamefont {S.}~\bibnamefont {Kasuga}}, \emph {et~al.}
  (\bibinfo {collaboration} {Super-Kamiokande Collaboration}),\ }\href@noop {}
  {\bibfield  {journal} {\bibinfo  {journal} {Phys. Rev. Lett.}\ }\textbf
  {\bibinfo {volume} {81}},\ \bibinfo {pages} {1562} (\bibinfo {year}
  {1998})}\BibitemShut {NoStop}%
\bibitem [{\citenamefont {Ahmad}\ \emph {et~al.}(2002)\citenamefont {Ahmad},
  \citenamefont {Allen}, \citenamefont {Andersen}, \citenamefont {Anglin},
  \citenamefont {Barton}, \citenamefont {Beier}, \citenamefont {Bercovitch},
  \citenamefont {Bigu}, \citenamefont {Biller}, \citenamefont {Black} \emph
  {et~al.}}]{Ahm02}%
  \BibitemOpen
  \bibfield  {author} {\bibinfo {author} {\bibfnamefont {Q.~R.}\ \bibnamefont
  {Ahmad}}, \bibinfo {author} {\bibfnamefont {R.~C.}\ \bibnamefont {Allen}},
  \bibinfo {author} {\bibfnamefont {T.~C.}\ \bibnamefont {Andersen}}, \bibinfo
  {author} {\bibfnamefont {J.~D.}\ \bibnamefont {Anglin}}, \bibinfo {author}
  {\bibfnamefont {J.~C.}\ \bibnamefont {Barton}}, \bibinfo {author}
  {\bibfnamefont {E.}~\bibnamefont {Beier}}, \bibinfo {author} {\bibfnamefont
  {M.}~\bibnamefont {Bercovitch}}, \bibinfo {author} {\bibfnamefont
  {J.}~\bibnamefont {Bigu}}, \bibinfo {author} {\bibfnamefont {S.~D.}\
  \bibnamefont {Biller}}, \bibinfo {author} {\bibfnamefont {R.~A.}\
  \bibnamefont {Black}}, \emph {et~al.} (\bibinfo {collaboration} {SNO
  Collaboration}),\ }\href@noop {} {\bibfield  {journal} {\bibinfo  {journal}
  {Phys. Rev. Lett.}\ }\textbf {\bibinfo {volume} {89}},\ \bibinfo {pages}
  {011301} (\bibinfo {year} {2002})}\BibitemShut {NoStop}%
\bibitem [{\citenamefont {Aghanim}\ \emph {et~al.}(2020)\citenamefont
  {Aghanim}, \citenamefont {Akrami}, \citenamefont {Ashdown}, \citenamefont
  {Aumont}, \citenamefont {Baccigalupi}, \citenamefont {Ballardini},
  \citenamefont {Banday}, \citenamefont {Barreiro}, \citenamefont {Bartolo},
  \citenamefont {Basak} \emph {et~al.}}]{Agh20}%
  \BibitemOpen
  \bibfield  {author} {\bibinfo {author} {\bibfnamefont {N.}~\bibnamefont
  {Aghanim}}, \bibinfo {author} {\bibfnamefont {Y.}~\bibnamefont {Akrami}},
  \bibinfo {author} {\bibfnamefont {M.}~\bibnamefont {Ashdown}}, \bibinfo
  {author} {\bibfnamefont {J.}~\bibnamefont {Aumont}}, \bibinfo {author}
  {\bibfnamefont {C.}~\bibnamefont {Baccigalupi}}, \bibinfo {author}
  {\bibfnamefont {M.}~\bibnamefont {Ballardini}}, \bibinfo {author}
  {\bibfnamefont {A.~J.}\ \bibnamefont {Banday}}, \bibinfo {author}
  {\bibfnamefont {R.~B.}\ \bibnamefont {Barreiro}}, \bibinfo {author}
  {\bibfnamefont {N.}~\bibnamefont {Bartolo}}, \bibinfo {author} {\bibfnamefont
  {S.}~\bibnamefont {Basak}}, \emph {et~al.} (\bibinfo {collaboration} {Planck
  Collaboration}),\ }\href@noop {} {\bibfield  {journal} {\bibinfo  {journal}
  {Astron. Astrophy.}\ }\textbf {\bibinfo {volume} {641}},\ \bibinfo {pages}
  {A6} (\bibinfo {year} {2020})}\BibitemShut {NoStop}%
\bibitem [{\citenamefont {Aker}\ \emph {et~al.}()\citenamefont {Aker},
  \citenamefont {Batzler}, \citenamefont {Beglarian}, \citenamefont {Behrens},
  \citenamefont {Beisenkotter}, \citenamefont {Biassoni}, \citenamefont
  {Bieringer}, \citenamefont {Biondi}, \citenamefont {Block}, \citenamefont
  {Bobien} \emph {et~al.}}]{Ake24}%
  \BibitemOpen
  \bibfield  {author} {\bibinfo {author} {\bibfnamefont {M.}~\bibnamefont
  {Aker}}, \bibinfo {author} {\bibfnamefont {D.}~\bibnamefont {Batzler}},
  \bibinfo {author} {\bibfnamefont {A.}~\bibnamefont {Beglarian}}, \bibinfo
  {author} {\bibfnamefont {J.}~\bibnamefont {Behrens}}, \bibinfo {author}
  {\bibfnamefont {J.}~\bibnamefont {Beisenkotter}}, \bibinfo {author}
  {\bibfnamefont {M.}~\bibnamefont {Biassoni}}, \bibinfo {author}
  {\bibfnamefont {B.}~\bibnamefont {Bieringer}}, \bibinfo {author}
  {\bibfnamefont {Y.}~\bibnamefont {Biondi}}, \bibinfo {author} {\bibfnamefont
  {F.}~\bibnamefont {Block}}, \bibinfo {author} {\bibfnamefont
  {S.}~\bibnamefont {Bobien}}, \emph {et~al.} (\bibinfo {collaboration} {KATRIN
  Collaboration}),\ }\href@noop {} {\ }\Eprint
  {https://arxiv.org/abs/2406.13516 (2024)} {arXiv:2406.13516 (2024)}
  \BibitemShut {NoStop}%
\bibitem [{\citenamefont {Agostini}\ \emph {et~al.}(2023)\citenamefont
  {Agostini}, \citenamefont {Benato}, \citenamefont {Detwiler}, \citenamefont
  {Menéndez},\ and\ \citenamefont {Vissani}}]{Ago23}%
  \BibitemOpen
  \bibfield  {author} {\bibinfo {author} {\bibfnamefont {M.}~\bibnamefont
  {Agostini}}, \bibinfo {author} {\bibfnamefont {G.}~\bibnamefont {Benato}},
  \bibinfo {author} {\bibfnamefont {J.~A.}\ \bibnamefont {Detwiler}}, \bibinfo
  {author} {\bibfnamefont {J.}~\bibnamefont {Menéndez}},\ and\ \bibinfo
  {author} {\bibfnamefont {F.}~\bibnamefont {Vissani}},\ }\href@noop {}
  {\bibfield  {journal} {\bibinfo  {journal} {Rev. Mod. Phys.}\ }\textbf
  {\bibinfo {volume} {95}},\ \bibinfo {pages} {025002} (\bibinfo {year}
  {2023})}\BibitemShut {NoStop}%
\bibitem [{\citenamefont {Terasaki}\ and\ \citenamefont {Iwata}(2019)}]{Ter19}%
  \BibitemOpen
  \bibfield  {author} {\bibinfo {author} {\bibfnamefont {J.}~\bibnamefont
  {Terasaki}}\ and\ \bibinfo {author} {\bibfnamefont {Y.}~\bibnamefont
  {Iwata}},\ }\href@noop {} {\bibfield  {journal} {\bibinfo  {journal} {Phys.
  Rev. C}\ }\textbf {\bibinfo {volume} {100}},\ \bibinfo {pages} {034325}
  (\bibinfo {year} {2019})}\BibitemShut {NoStop}%
\bibitem [{\citenamefont {Suhonen}(2017)}]{Suh17}%
  \BibitemOpen
  \bibfield  {author} {\bibinfo {author} {\bibfnamefont {J.}~\bibnamefont
  {Suhonen}},\ }\href@noop {} {\bibfield  {journal} {\bibinfo  {journal} {Phys.
  Rev. C}\ }\textbf {\bibinfo {volume} {96}},\ \bibinfo {pages} {055501}
  (\bibinfo {year} {2017})}\BibitemShut {NoStop}%
\bibitem [{\citenamefont {Arima}\ \emph {et~al.}(1987)\citenamefont {Arima},
  \citenamefont {Shimizu},\ and\ \citenamefont {Bentz}}]{Ari87}%
  \BibitemOpen
  \bibfield  {author} {\bibinfo {author} {\bibfnamefont {A.}~\bibnamefont
  {Arima}}, \bibinfo {author} {\bibfnamefont {K.}~\bibnamefont {Shimizu}},\
  and\ \bibinfo {author} {\bibfnamefont {W.}~\bibnamefont {Bentz}},\
  }\href@noop {} {\bibfield  {journal} {\bibinfo  {journal} {Adv. Nucl. Phys}\
  }\textbf {\bibinfo {volume} {18}},\ \bibinfo {pages} {1} (\bibinfo {year}
  {1987})}\BibitemShut {NoStop}%
\bibitem [{\citenamefont {Towner}(1987)}]{Tow87}%
  \BibitemOpen
  \bibfield  {author} {\bibinfo {author} {\bibfnamefont {I.~S.}\ \bibnamefont
  {Towner}},\ }\href@noop {} {\bibfield  {journal} {\bibinfo  {journal} {Phys.
  Rep.}\ }\textbf {\bibinfo {volume} {155}},\ \bibinfo {pages} {263} (\bibinfo
  {year} {1987})}\BibitemShut {NoStop}%
\bibitem [{\citenamefont {Gysbers}\ \emph {et~al.}(2019)\citenamefont
  {Gysbers}, \citenamefont {Hagen}, \citenamefont {Holt}, \citenamefont
  {Jansen}, \citenamefont {Morris}, \citenamefont {Navrátil}, \citenamefont
  {Papenbrock}, \citenamefont {Quaglioni}, \citenamefont {Schwenk},
  \citenamefont {Stroberg},\ and\ \citenamefont {Wendt}}]{Gys19}%
  \BibitemOpen
  \bibfield  {author} {\bibinfo {author} {\bibfnamefont {P.}~\bibnamefont
  {Gysbers}}, \bibinfo {author} {\bibfnamefont {G.}~\bibnamefont {Hagen}},
  \bibinfo {author} {\bibfnamefont {J.~D.}\ \bibnamefont {Holt}}, \bibinfo
  {author} {\bibfnamefont {G.~R.}\ \bibnamefont {Jansen}}, \bibinfo {author}
  {\bibfnamefont {T.~D.}\ \bibnamefont {Morris}}, \bibinfo {author}
  {\bibfnamefont {P.}~\bibnamefont {Navrátil}}, \bibinfo {author}
  {\bibfnamefont {T.}~\bibnamefont {Papenbrock}}, \bibinfo {author}
  {\bibfnamefont {S.}~\bibnamefont {Quaglioni}}, \bibinfo {author}
  {\bibfnamefont {A.}~\bibnamefont {Schwenk}}, \bibinfo {author} {\bibfnamefont
  {S.~R.}\ \bibnamefont {Stroberg}},\ and\ \bibinfo {author} {\bibfnamefont
  {K.~A.}\ \bibnamefont {Wendt}},\ }\href@noop {} {\bibfield  {journal}
  {\bibinfo  {journal} {Nat. Phys.}\ }\textbf {\bibinfo {volume} {15}},\
  \bibinfo {pages} {428} (\bibinfo {year} {2019})}\BibitemShut {NoStop}%
\bibitem [{\citenamefont {Coraggio}\ \emph {et~al.}(2024)\citenamefont
  {Coraggio}, \citenamefont {Itaco}, \citenamefont {Gregorio}, \citenamefont
  {Gargano}, \citenamefont {Cheng}, \citenamefont {Ma}, \citenamefont {Xu},\
  and\ \citenamefont {Viviani}}]{Cor24}%
  \BibitemOpen
  \bibfield  {author} {\bibinfo {author} {\bibfnamefont {L.}~\bibnamefont
  {Coraggio}}, \bibinfo {author} {\bibfnamefont {N.}~\bibnamefont {Itaco}},
  \bibinfo {author} {\bibfnamefont {G.~D.}\ \bibnamefont {Gregorio}}, \bibinfo
  {author} {\bibfnamefont {A.}~\bibnamefont {Gargano}}, \bibinfo {author}
  {\bibfnamefont {Z.~H.}\ \bibnamefont {Cheng}}, \bibinfo {author}
  {\bibfnamefont {Y.~Z.}\ \bibnamefont {Ma}}, \bibinfo {author} {\bibfnamefont
  {F.~R.}\ \bibnamefont {Xu}},\ and\ \bibinfo {author} {\bibfnamefont
  {M.}~\bibnamefont {Viviani}},\ }\href@noop {} {\bibfield  {journal} {\bibinfo
   {journal} {Phys. Rev. C}\ }\textbf {\bibinfo {volume} {109}},\ \bibinfo
  {pages} {014301} (\bibinfo {year} {2024})}\BibitemShut {NoStop}%
\bibitem [{\citenamefont {Castillo}\ \emph {et~al.}()\citenamefont {Castillo},
  \citenamefont {Jokiniemi}, \citenamefont {Soriano},\ and\ \citenamefont
  {Men\'{e}ndez}}]{Cas24}%
  \BibitemOpen
  \bibfield  {author} {\bibinfo {author} {\bibfnamefont {D.}~\bibnamefont
  {Castillo}}, \bibinfo {author} {\bibfnamefont {L.}~\bibnamefont {Jokiniemi}},
  \bibinfo {author} {\bibfnamefont {P.}~\bibnamefont {Soriano}},\ and\ \bibinfo
  {author} {\bibfnamefont {J.}~\bibnamefont {Men\'{e}ndez}},\ }\href@noop {}
  {}\Eprint {https://arxiv.org/abs/2408.03373 (2024)} {arXiv:2408.03373 (2024)}
  \BibitemShut {NoStop}%
\bibitem [{\citenamefont {Belley}\ \emph {et~al.}(2021)\citenamefont {Belley},
  \citenamefont {Payne}, \citenamefont {Stroberg}, \citenamefont {Miyagi},\
  and\ \citenamefont {Holt}}]{Bel21}%
  \BibitemOpen
  \bibfield  {author} {\bibinfo {author} {\bibfnamefont {A.}~\bibnamefont
  {Belley}}, \bibinfo {author} {\bibfnamefont {C.~G.}\ \bibnamefont {Payne}},
  \bibinfo {author} {\bibfnamefont {S.~R.}\ \bibnamefont {Stroberg}}, \bibinfo
  {author} {\bibfnamefont {T.}~\bibnamefont {Miyagi}},\ and\ \bibinfo {author}
  {\bibfnamefont {J.~D.}\ \bibnamefont {Holt}},\ }\href@noop {} {\bibfield
  {journal} {\bibinfo  {journal} {Phys. Rev. Lett.}\ }\textbf {\bibinfo
  {volume} {126}},\ \bibinfo {pages} {042502} (\bibinfo {year}
  {2021})}\BibitemShut {NoStop}%
\bibitem [{\citenamefont {Cirigliano}\ \emph {et~al.}(2019)\citenamefont
  {Cirigliano}, \citenamefont {Dekens}, \citenamefont {de~Vries}, \citenamefont
  {Graesser}, \citenamefont {Mereghetti}, \citenamefont {Pastore},
  \citenamefont {Piarulli}, \citenamefont {van Kolck},\ and\ \citenamefont
  {Wiringa}}]{Cir19}%
  \BibitemOpen
  \bibfield  {author} {\bibinfo {author} {\bibfnamefont {V.}~\bibnamefont
  {Cirigliano}}, \bibinfo {author} {\bibfnamefont {W.}~\bibnamefont {Dekens}},
  \bibinfo {author} {\bibfnamefont {J.}~\bibnamefont {de~Vries}}, \bibinfo
  {author} {\bibfnamefont {M.~L.}\ \bibnamefont {Graesser}}, \bibinfo {author}
  {\bibfnamefont {E.}~\bibnamefont {Mereghetti}}, \bibinfo {author}
  {\bibfnamefont {S.}~\bibnamefont {Pastore}}, \bibinfo {author} {\bibfnamefont
  {M.}~\bibnamefont {Piarulli}}, \bibinfo {author} {\bibfnamefont
  {U.}~\bibnamefont {van Kolck}},\ and\ \bibinfo {author} {\bibfnamefont
  {R.~B.}\ \bibnamefont {Wiringa}},\ }\href@noop {} {\bibfield  {journal}
  {\bibinfo  {journal} {Phys. Rev. C}\ }\textbf {\bibinfo {volume} {100}},\
  \bibinfo {pages} {055504} (\bibinfo {year} {2019})}\BibitemShut {NoStop}%
\bibitem [{\citenamefont {Vergados}(2002)}]{Ver02}%
  \BibitemOpen
  \bibfield  {author} {\bibinfo {author} {\bibfnamefont {J.~D.}\ \bibnamefont
  {Vergados}},\ }\href@noop {} {\bibfield  {journal} {\bibinfo  {journal}
  {Phys. Rep.}\ }\textbf {\bibinfo {volume} {361}},\ \bibinfo {pages} {1}
  (\bibinfo {year} {2002})}\BibitemShut {NoStop}%
\bibitem [{\citenamefont {Cirigliano}\ \emph {et~al.}(2018)\citenamefont
  {Cirigliano}, \citenamefont {Dekens}, \citenamefont {Mereghetti},\ and\
  \citenamefont {A.Walker-Loud}}]{Cir18}%
  \BibitemOpen
  \bibfield  {author} {\bibinfo {author} {\bibfnamefont {V.}~\bibnamefont
  {Cirigliano}}, \bibinfo {author} {\bibfnamefont {W.}~\bibnamefont {Dekens}},
  \bibinfo {author} {\bibfnamefont {E.}~\bibnamefont {Mereghetti}},\ and\
  \bibinfo {author} {\bibnamefont {A.Walker-Loud}},\ }\href@noop {} {\bibfield
  {journal} {\bibinfo  {journal} {Phys. Rev. C}\ }\textbf {\bibinfo {volume}
  {97}},\ \bibinfo {pages} {065501} (\bibinfo {year} {2018})}\BibitemShut
  {NoStop}%
\bibitem [{\citenamefont {Commins}\ and\ \citenamefont
  {Bucksbaum}(1983)}]{Com83}%
  \BibitemOpen
  \bibfield  {author} {\bibinfo {author} {\bibfnamefont {E.~D.}\ \bibnamefont
  {Commins}}\ and\ \bibinfo {author} {\bibfnamefont {P.~H.}\ \bibnamefont
  {Bucksbaum}},\ }\href@noop {} {\emph {\bibinfo {title} {Weak Interactions of
  Leptons and Quarks}}}\ (\bibinfo  {publisher} {Cambridge Univ. Press},\
  \bibinfo {address} {Cambridge},\ \bibinfo {year} {1983})\BibitemShut
  {NoStop}%
\bibitem [{\citenamefont {Branco}\ and\ \citenamefont {Rebelo}(2009)}]{Bra09}%
  \BibitemOpen
  \bibfield  {author} {\bibinfo {author} {\bibfnamefont {G.~C.}\ \bibnamefont
  {Branco}}\ and\ \bibinfo {author} {\bibfnamefont {M.~N.}\ \bibnamefont
  {Rebelo}},\ }\href@noop {} {\bibfield  {journal} {\bibinfo  {journal} {Phys.
  Rev. D}\ }\textbf {\bibinfo {volume} {79}},\ \bibinfo {pages} {013001}
  (\bibinfo {year} {2009})}\BibitemShut {NoStop}%
\bibitem [{\citenamefont {Horoi}\ and\ \citenamefont {Stoica}(2010)}]{Hor10}%
  \BibitemOpen
  \bibfield  {author} {\bibinfo {author} {\bibfnamefont {M.}~\bibnamefont
  {Horoi}}\ and\ \bibinfo {author} {\bibfnamefont {S.}~\bibnamefont {Stoica}},\
  }\href@noop {} {\bibfield  {journal} {\bibinfo  {journal} {Phys. Rev. C}\
  }\textbf {\bibinfo {volume} {81}},\ \bibinfo {pages} {024321} (\bibinfo
  {year} {2010})}\BibitemShut {NoStop}%
\bibitem [{\citenamefont {\v{S}imkovic}\ \emph {et~al.}(2011)\citenamefont
  {\v{S}imkovic}, \citenamefont {Hod\'{a}k}, \citenamefont {Faessler},\ and\
  \citenamefont {Vogel}}]{Sim11}%
  \BibitemOpen
  \bibfield  {author} {\bibinfo {author} {\bibfnamefont {F.}~\bibnamefont
  {\v{S}imkovic}}, \bibinfo {author} {\bibfnamefont {R.}~\bibnamefont
  {Hod\'{a}k}}, \bibinfo {author} {\bibfnamefont {A.}~\bibnamefont
  {Faessler}},\ and\ \bibinfo {author} {\bibfnamefont {P.}~\bibnamefont
  {Vogel}},\ }\href@noop {} {\bibfield  {journal} {\bibinfo  {journal} {Phys.
  Rev. C}\ }\textbf {\bibinfo {volume} {83}},\ \bibinfo {pages} {015502}
  (\bibinfo {year} {2011})}\BibitemShut {NoStop}%
\bibitem [{\citenamefont {Kotila}\ and\ \citenamefont
  {Iachello}(2012)}]{Kot12}%
  \BibitemOpen
  \bibfield  {author} {\bibinfo {author} {\bibfnamefont {J.}~\bibnamefont
  {Kotila}}\ and\ \bibinfo {author} {\bibfnamefont {F.}~\bibnamefont
  {Iachello}},\ }\href@noop {} {\bibfield  {journal} {\bibinfo  {journal}
  {Phys. Rev. C}\ }\textbf {\bibinfo {volume} {85}},\ \bibinfo {pages} {034316}
  (\bibinfo {year} {2012})}\BibitemShut {NoStop}%
\bibitem [{\citenamefont {\v{S}imkovic}\ \emph {et~al.}(1999)\citenamefont
  {\v{S}imkovic}, \citenamefont {Pantis}, \citenamefont {Vergados},\ and\
  \citenamefont {Faessler}}]{Sim99}%
  \BibitemOpen
  \bibfield  {author} {\bibinfo {author} {\bibfnamefont {F.}~\bibnamefont
  {\v{S}imkovic}}, \bibinfo {author} {\bibfnamefont {G.}~\bibnamefont
  {Pantis}}, \bibinfo {author} {\bibfnamefont {J.~D.}\ \bibnamefont
  {Vergados}},\ and\ \bibinfo {author} {\bibfnamefont {A.}~\bibnamefont
  {Faessler}},\ }\href@noop {} {\bibfield  {journal} {\bibinfo  {journal}
  {Phys. Rev. C}\ }\textbf {\bibinfo {volume} {60}},\ \bibinfo {pages} {055502}
  (\bibinfo {year} {1999})}\BibitemShut {NoStop}%
\bibitem [{\citenamefont {Ring}\ and\ \citenamefont {Schuck}(1980)}]{Rin80}%
  \BibitemOpen
  \bibfield  {author} {\bibinfo {author} {\bibfnamefont {P.}~\bibnamefont
  {Ring}}\ and\ \bibinfo {author} {\bibfnamefont {P.}~\bibnamefont {Schuck}},\
  }\href@noop {} {\emph {\bibinfo {title} {The Nuclear Many-body Problem}}}\
  (\bibinfo  {publisher} {Springer-Verlag},\ \bibinfo {address} {Berlin},\
  \bibinfo {year} {1980})\BibitemShut {NoStop}%
\bibitem [{\citenamefont {Terasaki}(2013)}]{Ter13}%
  \BibitemOpen
  \bibfield  {author} {\bibinfo {author} {\bibfnamefont {J.}~\bibnamefont
  {Terasaki}},\ }\href@noop {} {\bibfield  {journal} {\bibinfo  {journal}
  {Phys. Rev. C}\ }\textbf {\bibinfo {volume} {87}},\ \bibinfo {pages} {024316}
  (\bibinfo {year} {2013})}\BibitemShut {NoStop}%
\bibitem [{\citenamefont {Terasaki}\ \emph {et~al.}(2005)\citenamefont
  {Terasaki}, \citenamefont {Engel}, \citenamefont {Bender}, \citenamefont
  {Dobaczewski}, \citenamefont {Nazarewicz},\ and\ \citenamefont
  {Stoitsov}}]{Ter05}%
  \BibitemOpen
  \bibfield  {author} {\bibinfo {author} {\bibfnamefont {J.}~\bibnamefont
  {Terasaki}}, \bibinfo {author} {\bibfnamefont {J.}~\bibnamefont {Engel}},
  \bibinfo {author} {\bibfnamefont {M.}~\bibnamefont {Bender}}, \bibinfo
  {author} {\bibfnamefont {J.}~\bibnamefont {Dobaczewski}}, \bibinfo {author}
  {\bibfnamefont {W.}~\bibnamefont {Nazarewicz}},\ and\ \bibinfo {author}
  {\bibfnamefont {M.}~\bibnamefont {Stoitsov}},\ }\href@noop {} {\bibfield
  {journal} {\bibinfo  {journal} {Phys. Rev. C}\ }\textbf {\bibinfo {volume}
  {71}},\ \bibinfo {pages} {034310} (\bibinfo {year} {2005})}\BibitemShut
  {NoStop}%
\bibitem [{\citenamefont {Terasaki}(2020)}]{Ter20}%
  \BibitemOpen
  \bibfield  {author} {\bibinfo {author} {\bibfnamefont {J.}~\bibnamefont
  {Terasaki}},\ }\href@noop {} {\bibfield  {journal} {\bibinfo  {journal}
  {Phys. Rev. C}\ }\textbf {\bibinfo {volume} {102}},\ \bibinfo {pages}
  {044303} (\bibinfo {year} {2020})}\BibitemShut {NoStop}%
\bibitem [{\citenamefont {Bartel}\ \emph {et~al.}(1982)\citenamefont {Bartel},
  \citenamefont {Quentin}, \citenamefont {Brack}, \citenamefont {Guet},\ and\
  \citenamefont {H{\aa}kansson}}]{Bar82}%
  \BibitemOpen
  \bibfield  {author} {\bibinfo {author} {\bibfnamefont {J.}~\bibnamefont
  {Bartel}}, \bibinfo {author} {\bibfnamefont {P.}~\bibnamefont {Quentin}},
  \bibinfo {author} {\bibfnamefont {M.}~\bibnamefont {Brack}}, \bibinfo
  {author} {\bibfnamefont {C.}~\bibnamefont {Guet}},\ and\ \bibinfo {author}
  {\bibfnamefont {H.-B.}\ \bibnamefont {H{\aa}kansson}},\ }\href@noop {}
  {\bibfield  {journal} {\bibinfo  {journal} {Nucl. Phys. A}\ }\textbf
  {\bibinfo {volume} {386}},\ \bibinfo {pages} {79} (\bibinfo {year}
  {1982})}\BibitemShut {NoStop}%
\bibitem [{\citenamefont {\hbox{Van Giai}}\ and\ \citenamefont
  {Sagawa}(1981)}]{Gia81}%
  \BibitemOpen
  \bibfield  {author} {\bibinfo {author} {\bibfnamefont {N.}~\bibnamefont
  {\hbox{Van Giai}}}\ and\ \bibinfo {author} {\bibfnamefont {H.}~\bibnamefont
  {Sagawa}},\ }\href@noop {} {\bibfield  {journal} {\bibinfo  {journal} {Phys.
  Lett. B}\ }\textbf {\bibinfo {volume} {106}},\ \bibinfo {pages} {379}
  (\bibinfo {year} {1981})}\BibitemShut {NoStop}%
\bibitem [{\citenamefont {Ter\'{a}n}\ \emph {et~al.}(2003)\citenamefont
  {Ter\'{a}n}, \citenamefont {Oberacker},\ and\ \citenamefont {Umar}}]{Ter03}%
  \BibitemOpen
  \bibfield  {author} {\bibinfo {author} {\bibfnamefont {E.}~\bibnamefont
  {Ter\'{a}n}}, \bibinfo {author} {\bibfnamefont {V.~E.}\ \bibnamefont
  {Oberacker}},\ and\ \bibinfo {author} {\bibfnamefont {A.~S.}\ \bibnamefont
  {Umar}},\ }\href@noop {} {\bibfield  {journal} {\bibinfo  {journal} {Phys.
  Rev. C}\ }\textbf {\bibinfo {volume} {67}},\ \bibinfo {pages} {064314}
  (\bibinfo {year} {2003})}\BibitemShut {NoStop}%
\bibitem [{\citenamefont {Blazkiewicz}\ \emph {et~al.}(2005)\citenamefont
  {Blazkiewicz}, \citenamefont {Oberacker}, \citenamefont {Umar},\ and\
  \citenamefont {Stoitsov}}]{Bla05}%
  \BibitemOpen
  \bibfield  {author} {\bibinfo {author} {\bibfnamefont {A.}~\bibnamefont
  {Blazkiewicz}}, \bibinfo {author} {\bibfnamefont {V.~E.}\ \bibnamefont
  {Oberacker}}, \bibinfo {author} {\bibfnamefont {A.~S.}\ \bibnamefont
  {Umar}},\ and\ \bibinfo {author} {\bibfnamefont {M.}~\bibnamefont
  {Stoitsov}},\ }\href@noop {} {\bibfield  {journal} {\bibinfo  {journal}
  {Phys. Rev. C}\ }\textbf {\bibinfo {volume} {71}},\ \bibinfo {pages} {054321}
  (\bibinfo {year} {2005})}\BibitemShut {NoStop}%
\bibitem [{\citenamefont {Oberacker}\ \emph {et~al.}(2007)\citenamefont
  {Oberacker}, \citenamefont {Blazkiewicz},\ and\ \citenamefont
  {Umar}}]{Obe07}%
  \BibitemOpen
  \bibfield  {author} {\bibinfo {author} {\bibfnamefont {V.~E.}\ \bibnamefont
  {Oberacker}}, \bibinfo {author} {\bibfnamefont {A.}~\bibnamefont
  {Blazkiewicz}},\ and\ \bibinfo {author} {\bibfnamefont {A.~S.}\ \bibnamefont
  {Umar}},\ }\href@noop {} {\bibfield  {journal} {\bibinfo  {journal} {Romanian
  Rep. Phys.}\ }\textbf {\bibinfo {volume} {59}},\ \bibinfo {pages} {559}
  (\bibinfo {year} {2007})}\BibitemShut {NoStop}%
\bibitem [{\citenamefont {Terasaki}\ and\ \citenamefont {Engel}(2010)}]{Ter10}%
  \BibitemOpen
  \bibfield  {author} {\bibinfo {author} {\bibfnamefont {J.}~\bibnamefont
  {Terasaki}}\ and\ \bibinfo {author} {\bibfnamefont {J.}~\bibnamefont
  {Engel}},\ }\href@noop {} {\bibfield  {journal} {\bibinfo  {journal} {Phys.
  Rev. C}\ }\textbf {\bibinfo {volume} {82}},\ \bibinfo {pages} {034326}
  (\bibinfo {year} {2010})}\BibitemShut {NoStop}%
\bibitem [{\citenamefont {Brown}\ and\ \citenamefont
  {Wildenthal}(1985)}]{Bro85}%
  \BibitemOpen
  \bibfield  {author} {\bibinfo {author} {\bibfnamefont {B.~A.}\ \bibnamefont
  {Brown}}\ and\ \bibinfo {author} {\bibfnamefont {B.~H.}\ \bibnamefont
  {Wildenthal}},\ }\href@noop {} {\bibfield  {journal} {\bibinfo  {journal}
  {Atom. Data and Nucl. Data Tab.}\ }\textbf {\bibinfo {volume} {33}},\
  \bibinfo {pages} {347} (\bibinfo {year} {1985})}\BibitemShut {NoStop}%
\bibitem [{\citenamefont {\v{S}imkovic}\ \emph
  {et~al.}(2018{\natexlab{a}})\citenamefont {\v{S}imkovic}, \citenamefont
  {Dvornick\'{y}}, \citenamefont {\v{S}tef\'{a}nik},\ and\ \citenamefont
  {Faessler}}]{Sim18}%
  \BibitemOpen
  \bibfield  {author} {\bibinfo {author} {\bibfnamefont {F.}~\bibnamefont
  {\v{S}imkovic}}, \bibinfo {author} {\bibfnamefont {R.}~\bibnamefont
  {Dvornick\'{y}}}, \bibinfo {author} {\bibfnamefont {D.}~\bibnamefont
  {\v{S}tef\'{a}nik}},\ and\ \bibinfo {author} {\bibfnamefont {A.}~\bibnamefont
  {Faessler}},\ }\href@noop {} {\bibfield  {journal} {\bibinfo  {journal}
  {Phys. Rev. C}\ }\textbf {\bibinfo {volume} {97}},\ \bibinfo {pages} {034315}
  (\bibinfo {year} {2018}{\natexlab{a}})}\BibitemShut {NoStop}%
\bibitem [{\citenamefont {Gando}\ \emph {et~al.}(2019)\citenamefont {Gando},
  \citenamefont {Gando}, \citenamefont {Hachiya}, \citenamefont {\hbox{Ha
  Minh}}, \citenamefont {Hayashida}, \citenamefont {Honda}, \citenamefont
  {Hosokawa}, \citenamefont {Ikeda}, \citenamefont {Inoue}, \citenamefont
  {Ishidoshiro} \emph {et~al.}}]{Gan19}%
  \BibitemOpen
  \bibfield  {author} {\bibinfo {author} {\bibfnamefont {A.}~\bibnamefont
  {Gando}}, \bibinfo {author} {\bibfnamefont {Y.}~\bibnamefont {Gando}},
  \bibinfo {author} {\bibfnamefont {T.}~\bibnamefont {Hachiya}}, \bibinfo
  {author} {\bibfnamefont {M.}~\bibnamefont {\hbox{Ha Minh}}}, \bibinfo
  {author} {\bibfnamefont {S.}~\bibnamefont {Hayashida}}, \bibinfo {author}
  {\bibfnamefont {Y.}~\bibnamefont {Honda}}, \bibinfo {author} {\bibfnamefont
  {K.}~\bibnamefont {Hosokawa}}, \bibinfo {author} {\bibfnamefont
  {H.}~\bibnamefont {Ikeda}}, \bibinfo {author} {\bibfnamefont
  {K.}~\bibnamefont {Inoue}}, \bibinfo {author} {\bibfnamefont
  {K.}~\bibnamefont {Ishidoshiro}}, \emph {et~al.} (\bibinfo {collaboration}
  {KamLAND-Zen Collaboration}),\ }\href@noop {} {\bibfield  {journal} {\bibinfo
   {journal} {Phys. Rev. Lett.}\ }\textbf {\bibinfo {volume} {122}},\ \bibinfo
  {pages} {192501} (\bibinfo {year} {2019})}\BibitemShut {NoStop}%
\bibitem [{nnd(2025)}]{nndc25}%
  \BibitemOpen
  \href@noop {} {}\bibinfo {howpublished} {National Nuclear Data Center,
  Brookhaven National Laboratory, \url{ http://www.nndc.bnl.gov }} (\bibinfo
  {year} {2025})\BibitemShut {NoStop}%
\bibitem [{\citenamefont {Men\'{e}ndez}(2018)}]{Men18}%
  \BibitemOpen
  \bibfield  {author} {\bibinfo {author} {\bibfnamefont {J.}~\bibnamefont
  {Men\'{e}ndez}},\ }\href@noop {} {\bibfield  {journal} {\bibinfo  {journal}
  {J. Phys. G}\ }\textbf {\bibinfo {volume} {45}},\ \bibinfo {pages} {014003}
  (\bibinfo {year} {2018})}\BibitemShut {NoStop}%
\bibitem [{\citenamefont {Horoi}\ and\ \citenamefont {Neacsu}(2016)}]{Hor16}%
  \BibitemOpen
  \bibfield  {author} {\bibinfo {author} {\bibfnamefont {M.}~\bibnamefont
  {Horoi}}\ and\ \bibinfo {author} {\bibfnamefont {A.}~\bibnamefont {Neacsu}},\
  }\href@noop {} {\bibfield  {journal} {\bibinfo  {journal} {Phys. Rev. C}\
  }\textbf {\bibinfo {volume} {93}},\ \bibinfo {pages} {024308} (\bibinfo
  {year} {2016})}\BibitemShut {NoStop}%
\bibitem [{\citenamefont {Coraggio}\ \emph {et~al.}(2020)\citenamefont
  {Coraggio}, \citenamefont {Gargano}, \citenamefont {Itaco}, \citenamefont
  {Mancino},\ and\ \citenamefont {Nowacki}}]{Cor20b}%
  \BibitemOpen
  \bibfield  {author} {\bibinfo {author} {\bibfnamefont {L.}~\bibnamefont
  {Coraggio}}, \bibinfo {author} {\bibfnamefont {A.}~\bibnamefont {Gargano}},
  \bibinfo {author} {\bibfnamefont {N.}~\bibnamefont {Itaco}}, \bibinfo
  {author} {\bibfnamefont {R.}~\bibnamefont {Mancino}},\ and\ \bibinfo {author}
  {\bibfnamefont {F.}~\bibnamefont {Nowacki}},\ }\href@noop {} {\bibfield
  {journal} {\bibinfo  {journal} {Phys. Rev. C}\ }\textbf {\bibinfo {volume}
  {101}},\ \bibinfo {pages} {044315} (\bibinfo {year} {2020})}\BibitemShut
  {NoStop}%
\bibitem [{\citenamefont {Coraggio}\ \emph {et~al.}(2022)\citenamefont
  {Coraggio}, \citenamefont {Itaco}, \citenamefont {Gregorio}, \citenamefont
  {Gargano}, \citenamefont {Mancino},\ and\ \citenamefont {Nowacki}}]{Cor22}%
  \BibitemOpen
  \bibfield  {author} {\bibinfo {author} {\bibfnamefont {L.}~\bibnamefont
  {Coraggio}}, \bibinfo {author} {\bibfnamefont {N.}~\bibnamefont {Itaco}},
  \bibinfo {author} {\bibfnamefont {G.~D.}\ \bibnamefont {Gregorio}}, \bibinfo
  {author} {\bibfnamefont {A.}~\bibnamefont {Gargano}}, \bibinfo {author}
  {\bibfnamefont {R.}~\bibnamefont {Mancino}},\ and\ \bibinfo {author}
  {\bibfnamefont {F.}~\bibnamefont {Nowacki}},\ }\href@noop {} {\bibfield
  {journal} {\bibinfo  {journal} {Phys. Rev. C}\ }\textbf {\bibinfo {volume}
  {105}},\ \bibinfo {pages} {034312} (\bibinfo {year} {2022})}\BibitemShut
  {NoStop}%
\bibitem [{\citenamefont {Mustonen}\ and\ \citenamefont {Engel}(2013)}]{Mus13}%
  \BibitemOpen
  \bibfield  {author} {\bibinfo {author} {\bibfnamefont {M.~T.}\ \bibnamefont
  {Mustonen}}\ and\ \bibinfo {author} {\bibfnamefont {J.}~\bibnamefont
  {Engel}},\ }\href@noop {} {\bibfield  {journal} {\bibinfo  {journal} {Phys.
  Rev}\ }\textbf {\bibinfo {volume} {87}},\ \bibinfo {pages} {064302} (\bibinfo
  {year} {2013})}\BibitemShut {NoStop}%
\bibitem [{\citenamefont {Hyv{\"{a}}rinen}\ and\ \citenamefont
  {Suhonen}(2015)}]{Hyv15}%
  \BibitemOpen
  \bibfield  {author} {\bibinfo {author} {\bibfnamefont {J.}~\bibnamefont
  {Hyv{\"{a}}rinen}}\ and\ \bibinfo {author} {\bibfnamefont {J.}~\bibnamefont
  {Suhonen}},\ }\href@noop {} {\bibfield  {journal} {\bibinfo  {journal} {Phys.
  Rev. C}\ }\textbf {\bibinfo {volume} {91}},\ \bibinfo {pages} {024613}
  (\bibinfo {year} {2015})}\BibitemShut {NoStop}%
\bibitem [{\citenamefont {\v{S}imkovic}\ \emph
  {et~al.}(2018{\natexlab{b}})\citenamefont {\v{S}imkovic}, \citenamefont
  {Smetana},\ and\ \citenamefont {Vogel}}]{Sim18b}%
  \BibitemOpen
  \bibfield  {author} {\bibinfo {author} {\bibfnamefont {F.}~\bibnamefont
  {\v{S}imkovic}}, \bibinfo {author} {\bibfnamefont {A.}~\bibnamefont
  {Smetana}},\ and\ \bibinfo {author} {\bibfnamefont {P.}~\bibnamefont
  {Vogel}},\ }\href@noop {} {\bibfield  {journal} {\bibinfo  {journal} {Phys.
  Rev. C}\ }\textbf {\bibinfo {volume} {98}},\ \bibinfo {pages} {064325}
  (\bibinfo {year} {2018}{\natexlab{b}})}\BibitemShut {NoStop}%
\bibitem [{\citenamefont {Fang}\ \emph {et~al.}(2018)\citenamefont {Fang},
  \citenamefont {Faessler},\ and\ \citenamefont {\v{S}imkovic}}]{Fan18}%
  \BibitemOpen
  \bibfield  {author} {\bibinfo {author} {\bibfnamefont {D.-L.}\ \bibnamefont
  {Fang}}, \bibinfo {author} {\bibfnamefont {A.}~\bibnamefont {Faessler}},\
  and\ \bibinfo {author} {\bibfnamefont {F.}~\bibnamefont {\v{S}imkovic}},\
  }\href@noop {} {\bibfield  {journal} {\bibinfo  {journal} {Phys. Rev. C}\
  }\textbf {\bibinfo {volume} {97}},\ \bibinfo {pages} {045503} (\bibinfo
  {year} {2018})}\BibitemShut {NoStop}%
\bibitem [{\citenamefont {Rodr\'{i}guez}\ and\ \citenamefont
  {Mart\'{i}nez-Pinedo}(2010)}]{Rod10}%
  \BibitemOpen
  \bibfield  {author} {\bibinfo {author} {\bibfnamefont {T.~R.}\ \bibnamefont
  {Rodr\'{i}guez}}\ and\ \bibinfo {author} {\bibfnamefont {G.}~\bibnamefont
  {Mart\'{i}nez-Pinedo}},\ }\href@noop {} {\bibfield  {journal} {\bibinfo
  {journal} {Phys. Rev. Lett.}\ }\textbf {\bibinfo {volume} {105}},\ \bibinfo
  {pages} {252503} (\bibinfo {year} {2010})}\BibitemShut {NoStop}%
\bibitem [{\citenamefont {Vaquero}\ \emph {et~al.}(2013)\citenamefont
  {Vaquero}, \citenamefont {Rodr{\'i}guez},\ and\ \citenamefont
  {Egido}}]{Vaq13}%
  \BibitemOpen
  \bibfield  {author} {\bibinfo {author} {\bibfnamefont {N.~L.}\ \bibnamefont
  {Vaquero}}, \bibinfo {author} {\bibfnamefont {T.~R.}\ \bibnamefont
  {Rodr{\'i}guez}},\ and\ \bibinfo {author} {\bibfnamefont {J.~L.}\
  \bibnamefont {Egido}},\ }\href@noop {} {\bibfield  {journal} {\bibinfo
  {journal} {Phys. Rev. Lett.}\ }\textbf {\bibinfo {volume} {111}},\ \bibinfo
  {pages} {142501} (\bibinfo {year} {2013})}\BibitemShut {NoStop}%
\bibitem [{\citenamefont {Song}\ \emph {et~al.}(2017)\citenamefont {Song},
  \citenamefont {Yao}, \citenamefont {Ring},\ and\ \citenamefont
  {Meng}}]{Son17}%
  \BibitemOpen
  \bibfield  {author} {\bibinfo {author} {\bibfnamefont {L.~S.}\ \bibnamefont
  {Song}}, \bibinfo {author} {\bibfnamefont {J.~M.}\ \bibnamefont {Yao}},
  \bibinfo {author} {\bibfnamefont {P.}~\bibnamefont {Ring}},\ and\ \bibinfo
  {author} {\bibfnamefont {J.}~\bibnamefont {Meng}},\ }\href@noop {} {\bibfield
   {journal} {\bibinfo  {journal} {Phys. Rev. C}\ }\textbf {\bibinfo {volume}
  {95}},\ \bibinfo {pages} {024305} (\bibinfo {year} {2017})}\BibitemShut
  {NoStop}%
\bibitem [{\citenamefont {Barea}\ \emph {et~al.}(2015)\citenamefont {Barea},
  \citenamefont {Kotila},\ and\ \citenamefont {Iachello}}]{Bar15}%
  \BibitemOpen
  \bibfield  {author} {\bibinfo {author} {\bibfnamefont {J.}~\bibnamefont
  {Barea}}, \bibinfo {author} {\bibfnamefont {J.}~\bibnamefont {Kotila}},\ and\
  \bibinfo {author} {\bibfnamefont {F.}~\bibnamefont {Iachello}},\ }\href@noop
  {} {\bibfield  {journal} {\bibinfo  {journal} {Phys. Rev. C}\ }\textbf
  {\bibinfo {volume} {91}},\ \bibinfo {pages} {034304} (\bibinfo {year}
  {2015})}\BibitemShut {NoStop}%
\bibitem [{\citenamefont {Deppisch}\ \emph {et~al.}(2020)\citenamefont
  {Deppisch}, \citenamefont {Graf}, \citenamefont {Rodejohann},\ and\
  \citenamefont {Xu}}]{Dep20}%
  \BibitemOpen
  \bibfield  {author} {\bibinfo {author} {\bibfnamefont {F.~F.}\ \bibnamefont
  {Deppisch}}, \bibinfo {author} {\bibfnamefont {L.}~\bibnamefont {Graf}},
  \bibinfo {author} {\bibfnamefont {W.}~\bibnamefont {Rodejohann}},\ and\
  \bibinfo {author} {\bibfnamefont {X.-J.}\ \bibnamefont {Xu}},\ }\href@noop {}
  {\bibfield  {journal} {\bibinfo  {journal} {Phys. Rev. D}\ }\textbf {\bibinfo
  {volume} {102}},\ \bibinfo {pages} {051701(R)} (\bibinfo {year}
  {2020})}\BibitemShut {NoStop}%
\bibitem [{\citenamefont {Men\'{e}ndez}\ \emph {et~al.}(2011)\citenamefont
  {Men\'{e}ndez}, \citenamefont {Gazit},\ and\ \citenamefont
  {Schwenk}}]{Men11}%
  \BibitemOpen
  \bibfield  {author} {\bibinfo {author} {\bibfnamefont {J.}~\bibnamefont
  {Men\'{e}ndez}}, \bibinfo {author} {\bibfnamefont {D.}~\bibnamefont
  {Gazit}},\ and\ \bibinfo {author} {\bibfnamefont {A.}~\bibnamefont
  {Schwenk}},\ }\href@noop {} {\bibfield  {journal} {\bibinfo  {journal} {Phys.
  Rev. Lett.}\ }\textbf {\bibinfo {volume} {107}},\ \bibinfo {pages} {062501}
  (\bibinfo {year} {2011})}\BibitemShut {NoStop}%
\end{thebibliography}
%
\end{document}